\documentclass[prx,twocolumn,superscriptaddress,aps,nofootinbib,10pt,floatfix]{revtex4-2}
\usepackage{hyperref}
\usepackage{amsmath,amsfonts,amssymb,amsthm,bbm,graphicx,enumerate,times}
\usepackage{mathtools}
\usepackage{graphicx}
\usepackage{float}
\usepackage{xcolor}
\usepackage{tikz}
\usepackage{tikz-cd}
\usepackage{dsfont}
\usepackage{physics}
\usepackage{braket}
\UseRawInputEncoding
\usepackage{inputenc}

\def\tr{\operatorname{tr}}   
\newcommand{\id}{\mathds{1}}
\newcommand{\cA}{{\mathcal A}}
\newcommand{\cB}{{\mathcal B}}

\newcommand{\cH}{{\mathcal H}}

\newcommand{\cK}{{\mathcal K}}

\newcommand{\cM}{{\mathcal M}}
\newcommand{\cN}{{\mathcal N}}
\newcommand{\cO}{{\mathcal O}}
\newcommand{\cP}{{\mathcal P}}

\newcommand{\cS}{{\mathcal S}}

\newcommand{\cV}{{\mathcal V}}

\newcommand{\cX}{{\mathcal X}}

\newcommand{\bC}{{\mathbb C}}

\newcommand{\ii}{{\mathrm{i}\,}}

\def\a{\alpha}

\def\c{\chi}

\def\d{\delta}

\def\g{\gamma}
\def\G{\Gamma}

\def\l{\lambda}
\def\L{\Lambda}

\def\p{\phi}
\def\P{\Phi}

\def\Q{\Theta}

\def\W{\Omega}
\def\x{\xi}

\def\y{\psi}
\def\Y{\Psi}

\newcommand{\be}{\begin{equation}}
	\newcommand{\ee}{\end{equation}}
\newcommand{\ba}{\begin{align}}
	\newcommand{\ea}{\end{align}}

\definecolor{alex}{rgb}{.2,.7,.1}

\newtheorem{theorem}{Theorem}
\newtheorem{lemma}{Lemma}
\newtheorem{corollary}{Corollary}
\begin{document}
	
\title{On Infinite Tensor Networks, Complementary Recovery and Type II Factors}

\author{Wissam Chemissany}
\affiliation{%
David Rittenhouse Laboratory, University of Pennsylvania, Philadelphia, PA 19104, U.S.A. 
}%

\author{Elliott Gesteau}
\affiliation{%
 Division of Physics, Mathematics, and Astronomy,
  California Institute of Technology,
  Pasadena, CA 91125, U.S.A.
}%
\affiliation{%
Kavli Institute for Theoretical Physics, Santa Barbara, CA 93106, U.S.A.}%

\author{Alexander Jahn}
\affiliation{%
Department of Physics,
 Freie Universit\"at Berlin,
 14195 Berlin, Germany.
}%

\author{Daniel Murphy}
\affiliation{%
 Institute for Quantum Information and Matter,
  California Institute of Technology, Pasadena, CA 91125, U.S.A. 
}%

\author{Leo Shaposhnik}
\thanks{Primary and corresponding author, \href{mailto:leo.shaposhnik@fu-berlin.de}{leo.shaposhnik@fu-berlin.de}.}
\affiliation{%
Department of Physics,
 Freie Universit\"at Berlin,
 14195 Berlin, Germany. 
}%

	
\begin{abstract}
We initiate a study of local operator algebras at the boundary of infinite tensor networks, using the mathematical theory of inductive limits. In particular, we consider tensor networks in which each layer acts as a quantum code with complementary recovery, a property that features prominently in the bulk-to-boundary maps intrinsic to holographic quantum error-correcting codes. In this case, we decompose the limiting Hilbert space and the algebras of observables in a way that keeps track of the entanglement in the network. 
As a specific example, we describe this inductive limit for the holographic HaPPY code model and relate its algebraic and error-correction features.
We find that the local algebras in this model are given by the hyperfinite type II$_\infty$ factor. Next, we discuss other networks that build upon this framework and comment on a connection between type II factors and stabilizer circuits. We conclude with a discussion of MERA networks in which complementary recovery is broken. We argue that this breaking possibly permits a limiting type III von Neumann algebra, making them more suitable ans\"atze for approximating subregions of quantum field theories.
\end{abstract}

\maketitle

\section{Motivation and setting}

Tensor networks have become an ubiquitous tool in modern physics, ranging
from the description of ground states of many-body quantum-mechanical
systems and topological phases of matter \cite{Bridgeman:2016dhh,cirac2009renormalization} to the study of quantum
information aspects of holographic dualities
\cite{Swingle:2009bg,bao2017sitter, Bao:2019fpq, Bao:2015uaa, Jahn:2020ukq,
Miyaji:2016mxg, Brown:2019rox}.
Despite their success in describing physical systems, discussions of the
continuum limit of finite dimensional tensor networks have usually been limited to investigations of the
limiting correlation functions of local operators \cite{Pfeifer:2008jt,cirac2009renormalization} and a precise formulation of the continuum limit in terms of a concrete Hilbert space and operators
acting on it is often left implicit. Although some models of continuum tensor networks, such as continuous matrix product states (cMPS), can be understood in terms of a continuum limit \cite{Verstraete:2010ft} from a lattice-based model, other more heuristic models, such as the continuous multiscale entanglement renormalization ansatz (cMERA), do not directly correspond to such a limit \cite{Haegeman:2011uy}. Based on wavelet models, the convergence to a free quantum field theory of certain lattice models was shown in \cite{Witteveen:2019lsk, Stottmeister:2020ezd, Osborne:2021ppp}, where an explicit realization of such a limit in terms of a MERA circuit was given in \cite{Witteveen:2019lsk}.
However, tools to analyze the limits of more general tensor networks remain limited; in particular, the operator algebras of subsystems are poorly understood. 

In this work, we investigate such operator algebras in infinitely large tensor networks for a class of layered tensor networks that can be associated with quantum error-correcting codes, using tools from the theory of inductive systems and the description of the observables of the system using local algebras, borrowing the language of algebraic quantum field theory \cite{Haag:1996hvx}. 
The use of inductive limits as a mathematical tool to rigorously formulate the continuum limit of tensor networks has been inspired by analogous studies in the context of Banach spaces and operator algebras \cite{van2024convergence}. This perspective has direct parallels with the layered tensor networks discussed here, particularly in how local algebras and Hilbert spaces grow iteratively to form a limiting theory. Similar constructions were given in \cite{Osborne:2017woa, Kang:2019dfi, Gesteau:2020hoz,Gesteau:2022hss, Gesteau:2020rtg} which focused on the formulation of the limiting systems and assignment of the limiting Hilbert space but, with the exception of \cite{Kang:2019dfi}, did not discuss the type of the resulting algebras. Furthermore, in \cite{Morinelli:2020uea, Stottmeister:2022ptp, Stottmeister:2020ezd, Witteveen:2019lsk} similar methods were used to prove convergence of certain lattice systems of free quantum field theories based on wavelet models. In contrast to these studies aiming to obtain quantum field theories that always have local algebras of type III$_1$, we focus on tensor networks on finite-dimensional Hilbert spaces that implement holographic quantum error-correcting codes that a priori do not have to converge to a quantum field theory. In particular, our goal is to determine the type of local algebra for these networks from an entanglement-based perspective. We are able to compute the type because we restrict ourselves to a very specific class of networks, namely networks that implement quantum error-correcting codes with complementary recovery, also known as holographic quantum error-correcting codes \cite{almheiri2015bulk}. We will focus primarily on the Harlow-Pastawski-Preskill-Yoshida (HaPPY) code model, which achieves complementary recovery with a hyperbolic tensor network of \emph{perfect tensors} \cite{Pastawski:2015qua}. 
As we will describe in the following, this property gives us strong control over the structure of the state of the network during the iteration process and allows for a direct mapping to the standard form of hyperfinite factors, the \emph{Araki-Woods-Powers factors} \cite{Araki1968ACO,Powers1967RepresentationsOU}, a possibility that was not made manifest in earlier studies of limits of infinitely large instances of such codes \cite{Gesteau:2020hoz, Gesteau:2022hss, Gesteau:2020rtg}.

\begin{figure*}
    \includegraphics[width=0.9\textwidth]{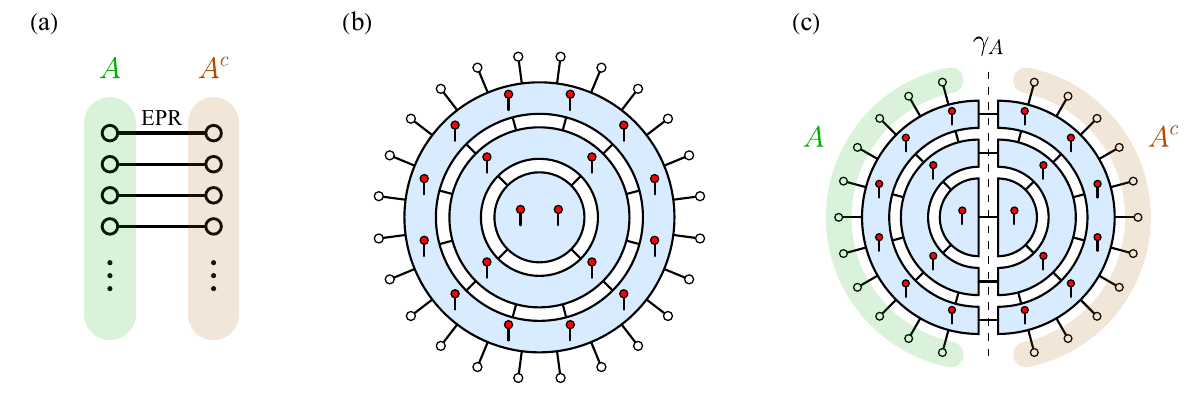}
    \caption{(a) In the Araki-Woods construction of a type II von Neumann algebra $\mathcal{A}$, one constructs an infinite series of maximally entangled pairs of qubits (EPR pairs), one side of which constitutes a subsystem $A$ (with complement $A^c$) on which operators in $\mathcal{A}$ act.
    (b) A layered tensor-network code forms an isometric map between bulk qubits (red) and boundary qubits (white). Layers can be added iteratively until both the number of bulk and boundary qubits become infinite. The tensor network contraction (black connecting lines) itself acts as a projection onto EPR pairs.
    (c) For a holographic tensor-network code with complementary recovery, a bipartition of the boundary qubits induces a clean bipartition of the bulk qubits along a Ryu-Takayanagi surface $\gamma_A$. Adding more layers increases EPR-like entanglement across $\gamma_A$, again ultimately leading to a type II von Neumann algebra for operators acting on $A$ in the limit of infinitely many layers, provided that boundary states contain only finite bulk entanglement.
    }
    \label{FIG_INTRO}
\end{figure*}

The main result of our work concerns the appearance of \emph{type II von Neumann algebras} associated to boundary regions of the infinitely large HaPPY code that contain infinite entanglement with their complementary region, but whose underlying entanglement pattern has the structure of maximally entangled \emph{Einstein-Podolski-Rosen} (EPR) pairs (Fig.\ \ref{FIG_INTRO}(a)). These algebras famously allow for the definition of a trace and reduced density matrices, notions which become ill-defined in systems with more complicated entanglement divergences, such as the type III algebras found in causally complete subregions of quantum field theories. As we shall show, type II algebras appear naturally in the scaling limits of layered tensor networks with a property known as \emph{complementary recovery}.
These were first considered in tensor networks that model holographic bulk/boundary dualities and act as encoding isometries of quantum error-correcting codes, known as \emph{holographic codes}
\cite{almheiri2015bulk,Pastawski:2015qua,Jahn:2021uqr,Pollack:2021yij}.
As visualized in Fig.\ \ref{FIG_INTRO}(b), holographic codes provide an isometric map from a bulk to a boundary Hilbert space, and we consider those with a layered structure such that the dimension of both Hilbert spaces diverges in the scaling limit of infinitely many layers.
Complementary recovery ensures that a bipartition of the ``physical'' boundary qubits into the $A$ and $A^c$ subregions induces a clean bipartition among the ``logical'' bulk qubits (Fig.\ \ref{FIG_INTRO}(c)).
Given this property, the tensor network also splits into two parts connected by a contraction (projection onto EPR pairs) that contributes to the entanglement between $A$ and $A^c$. By analogy with continuum holography, this tensor network cut is commonly referred to as a (discrete) \emph{Ryu-Takayanagi (RT) surface} \cite{Ryu:2006bv}. 
An operator-algebraic formulation of complementary recovery for the finite-dimensional type I setting has already been established in \cite{Harlow:2016vwg}, with the appearance of $C^\star$ and von Neumann algebras in the scaling limit studied in subsequent works \cite{Kang:2019dfi,Gesteau:2020rtg,Faulkner:2020hzi,Faulkner:2022ada,Gesteau:2023hbq}. The contribution of our work is to show the precise decomposition of these algebras in holographic codes in the inductive limit and how the ``geometrical'' entanglement in these models leads to a type II von Neumann algebra.


The starting point of our work are tensor networks with a layered structure, i.e., iterative maps 
\begin{equation}
    \ket{\Y^1} \mapsto \ket{\Y^2} \mapsto \ket{\Y^3} \mapsto \hdots
\end{equation}
between states of the network at different layers, each described by
finite dimensional Hilbert spaces 
\begin{equation}
    \cH^1 \rightarrow \cH^2 \rightarrow \cH^3 \rightarrow \hdots
\end{equation}
We assume that each of these Hilbert spaces spaces at layer $\L$ can be written as a bipartition 
\begin{equation}
\label{EQ_HILBERT_BIPART}
    \cH^\L = \cH^\L_A \otimes \cH^\L_{A^c}, 
\end{equation}
which can be thought of as a decomposition of the space into ``local'' subsystems. The sequences of networks are obtained by taking the network at layer $\L$
and contracting it with a new layer of the network, which we encode by an
isometric operator $\g^{\L,\L+1}: \cH^{\L} \rightarrow \cH^{\L+1}$, such that
\begin{equation}
\ket{\Y^{\L+1}} = \g^{\L,\L+1} \ket{\Y^{\L}}.
\end{equation}
$\g^{\L,\L+1}$ has to be isometric, so we map normalized states to
normalized states.
The collection of Hilbert spaces $\cH^\L$ together with the maps
$\g^{\L,\L+1}$ defined by the tensor network is called an \textit{inductive
system}. As described later, this allows one to define a limiting Hilbert
space $\cH$ that can intuitively be thought of as having a vacuum described by a reference state of the infinite tensor network, together with states that arise
from acting with a local operator on a subregion of the tensor network during
the iteration procedure.
Additionally, we demand of our network that, together with the isometry $\g^{\L,\L+1}$ that translates between states at different layers, we can identify a bounded 
operator $O^{\L} \in \cB(\cH^{\L})$ at layer $\L$ with an operator in the
next layer. We denote this identification as an operator-pushing map
$\phi^{\L,\L+1}$
\begin{equation}\label{eq:operator_pushing_1}
    O^{\L+1} = \phi^{\L,\L+1}(O^\L).
\end{equation}
and we demand that these are unital $*$-homomorphisms, i.e., they are linear maps mapping the identity to the identity that respect multiplication of operators and satisfy $\phi^{\L,\L+1}(O^\dagger) = \phi^{\L,\L+1}(O)^\dagger$, so that they preserve the algebraic structure.
Note that one might be tempted to implement $\phi^{\L,\L+1}$ by a conjugation with isometry $\g^{\L,\L+1}$, which may appear natural in the tensor network but will not allow for good reconstruction properties to study the inductive limit, as we shall see below.
Instead, we will define $\phi^{\L,\L+1}$ in the HaPPY code as a conjugation with a unitary 
\begin{equation}
    \phi^{\L,\L+1}(O) = U^\dagger (O\otimes \id) U^\dagger,
\end{equation}
where the unitary exists in an enlarged space on which the additional identity acts.\footnote{In the context of entanglement renormalization, such maps between layers are often referred to as \emph{ascending superoperators} \cite{Vidal.ERintro}.} 
We think of a subsystem at layer $\L$ as given by a subset of the legs $A$ of the tensor network state $\ket{\Y^{\L}}$ with decomposition \eqref{EQ_HILBERT_BIPART} and a global algebra 
\begin{equation}
    \cB(\cH^\L) = \cB(\cH_A^\L) \otimes \cB(\cH_{A^c}^\L),
\end{equation}
so that the operators belonging to $A$ are represented as elements of the local algebra
\begin{equation}
    \cA^\L_A = \cB(\cH_A^\L) \otimes \id_{\cH_{A^c}^\L}.
\end{equation}
Using the operator pushing map \eqref{eq:operator_pushing_1}, we can push the
whole algebra $\cA^\L_A$ to the next layer, thus also obtaining an
inductive system of algebras $(\cA^\L_A,\phi^{\L,\L+1})$. Therefore, we
can define its inductive limit $\cA_A$ which is an abstract $C^\star$-algebra
that defines the operators that are associated to $A$ in the limit $\L \rightarrow \infty$. This algebra can be represented on the limiting Hilbert space $\cH$
and the image can be completed to a von Neumann algebra. The collection of
all such von Neumann algebras then defines a \emph{net} of algebras $\cN$. We
define the limiting theory to then be given by the tuple $(\cH,\cN,\Y)$, where $\Y$ is the state of the infinite-dimensional tensor network that, given operators $O^\L$ that exist at layer $\L$, is defined by  
\begin{equation}
    \Y(O) := \braket{\Psi^{\L}|O^\L|\Y^\L}.   
\end{equation}
and by a limiting procedure for more general operators.\footnote{Note that this definition only makes sense if the operator pushing map is compatible with the isometry, so that the value is independent of the layer at which one evaluates the state. We comment further on this below.}
Performing this construction and an analysis of the local algebras for tensor networks that
represent holographic quantum error-correcting codes is the main goal of
this paper. In particular, we demonstrate that for the HaPPY code, the
von Neumann algebras associated to boundary subregions that satisfy complementary recovery are type II$_\infty$ factors.
We begin by providing a recap of von Neumann algebras and inductive limits in Sec.\ \ref{sec:Background}. In
Sec.\ \ref{sec:intuition_happy}, we take the inductive limit of the HaPPY code and show that II$_\infty$ factors emerge for subregions satisfying complementary recovery. Then in Sec.\ \ref{sec:inductive_limits} we provide a summary of our construction in the
general case and explain more generally how to determine the von Neumann algebra type. In Sec.\ \ref{sec:examples}, we discuss examples and generalizations of this construction, including other tensor networks such as the MERA and those based on Majorana dimers. We end with a general discussion in Sec.\
\ref{sec:discussion}.

\section{Background and preliminaries}\label{sec:Background}
We begin by laying the theoretical foundations for our analysis. This includes an overview of von Neumann algebras, tensor networks, and the basics of the theory of inductive limits.

 \subsection{A von Neumann algebra primer}

    In this section, we provide a brief introduction to the key concepts of von Neumann algebras necessary to understand the following sections.

    \subsubsection{General von Neumann Algebras}
    To define a von Neumann algebra, one needs to first introduce
    $C^\star$-algebras. A $C^\star$-algebra $\cA$ is built on a vector space over the complex     numbers together with a norm $\norm{.}$, a multiplication, an addition, and an involution $\dagger: a \rightarrow a^\dagger$.
    The algebra has to be complete for the norm, i.e.,  every Cauchy sequence $a_n \in \cA$ with respect to $\norm{.}$ has a limit in $\cA$. A sequence $(a_{\mu})^{\infty}_{1}$ is called a Cauchy sequence if, for every $\delta$, there exists an $M\in \mathbb{N}$ such that for all $\mu, \nu \geq M$, $\|{a_{\mu}-a_{\nu}}\|<\delta$. 
    The involution is an antilinear map that additionally satisfies 
    \begin{align}
        a &= (a^\dagger)^\dagger \ , & (ab)^\dagger &= b^\dagger a^\dagger \ . 
    \end{align}
    Elements of a  $C^\star$-algebra satisfy the usual rules of addition and multiplication, but additionally the norm satisfies the condition
    \begin{equation}
        \norm{a a^\dagger} = \norm{a}^2.
    \end{equation}
    The simplest example is the set of bounded linear operators $\cB(\cH)$ acting on a Hilbert
    space $\cH$.\footnote{In general, a $C^\star$- algebra can be defined without making an explicit reference to an underlying Hilbert space.}
    In the following, we fix a Hilbert space $\cH$ and  restrict our analysis to unital
    algebras, i.e.,  those that contain the identity $\id$. Defining a von Neumann algebra involves introducing the commutant of a subset $S \subset \cB(\cH)$,
    which is defined as the set of bounded operators that commute with all of
    $S$, i.e.,  
    \begin{equation}
        S' := \{ a \in \cB(\cH): [a,s] = 0\ \forall s \in S \}.
    \end{equation}
    A von Neumann algebra is then defined as a subalgebra of $\cB(\cH)$ that is closed under Hermitian conjugation and equal to its double commutant, i.e., 
    \begin{equation}
        \cA = \cA''.
    \end{equation}
    Now, given a self-adjoint subset $A \subset \cB(\cH)$, the double commutant
    \begin{equation}
        \cA := A''
    \end{equation}
    is always a von Neumann algebra, and is called the von Neumann algebra \emph{generated} by $A$.
    Von Neumann's double commutant theorem \cite{takesaki1979theory, sunder_1987} establishes that this is
    equivalent to the algebra $\cA$ being closed in the \emph{weak operator topology} (WOT). This means that for a sequence $a_n \in \cA$, there exists
    a bounded operator $a$ with 
    \begin{equation}
       \lim_{n \rightarrow \infty} \braket{\x|a_n|\y} = \braket{\x|a|\y}
    \end{equation}
    for all states $\ket{\x}$ and $\ket{\y}$, then $a \in \cA$. This implies that a von
    Neumann algebra is automatically a $C^\star$-algebra, where the norm refers to the
    usual operator norm. This is because if a sequence of
    operators converges in the operator norm, then it also converges in the weak operator topology.
    Finally, a von Neumann algebra is called a factor if 
    \begin{equation}
        \cA \cap \cA' = \mathbb{C} \id.
    \end{equation}
    A simple example of a factor can be found in a bipartite system whose Hilbert space $ \cH$ takes on the form
    \begin{equation}
        \cH = \cH_A \otimes \cH_B.
    \end{equation}
    Consider the algebra
    \begin{equation}
        \cA = \cB(\cH_A) \otimes \id,
    \end{equation}
    whose commutant reads
    \begin{equation}
        \cA' = \id \otimes \cB(\cH_B),
    \end{equation}
    and is therefore clearly a factor.
   \subsection{Layered tensor networks}
   The main object of study in this paper are limits of tensor networks that can be constructed by an iteration across layers. Roughly speaking, given a Hilbert space $\cH$ that has the form $\cH = \bigotimes_{i=1}^{N} \cH_i$, a tensor network is a representation of a state $\ket{\Psi} \in \cH$ that has a graph $\G$ associated to it. The graph has a set of vertices $V$ and edges $E$, where $E$ is subdivided into a set of ``bond'' edges $B$ that connect two vertices and ``physical'' edges $P$ that are only attached to one vertex. 
   We associate a Hilbert space $\cH_{(e,v)}$ to each edge $e \in E$ adjacent to a vertex $v \in V$, i.e., for each bond edge we have two Hilbert spaces, one for each vertex it connects to. We assume that for any bond $b$ that connects vertices $v_1,v_2$, the Hilbert spaces $\cH_{(b,v_1)} \cong \cH_{(b,v_2)}$ are isomorphic, so that their dimensions match.  The physical Hilbert spaces $\cH_i$ are attached to physical edges in $P$. To obtain a tensor network state $\ket{\Psi}$ one associates to each vertex $v$ a state
   \begin{align}
   \ket{\psi}_v &\in \bigotimes_{\{(e,v)\}}
   \cH_{(e,v)},
   \end{align}
   where the product runs over all edges connected to $v$. Given a collection of such states for each vertex, one obtains the tensor network state by projecting the states on the two sides of each bond edge $b \in B$ onto the maximally entangled state, which contracts the tensors characterizing the state at the vertex $v$ along the indices associated to the bond, i.e., for any $b \in B$ we define 
   \begin{equation}
       \ket{\c}_b = \sum_k \frac{1}{\sqrt{\text{dim}(\cH_b)}} \ket{k}_{v_1}\otimes \ket{k}_{v_2} \in \cH_{(b,v_1)}\otimes \cH_{(b,v_2)},
   \end{equation}
   where $\ket{k}_{v_i}$ is an orthonormal basis for $\cH_{(b,v_i)}$. Then the tensor network state $\ket{\Y}$ is given by 
   \begin{equation}
       \ket{\Y} = \bigotimes_{b \in B} \bra{\c}_b \Big( \bigotimes_v \ket{\y}_v\Big).
   \end{equation}
   In a graphical representation, each of the above states lives on
   the node of a graph, where the edges connecting two nodes represent the maximally entangled state contracted into the states of the respective vertex and the open edges represent the information associated to the ``physical'' Hilbert spaces $\cH_i$. For more details see \cite{Bridgeman:2016dhh}.
    Using the isomorphism between linear maps $O : \cH \rightarrow \cH'$
    between finite-dimensional Hilbert spaces and states $\ket{O} \in \cH'
    \otimes \cH^*$, where $\cH^*$ is the dual space, we can also represent
    linear maps as tensor networks.
    Here we will be interested in layered tensor networks, where we consider a
    sequence of tensor network states
    \begin{equation}
        \ket{\Y^{1}}\rightarrow \ket{\Y^{2}}\rightarrow \hdots
    \end{equation}
    that are connected via isometries 
    \begin{equation}
        \ket{\Y^{\L+1}} = \g^{\L,\L+1}\ket{\Y^{\L}},
    \end{equation}
    where the isometries are themselves given by tensor networks. One can think
    of it as a graph that is built in an iterative procedure where in each
    iteration the open edges of the previous step are contracted with edges of
    another graph. Furthermore, we consider networks where to each isometry $\g^{\L,\L+1}$ one assigns a layer-to-layer operator pushing map $\p^{\L,\L+1}$ (which is a unital $\star$-homomorphism) that associates to each operator living at layer $\L$ an operator that lives at layer $\L+1$ such that expectation values
    \begin{equation}
    \braket{\Y^{\L+1}|\phi^{\L,\L+1}(O^\L)|\Y^{\L+1}} =\braket{\Y^{\L}|O^\L|\Y^{\L}}
    \end{equation}
    are preserved and the identity is mapped to the identity.
    
    \subsection{Complementary recovery}
    As we will explain in Sec.\ \ref{sec:inductive_limits}, we focus on layered tensor networks which
    satisfy an inductive version of complementary recovery, which is a fundamental property of holographic quantum error-correcting codes \cite{Harlow:2016vwg} that gives us control over the entanglement pattern in the network. Complementary recovery can be defined as follows. Let $V$ be an isometry $V:\cH_{\text{bulk}} \rightarrow \cH_{\text{bdy}}$, that maps from a ``bulk'' to a ``boundary'' Hilbert space in the language of holography or from a ``logical'' to a ``physical'' Hilbert space in the language of quantum error correction. Now consider a subalgebra $\cA_a$ of $\cB(\cH_{\text{bulk}})$. If the boundary has a bipartition $\cH_{\text{bdy}} = \cH_A \otimes \cH_{A^c}$, we say that $\cA_a$ is \textit{recoverable} from $A$ if $\forall \cO \in \cA_a$ there exists an operator $\iota(O_a) \in \cB(\cH_A)\otimes \id_{A^c}$ such that for all $\ket{\psi}\in \cH_{\text{bulk}}$ one has 
    \begin{equation}\label{eq:bulk_reconstruction}
        \iota(O_a) V \ket{\psi} = V O_a \ket{\psi},
    \end{equation}
    We assume furthermore that the map $\iota:\cA_a \rightarrow \cB(\cH_A)\otimes \id$ is a faithful, unital $\star$-homomorphism. We call such a map $\iota$ a bulk-to-boundary operator pushing map. Note that we have assumed that $\cH$ manifestly factorizes into $A$ and its complement. A slightly more general perspective is to consider instead an abstract boundary subalgebra $\cA_A$ such that $\iota(\cA_a)\subset \cA_A$ and eq. \eqref{eq:bulk_reconstruction} hold. This way one removes oneself from the geometric picture and in case that $\cA_A$ is a factor recovers the geometric decomposition after a suitable isomorphism. Now we say that the code $V$ satisfies \emph{complementary recovery} for $\cA_a$ in the boundary region $A$ if $\cA_a$ is recoverable from $A$ and its commutant $\cA'_a$ is recoverable from $A^c$. A bulk region $a$ anchored in a boundary region $A$ that satisfies complementary recovery is called an \textit{entanglement wedge}.
    Note that the above does not assume that $\cA_a$ is a factor but in the following we will usually restrict to factors, i.e., that $\cH = \cH_a \otimes \cH_{a^c}$ and $\cA_a = \cB(\cH_a) \otimes \id$. Now Harlow proved \cite{Harlow:2016vwg} that if $V$ is a code with complementary recovery for $\cA_a$ and one considers a product state $\ket{ij}$, where $\ket{i}\in \cH_a,\ket{j}\in \cH_{a^c}$, that there exist a pair of local unitaries $U_A,U_{A^c}$ in $A,A^c$ such that 
    \begin{equation}\label{eq:Harlow's_thm}
        V \ket{ij} = U_A U_{A^c} \big(\ket{ij}\otimes \ket{\c}\big),
    \end{equation}
    where the state $\ket{\c}$, independent of $\ket{ij}$, determines the entanglement between $A$ and $A^c$. We will refer to this statement in the following as \textit{Harlow's theorem}. 
    Note that we have here presented an algebraic view on operator reconstruction. This does not have to fit into a geometric picture where $\cA_a$ is a ``set of bulk qubits'' and $\cA_{a^c}$ is the complementary set of bulk qubits as in the HaPPY code. The mathematical reason is that, if one considers $\cA_{a}$ to be the operators that act on a ``set of bulk qubits'' it automatically is a factor. We see that if $\cA_{a}$ is not a factor, such a geometric picture has to break down. An explicit example of such a situation is given in \cite{Steinberg:2023wll} where the entanglement wedges of boundary regions do not have a simple geometric picture. On the other hand, if $\cA_a$ is a factor, one can always find a unitary $U$ such that 
    \begin{equation}
        \cH = U(\cH_a \otimes \cH_{a^c}),\ \cA_a = U \big( \cB(\cH_a)\otimes \id \big) U^\dagger,
    \end{equation}
    so that with respect to the decomposition induced by $U$, the reconstruction is ``geometric".
    
    \subsection{Inductive limits}
    \label{sec:inductive_limits_intro}
    Inductive limits provide a mathematical framework for constructing infinite-dimensional structures from sequences of finite-dimensional ones, which we employ to construct the limiting system of a tensor network as the inductive limit of $C^{\star}$-algebras and Hilbert spaces induced by the tensor network. For $C^{\star}$-algebras, this involves a directed system $(\cA_{n},\phi_{mn})$, where $\cA_{n}$ are $C^{\star}$-algebras and 
    \be \phi_{mn}:\quad \cA_{n}\to \cA_{m}\ee
    are $\star$-homomorphism for $n\leq m,$ satisfying compatibility conditions \cite{blackadar2006operator} defined below.  
    The inductive limit algebra $\cA$ is then a $C^{\star}$-algebra that encodes the
    structure of the entire sequence. Similarly, for Hilbert spaces, an
    inductive system consists of a sequence of Hilbert spaces $\{\cH_{n}\}$ and
    isometric embeddings $\{\iota_{nm}\}$ \cite{blackadar2006operator}. 
    In the following, we present an overview of inductive limits. We begin by describing the limits of general vector spaces, which will directly translate to the limiting Hilbert spaces generated by tensor networks, and then proceed to describe limits of algebras, laying the groundwork for our later discussion of von Neumann algebras.

    \subsubsection{Inductive Limits of Vector spaces}
    Here we provide an introduction to inductive limits
    \cite{blackadar2006operator, kadison1986fundamentals, Roerdam, brown_2008}
    in the category of Banach spaces, i.e., normed, complete vector spaces. 
    $C^\star$-algebras and Hilbert spaces carry the structure of a Banach space so it serves as an example for inductive limits that illustrates the procedure. To obtain limits in the category of $C^\star$-algebras and Hilbert spaces, one has to define an additional structure such as a multiplication and adjoint for $C^\star$-algebras and an inner product for Hilbert spaces, which modify the exact construction. Since the main steps, up to the additional structure, are conceptually the same, we describe the procedure for Banach spaces here.
    
    Assume that we have a sequence of normed vector spaces $\cV_i$, indexed by some set
    $\W$, such as the Hilbert spaces $\cH_i$ in which a layered
    tensor network lives at each level,
    and linear maps $\p^{ij} : \cV_j
    \rightarrow \cV_i$ that are contractive, i.e.,
    \begin{equation}
    \norm{v}_{\cV_i} \geq \norm{\p^{ij}(v)}_{\cV_j}, 
    \end{equation}
    and are compatible between the indices in the sense that 
    \begin{equation}
        \p^{ij} = \p^{kj} \circ \p^{ik}, \forall i \leq k \leq j.
    \end{equation}
    One can then identify vectors between the layers via their image under
the maps $\p^{ij}$, i.e., we identify two vectors $v_i \in \cV_i, v_j \in
    \cV_j$, where we assumed $j \geq i$, if $v_j$ is the image of $v_i$ under the
    embeddings $\p^{ij}$:
    \begin{equation}
        v_i \sim v_j \Leftrightarrow \p^{ij}(v_i) = v_j.
    \end{equation}
    We denote the equivalence class of $v_i$ as $[v_i]$.
    A family of examples is the actual tensor networks we consider in this paper: The state they represent at layer $i$ is identified with the state at layer $j\geq i$.
    We denote the set of such equivalence classes of vectors by $V$. 
    Given $V$, we can define sums of its elements
    directly via representatives, i.e., if $j \geq i$ then we define
    \begin{equation}
        [v_i] + [v_j] := [\phi^{ij}(v_i)+v_j].
    \end{equation}
    and multiplication via scalars
    \begin{equation}
        \alpha [v] := [\alpha v],\ \forall \alpha \in \bC.
    \end{equation}
    This definition does not depend on the choice of representative and promotes the set of
    equivalence classes to a vector space.
    Because the maps $\phi^{ij}$ are contractive, we can define the norm of
    $[v_i]$ via 
    \begin{equation}
        \norm{[v_i]} = \lim_{j \rightarrow \infty} \norm{\p^{ij}(v_i)}.
    \end{equation}
    This extends $(V,+,\norm{.})$ to a normed vector space. Now taking the
    completion with respect to this norm defines a Banach space $\cV$.
    $\cV$ is what we call the inductive limit of the inductive set
    $(\cV_i,\p^{ij})$ and we write it as 
    \begin{equation}
        \cV = \varinjlim V_i.
    \end{equation}

    \subsubsection{Inductive Limits of Hilbert spaces and Algebras in Layered Tensor Networks}\label{sec:inductive_limits_algebras}
    The above discussion defines the inductive limit of Banach spaces. This already allows us to associate a limiting object with both the tensor network and the local algebras, i.e., given the isometry
    $\g^{\L,\L+1}$ that embeds the network at layer $\L$ into the network at
    layer $\L+1$, we define equivalence classes for states via the identification
    \begin{equation}
        \g^{\L,\L+1} \ket{\Y^\L} \sim \ket{\Y^\L}.
    \end{equation}
    The set of equivalence classes again defines an inductive limit $\cV$.
    However, we have not equipped $\cV$ with the structure of a Hilbert space, namely a scalar product. We now explain how to do so.
    We first define an inner product between equivalence classes: For two equivalence classes $[\Psi^\L],[\Psi^{\L'}], \L \leq \L'$ we define the inner product as
    \begin{equation}
            \braket{[\Y^{\L}]|[\P^{\L'}]} =
            \braket{\g^{\L,\L'}(\Y^{\L})|\P^{\L'}},
    \end{equation}
    where we defined the multi-layer isometry
    \begin{equation}\label{eq:gllp}
        \g^{\L,\L'} = \g^{\L'-1,\L'}\circ \g^{\L'-2,\L'-1} \circ \hdots \circ \g^{\L+1,\L+2}\circ \g^{\L,\L+1}.
    \end{equation}
    This inner product is defined on a dense set of vectors in $\cV$ and since the embeddings $\g^{\L,\L'}$ are isometries, the norm defined by the inner product coincides with the norm of the representatives. Therefore, one can extend the scalar product to all vectors in $\cV$, which extends $\cV$ (upon completion) to a Hilbert space $\cH$.
    Having constructed the limiting Hilbert space, we want to identify
    operators, or more generally, operator algebras that arise from algebras at
    finite layers and survive the limiting procedure. Having the operator
    pushing map $\p^{\L,\L+1}$ associated to the layered network at hand, we define equivalence classes of operators in which we
    identify an operator $O$ that lives at layer $\L$ with its image under the
    pushing map
    \begin{equation}
        \phi^{\L,\L+1}(O) \sim O.
    \end{equation}
    This step is why we demanded in the introduction that the operator pushing map be unital, so that the identity of a given layer will be identified with the identity of the next. 
    Later on in Sec.\ \ref{sec:inductive_limits} we will need the unitality of $\phi$ also to have a good decomposition of the Hilbert space between layers that preserves the structure of the previous layers.
   Similarly, we can consider a local subalgebra $\cA^{\L}_A = \cB(\cH^{\L}_A) \otimes \id_{A^{C}}$ of $\cH^{\L}$ at
    layer $\L$ that corresponds to the bounded operators of a subset $A$ of the
    open legs of the network at layer $\L$. We can similarly identify it with its image in the next layer
    \begin{equation}
        \cA^{\L}_A \sim \phi^{\L,\L+1}(\cA^{\L}_A).
    \end{equation}
    When mapping between layers, the local algebra $\cA^{\L}_A$ will be mapped to a subalgebra of the lightcone of $A$, where the lightcone $J^+(A)$ is defined as the set of qubits on which $\phi^{\L,\L+1}(A)$ is supported. Note that we define the lightcone through the map $\phi$, which defines it in the sense of the connectivity of the underlying network rather that in a sense of time evolution.
    We therefore obtain a sequence of algebras $\cA^{\L}_A, \cA^{\L+1}_{J^+(A)},...$ that together with the operator pushing map between layers $\phi^{\L,\L+1}$ again form an inductive system.
    For this inductive system we also take the inductive limit and obtain a
    lightcone C$^\star$-algebra $\hat{\cA}_A$, i.e., 
    \begin{equation}
      \hat{\cA}_A = \lim_{\longrightarrow} \cA^\L_{J^+(A)} . 
    \end{equation} This will be an abstract $C^\star$-algebra because the embedding $\phi$
    is implemented by an isometry. We represent this algebra on the Hilbert
    space we just constructed by defining it on a dense set of states for each
    operator $O^{\L}$ that is representable at a finite layer $\L$
    using
    \begin{equation}\label{eq:limit_algebra_action}
                \pi([O^{\L}])\ket{[\Y^{\L'}]} := \begin{cases}
                \ket{[\phi^{\L,\L'}(O^{\L})\Y^{\L'}]}\ \text{if } \L \leq
                \L',\\
                \ket{[O^{\L} \g^{\L',\L}(\Y^{\L'})]}\ \text{if } \L' \leq
                \L.                    
                \end{cases}
    \end{equation}
    and extending the representation to all of $\hat{\mathcal{A}}_A$ by continuity. Note that for this definition to be well defined, the operator pushing map $\p^{\L,\L+1}$ has to be compatible with the Hilbert space isometry $\g^{\L,\L+1}$ in the sense that if one considers an operator $O^{\L}$ at layer $\L$ that is also the image of an operator $O^\L = \p^{\L',\L}(O^{\L'})$ at a lower layer $\L'$ and the same holds for the state, then both cases of eq. \eqref{eq:limit_algebra_action} have to coincide, i.e., 
    \begin{equation}\label{eq:compatibility}
        O^\L \g^{\L',\L}\ket{\Psi^{\L'}} = \phi^{\L',\L}(O^{\L'}) \g^{\L',\L}\ket{\Psi^{\L '}} \overset{!}{=} \g^{\L',\L}\ket{O^{\L'}\Psi^{\L'}}.
    \end{equation}
    Here we have for convenience let the layer associated to the operator $O^{\L'}$ be the same as for the state $\Psi^{\L'}$. This requirement can be straightforwardly generalized when these differ, the main point being that operator pushing and the isometry between Hilbert spaces have to be compatible.
    We then define the von Neumann algebra
    \begin{equation}
    \mathcal{A}_A := \pi(\hat{\cA}_A)''.
    \end{equation}
     Our main objective in this paper
    is to determine the type of the von Neumann algebra $\mathcal{A}_A$ for layered tensor networks. A recap of the type classification suitable for our needs is provided in Appendix \ref{sec:type_classification}. 
    Next we will provide the main intuition behind our study on the example of the HaPPY code.

\section{Intuition from the HaPPY code}\label{sec:intuition_happy}
\subsection{HaPPY codes at a fixed layer}\label{sec:happy_fixed_layer}
\begin{figure*}[ht]
    \centering
    \includegraphics[width=0.9\textwidth]{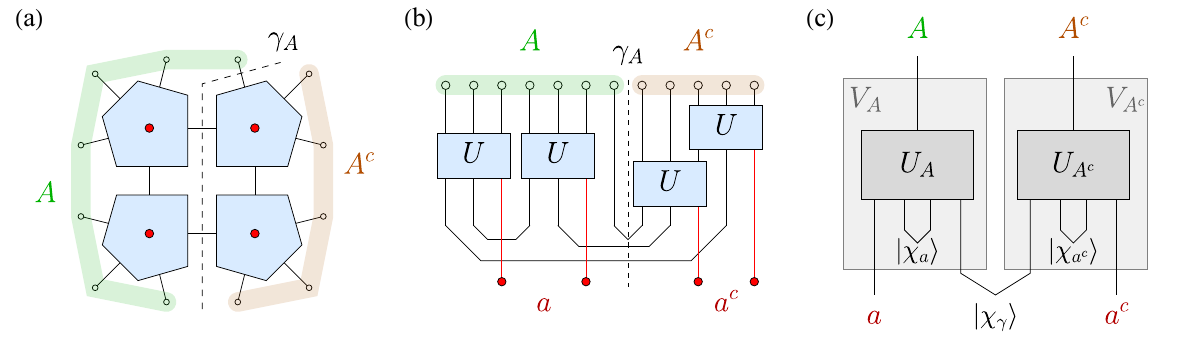}
    \caption{Turning a holographic tensor network into an encoding circuit. (a) We take a small HaPPY code with four contracted perfect tensors and consider a boundary bipartition into $A$ and $A^c$. From each region, two logical qubits (red dots) can be recovered, forming the ``bulk regions'' $a$ and $a^c$, separated by a cut $\gamma_A$ through the tensor network.
    (b) Using the property that the six-leg perfect tensor acts as a unitary $U$ from any three legs to the remaining three, we can reorganize the tensor network into a circuit from the logical qubits in $a$ and $a^c$ to the physical qubits in $A$ and $A^c$. In this circuit, some of the tensor contractions become insertions of maximally entangled pairs into the circuit. Three of such pairs cross between $A$ and $A^c$, leading to an entanglement entropy $S_A = \log 3 + S_a$.
    (c) The generic holographic encoding circuit in terms of two unitaries $U_A$ and $U_{A^c}$ (or equivalently, isometries $V_A$ and $V_{A^c}$), with resource states $|\chi_a\rangle$ and $|\chi_{a^c}\rangle$ contributing only to entanglement within each subregion and $|\chi_{\gamma}\rangle$ contributing to the entanglement between $A$ and $A^c$. For HaPPY codes, these resource states are copies of maximally entangled pairs.
    }
    \label{fig:happy_to_unitary}
\end{figure*}

\begin{figure*}[ht]
    \centering
    \includegraphics[width=1.0\textwidth]{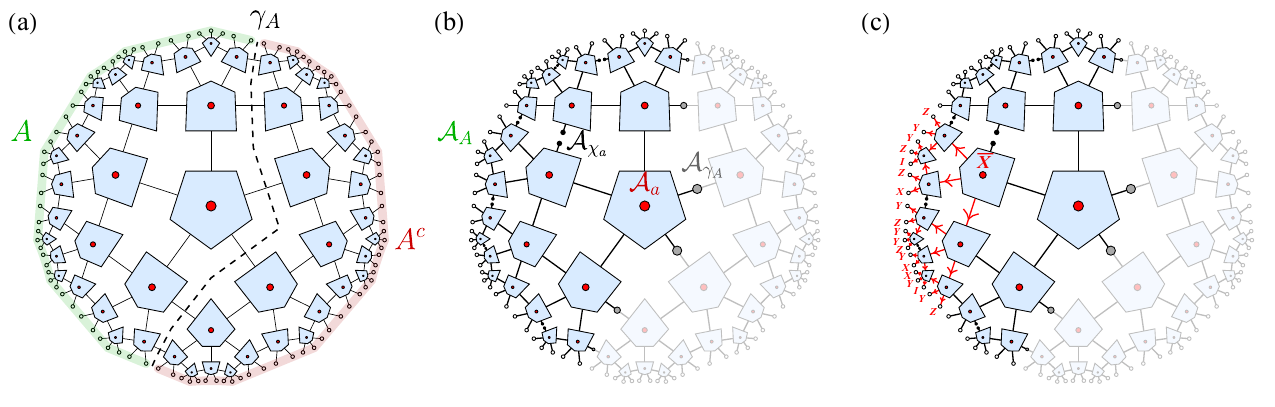}
    \caption{Subregion algebra reconstruction in the HaPPY model. (a) A boundary bipartition into $A$ and $A^c$ of the full $\{5,4\}$ HaPPY code. The Ryu-Takayanagi cut $\gamma_A$ separates the bulk into two wedges $a$ and $a^c$, logical qubits (red dots) in which are reconstructable (only) on $A$ and $A^c$ (white dots), respectively.
    (b) Mapping the full boundary subregion algebra $\mathcal{A}_A$ back into the bulk: Removing $a^c$ and bonds corresponding to (one choice of) ancillas $\ket{\chi_a}$ turns the remaining tensors into a unitary circuit (following Fig.\ \ref{fig:happy_to_unitary}). $\mathcal{A}_A$ is unitarily mapped to the bulk algebra $\mathcal{A}_a$ (red), the wedge ancilla algebra $\mathcal{A}_{\chi_a}$ (black), and the Ryu-Takayanagi algebra $\mathcal{A}_{\gamma_A}$ (gray).
    (c) With the ancilla bonds removed, operator-pushing a logical operator (here $\bar{X}$ acting on one bulk qubit) follows a unique flow towards the boundary, resulting in a unique boundary representation of the logical operator.
    }
    \label{fig:happy_to_unitary_algebras}
\end{figure*}

In this section we provide intuition behind the direct limit construction of the previous section using holographic tensor network codes.
We use the code structure of the network to identify operators between layers, allowing us to rigorously treat the inductive limit of the algebras. In particular, we demonstrate that the network can be written as a unitary map by opening contracted legs in the network, which enables us to find a decomposition of boundary subregion algebras by considering their analogous bulk decomposition.
We focus on the family of HaPPY codes \cite{Pastawski:2015qua}, built from perfect tensors with an even number of legs, one of which is associated with an encoded logical qubit. Such tensors mediate maximal entanglement between any bipartition into two sets of legs, thereby forming an isometry from the smaller set to the larger. In particular, any bipartition into equally many legs yields a unitary map. For the case of six-leg tensors (one bulk leg and five ``planar'' legs) and qubits associated with each leg (i.e.,  bond dimension $\chi=2$), such a perfect tensor is given by the encoding isometry of the five-qubit \emph{Laflamme code} \cite{Laflamme:1996iw}, which can correct one single-qubit error.
As we show in Fig.\ \ref{fig:happy_to_unitary}, one can use the perfect tensor property to decompose any bipartition $(A,A^c)$ of the open boundary legs of the full HaPPY code into a unitary circuit that prepares the physical boundary state starting from the logical bulk state and some maximally entangled ancillary states.\footnote{As noted in Ref.\ \cite{Pastawski:2015qua}, such a clear decomposition can fail for a small subset of boundary regions for which the bulk bipartition is not exactly complementary. Here we do not consider these cases.}
While some of the ancillae contribute to the entanglement between $A$ and $A^c$, others act only within one of the two regions. The latter type is necessary for the tensor network code to provide meaningful quantum error correction under bipartition: otherwise, the encoding unitary $U_A$ for a subregion $A$ (see Fig.\ \ref{fig:happy_to_unitary}(c) following \cite{Harlow:2016vwg}) would merely mix $n$ logical qubits and $m$ additional qubits that are maximally entangled with $A^c$ into $n+m$ physical qubits. This would imply that each operator $\mathcal{O}_a$ acting on the logical qubits in $a$ has only one representation on $A$, making it highly susceptible to errors. Introducing additional ancillae $|\chi_{a}\rangle$ and $|\chi_{a^c}\rangle$ into the circuit allows one to apply ``gauge'' operators on them, leading to different physical representations of $\mathcal{O}_a$.
We have two natural ways to encode a bulk operator $\cO_{\text{bulk}}$ in the boundary. The first
is to conjugate it with the isometries for each bulk \textit{entanglement
wedge} $a$ and $a^c$: $V_A = U_A \ket{\chi_a}, V_{A^c} = U_{A^c}\ket{\c_{a^c}}$,
where $U_A,U_{A^c}$ are the opened-up networks, so that 
\begin{align}
    \mathcal{O}_\text{bdy} = V_A V_{A^c} \ket{\chi_{\gamma}} \mathcal{O}_\text{bulk} \bra{\chi_{\gamma}} V_A^\dagger V_{A^c}^\dagger \ ,
\end{align}
which can be checked to have the correct action on code states $V \ket{ij}$, where $V = V_A V_{A^c} \ket{\chi_{\gamma}}$, essentially because of the isometric property $V^\dagger V = \id$. However, these representations $\mathcal{O}_\text{bdy}$ act as projectors onto the codespace. Rather than including the physical identity acting on $A$, they only include a \emph{logical} identity that acts as an identity on states within the codespace. 
For our purposes, a more suitable operator map is the natural \emph{operator pushing map} defined by the perfect tensors, whose stabilizers allow one to replace operators acting on a subset of $k \leq 2$ physical qubits by equivalently-acting operators on the other $5-k$ qubits \cite{Pastawski:2015qua}. In the notation where we have opened up some of the internal legs of the tensor network to extend the isometry $V_A$ into a unitary $U_A$, this takes the form 
\begin{equation}\label{eq:operator_pushing}
    \begin{aligned}
       \cO_{\text{bdy}} &= \iota(\cO) := U_A U_{A^c}\, \cO\, U_A^{\dagger} U_{A^c}^{\dagger} \ ,
    \end{aligned}
\end{equation}
which is the bulk-to-boundary map of Fig.\ \ref{fig:happy_to_unitary}(c), where
one encodes an operator $\cO=\cO_{\text{bulk}} \otimes \id_\g \otimes \id_{\chi_a} \otimes \id_{\chi_{a^c}}$ (here, $\chi$ refers to ancilla degrees of freedom while $\gamma$ refers to degrees of freedom on the RT surface).
For a bulk operator $\cO_{\text{bulk}} = \cO_a \otimes \id_{a^c}$ that only has support in the entanglement wedge $a$ of $A$, this further simplifies into a boundary operator $\cO_A$ that only has support on $A$:
\begin{align}
    \cO_A = \tr_{A^c} \cO_{\text{bdy}} = U_A \cO_a U_A^\dagger \ .
\end{align}
We thus find that the operator map $\iota(\cO)$ also satisfies complementary recovery.
Note that our construction of $U_A$ and $U_{A^c}$ is non-unique \cite{Gesteau:2022hss}, as one may open different pairs of contracted legs to construct such unitaries from the isometries $V_A$ and $V_{A^c}$. These different choices of unitaries lead to different bulk-to-boundary maps $\iota$, related to different logical representations that we discuss further below.

In the language of quantum error correction, $\mathcal{O}_{A}$ is a logical operator $\overline{\mathcal{O}_{\text{bulk}}}$ that acts on a subset of the qubits of a physical state.
As $\dim \mathcal{H}_a < \dim \mathcal{H}_A$, the boundary algebra generated by encoding with \eqref{eq:operator_pushing} every element of the wedge algebra $\mathcal{A}_a$ (operators acting on the logical qubits in $a$) is only a subalgebra of the full algebra $\mathcal{A}_A$ of all boundary operators. What are the other subalgebras? We find the answer by conjugating the unitary $U_A$, which acts on the Hilbert spaces
\begin{equation}\label{eq:boundary_unitary}
    U_A^\dagger : \mathcal{H}_A \to \mathcal{H}_a \otimes \mathcal{H}_{\chi_a} \otimes \mathcal{H}_{\gamma_A} \ ,
\end{equation}
whose algebras are visualized in Fig.\ \ref{fig:happy_to_unitary_algebras}(b).
Operators acting on the last piece of the tensor product form the \emph{Ryu-Takayanagi algebra} $\mathcal{A}_{\gamma_A}$. For the standard HaPPY code where $\ket{\chi_{\gamma}}$ is a set of EPR pairs, these operators act equivalently on $\mathcal{H}_{\gamma_A}$ and $\mathcal{H}_{\gamma_{A^c}}$.
The second piece of the tensor product is acted upon by the so-called \emph{wedge ancilla algebra} $\mathcal{A}_{\chi_a}$.
The geometric setting for this mapping of algebras in the HaPPY code is shown in Fig.\ \ref{fig:happy_to_unitary_algebras}.
By conjugating operators belonging to these algebras by $U_A, U_{A^c}$ we again
obtain their respective boundary representations.
 Note that the
above discussion gives a concrete realization of the bulk-to-boundary operator pushing map,
usually denoted by $\iota$, which maps logical bulk operators to a boundary
operator by explicit conjugation by an unitary
\eqref{eq:unitary_operator_pushing}. Repeating the same discussion for the
complementary region, we find that 
\begin{equation}
    \cH \cong \cH_a \otimes \cH_{\c_a} \otimes \cH_{\g_A} \otimes \cH_{a^c}
    \otimes \cH_{\c_{a^c}} \otimes \cH_{\g_{A^c}}\ ,
\end{equation}
for the full boundary Hilbert space $\cH$.

\subsection{Ancilla algebras and stabilizers}
\label{SUBSEC_ANCILLA_STAB}

In our construction of the operator pushing map $\iota$, we have extended the isometric map furnished by the HaPPY tensor network into a unitary. We now briefly comment on the nonuniqueness of such an extension and its relationship to quantum error correction.
In a quantum code, we map logical states to a subspace of the physical space, called \emph{codespace}, allowing different physical operators to have equal action on states in the codespace. We call such different but logically equivalent operators \emph{representations} of a logical operator.
In a \emph{stabilizer code} \cite{Gottesman:1997zz}, we can switch between different representations of logical operators by applying stabilizer operators, the $+1$ eigenspace of which forms the codespace.
The HaPPY code on a hyperbolic pentagon tiling is an example of a qubit stabilizer code.
By opening some legs of the tensor network to extend the isometries $V_A$ and $V_{A^c}$ into unitaries $U_A$ and $U_{A^c}$, we have effectively fixed all logical operators to a unique representation, or equivalently, fixed the operator-pushing flow from bulk to boundary (see Fig.\ \ref{fig:happy_to_unitary_algebras}(c)).
Suppose that there are three ways to extend this construction to produce different representations:
\begin{enumerate}
\setlength\itemsep{0pt}
    \item Act on the fixed representation with stabilizer operators, which act as identities on codestates.
    \item Open different legs of the tensor network that lead to different unitaries $U_A$ and $U_{A^c}$.
    \item Set a different operator-pushing flow to map bulk to boundary operators.
\end{enumerate}
We now show that approach 2 and 3 are equivalent and form a special case of approach 1, in which we take a particular choice of $U_A$ and $U_{A^c}$ (and correspondingly opened legs), which projected onto ancilla states within each bulk region form the isometries $V_A = U_A \ket{\chi_a}$ and $V_{A^c} = U_{A^c}\ket{\c_{a^c}}$. Any bulk state $\ket{\psi}$ in $\cH_a \otimes \cH_{a^c}$ is then mapped to the logical state
\begin{align}
    \ket{\bar{\psi}} \equiv V_A V_{A^c} \ket{\psi}\ket{\chi_\gamma} \ .
\end{align}
We now try to find an operator $\mathcal{O} = \mathcal{O}_a \mathcal{O}_{\chi_a} \mathcal{O}_{\chi_{\gamma_A}}$ (omitting identities on $a^c$ and $\chi_{a^c}$) that is mapped to a logical operator $\overline{\cO}$ with support only on $A$ that acts as a stabilizer $\overline{\id}$, i.e.,  leaves any $\ket{\bar{\psi}}$ invariant. We find
\begin{align}
    \bar{\mathcal{O}} \ket{\bar{\psi}} &= U_A \mathcal{O}_a \mathcal{O}_{\chi_a} \mathcal{O}_{\chi_{\gamma_A}} U_A^\dagger V_A V_{A^c} \ket{\psi} \ket{\chi_\gamma} \nonumber\\
    &= U_A V_{A^c} (\mathcal{O}_a \ket{\psi}) (\mathcal{O}_{\chi_a}\ket{\chi_a}) (\mathcal{O}_{\chi_{\gamma_A}} \ket{\chi_\gamma}) \ .
\end{align}
For this expression to reduce to $\ket{\bar{\psi}}$, we require three conditions
\begin{subequations}\label{eq:unitary_operator_pushing}
\begin{align}
    \cO_a \otimes \id_{a^c} \ket\psi &= \ket\psi \ , \\
    \mathcal{O}_{\chi_a}\ket{\chi_a} &= \ket{\chi_a} \ , \\
    \mathcal{O}_{\chi_{\gamma_A}} \otimes 
    \label{EQ_EPR_COND3}
    \id_{\gamma_{A^c}} \ket{\chi_\gamma} &= \ket{\chi_\gamma} \ ,
\end{align}
\end{subequations}
where we have restored identity operators. To fulfill the first condition for a general $\ket\psi$, we require $\cO_a = \id_a$. For the HaPPY code, where $\ket{\chi_a}$ and $\ket{\chi_\gamma}$ are sets of EPR pairs, solutions of the second and third condition are given by $\mathcal{O}_{\chi_{\gamma_A}} = \id_{\gamma_A}$ and such operators $\mathcal{O}_{\chi_a}$ that act equivalently on both ends of each EPR pair, e.g.\ Paulis $X \otimes X$ for a 2-qubit EPR pair. The stabilizers with support on $A$ are thus found by considering all such operators $\mathcal{O}_{\chi_a}$ and unitarily mapping them to the boundary using $\iota$.
Now consider approach 2, which fixes the input operator $\cO = \cO_a \id_{\chi_a} \id_{\chi_{\gamma_A}}$ and instead changes which legs to open up to define $\iota$, i.e.,  changing the sites on which $\cA_{\chi_a}$ acts. In the operator-pushing picture, opening up the legs fixes the local stabilizer on each tensor that can be used to push an operator from one layer to the next. Changing which leg is opened up is equivalent to changing which local stabilizer to push with, which shows the equivalence between approach 2 and 3. What is the operator that maps from a boundary representation of a logical operator that is pushed along one leg rather than another? We find that this is exactly an operator that fulfills \eqref{EQ_EPR_COND3}, acting on both of the newly opened legs with a pair of conjugate operators.
An example for the Pauli stabilizers of the Laflamme code used in the HaPPY model: Here valid stabilizers are cyclic permutations of either $XZZXI$, $YXXYI$, or $ZYYZI$ (the products of elements of any one set yield the other two). If we consider the first qubit as an input (part of $\cH_a$) and ``open'' the tensor leg corresponding to the last qubit (part of $\cH_{\chi_a}$), we can operator-push an input $X$ by applying $XZZXI$, thus mapping $XIIII\mapsto IZZXI$. 
More specifically, consider three pentagon tensors that jointly form a unitary map $\iota$ from three incoming physical legs, three logical legs, and a pair of legs formed by opening up a contraction. For the central pentagon, operator-pushing $X$ can then be visualized as
\begin{align}
\label{EQ_IOTA_LEFT_CUT}
    \begin{gathered}
        \includegraphics[width=0.2\textwidth]{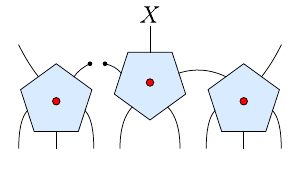}
    \end{gathered}
    &=
    \begin{gathered}
        \includegraphics[width=0.2\textwidth]{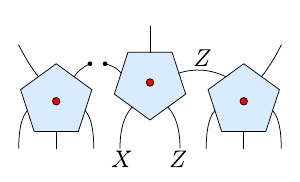}
    \end{gathered}
    \nonumber\\
    &=
    \begin{gathered}
        \includegraphics[width=0.2\textwidth]{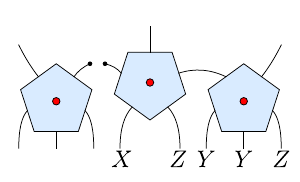}
    \end{gathered} \ ,
\end{align}
where we used the stabilizer $XZZXI$ on the central pentagon and $IZYYZ$ on the right-most pentagon. However, we could have also turned this contraction of three pentagon tensors into a unitary map by instead opening up the contraction between the central and the right-most pentagon. In that case, operator-pushing $X$ would take the form
\begin{align}
\label{EQ_IOTA_RIGHT_CUT}
    \begin{gathered}
        \includegraphics[width=0.2\textwidth]{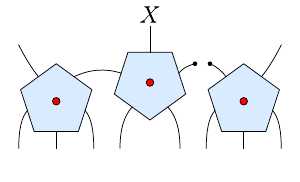}
    \end{gathered}
    &=
    \begin{gathered}
        \includegraphics[width=0.2\textwidth]{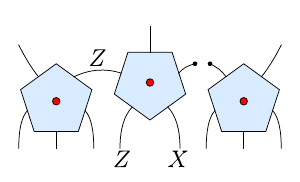}
    \end{gathered}
    \nonumber\\
    &=
    \begin{gathered}
        \includegraphics[width=0.2\textwidth]{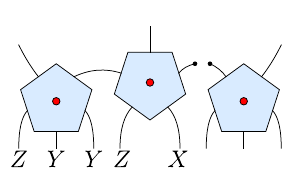}
    \end{gathered} \ .
\end{align}
On the ``boundary'' (the legs on the bottom), these two representations differ by 
\begin{equation}
\label{EQ_HAPPY_STAB_DIFF}
    (IIIXZYYZ)(ZYYZXIII) = ZYYYYYYZ \ ,
\end{equation}
which acts as a stabilizer on the full code. We can see that applying this stabilizer corresponds to changing the operator-pushing flow of $X$ from the right to the left (and vice-versa).
We can also generate this stabilizer term for a fixed operator pushing map $\iota$ (and associated opened legs) by pushing the operator corresponding to the product of the two stabilizers applied to the central qubit in both situations, 
\begin{equation}
    (XZZXI)(XIXZZ) = IZYYZ \ ,
\end{equation}
which is another stabilizer of the five-qubit code. It acts trivially on the first qubit, thus not performing any logical operation on $\cH_a$. 
This is equivalent to operator-pushing a pair of $Z$ operators applied on the newly-opened legs:
\begin{align}
\label{EQ_IOTA_ZZ_INSERTION}
    \begin{gathered}
        \includegraphics[width=0.2\textwidth]{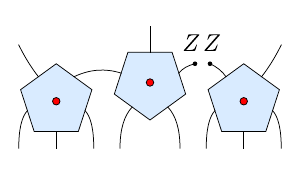}
    \end{gathered}
    &=
    \begin{gathered}
        \includegraphics[width=0.2\textwidth]{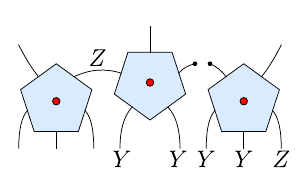}
    \end{gathered}
    \nonumber\\
    &=
    \begin{gathered}
        \includegraphics[width=0.2\textwidth]{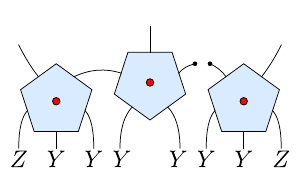}
    \end{gathered} \ ,
\end{align}
which yields the stabilizer \eqref{EQ_HAPPY_STAB_DIFF}. Note that this pair of $Z$s would cancel each other out if we contracted both legs (turning the unitary map into an isometry), which confirms that \eqref{EQ_HAPPY_STAB_DIFF} acts as a logical identity under the encoding isometry of the full code.
We can explicitly construct the map between the two operator pushing maps $\iota$ and $\iota^\prime$ (given by the tensor networks in \eqref{EQ_IOTA_LEFT_CUT} and \eqref{EQ_IOTA_RIGHT_CUT}, respectively) from the tensor network construction. Consider a logical operator $\cO$ in one configuration and $\cO^\prime$ in the other, which are assumed to map to the same boundary operator $\iota(\cO) = \iota^\prime(\cO^\prime)$. As each operator pushing map is associated with a unitary transformation $U$ and $U^\prime$, we find
\begin{equation}
    \cO^\prime = {U^\prime}^\dagger U\, \cO\, U^\dagger U^\prime \ ,
\end{equation}
resulting in a transformation by another unitary ${U^\prime}^\dagger U$ defining a unitary \emph{superoperator} $\mathcal{S}(\bullet) = {U^\prime}^\dagger U\, \bullet\, U^\dagger U^\prime$ acting on bulk operators, mapping between the logical representations corresponding to different operator-pushing flows.
The unitary map can again be expressed as a tensor network:
\begin{align}
\label{EQ_PUSHING_FLOW_TRANSFORMATION}
    {U^\prime}^\dagger U =
    \begin{gathered}
        \includegraphics[width=0.175\textwidth]{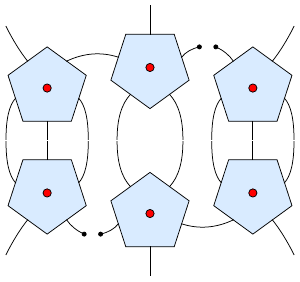}
    \end{gathered}
    =
    \begin{gathered}
        \includegraphics[width=0.175\textwidth]{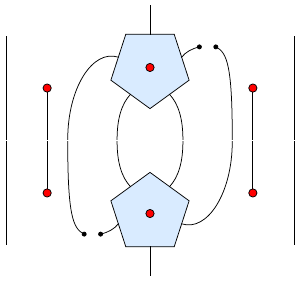}
    \end{gathered} \ .
\end{align}
Here the lower row of tensors form the adjoint ${U^\prime}^\dagger$, which is simply a mirrored version of the original tensors of the Laflamme code (which has a real-valued tensor representation). In the last step we used the perfect tensor property to reduce two tensor pairs into products of identities (in general, the non-reducible part will consist of tensors stretching between the two choices of ancilla openings). 
Using operator pushing with this extended map shows how ancilla-free operators in one configuration of opened legs get mapped to operators with nontrivial ancilla support in the other, e.g.\
\begin{align}
    \begin{gathered}
        \includegraphics[width=0.12\textwidth]{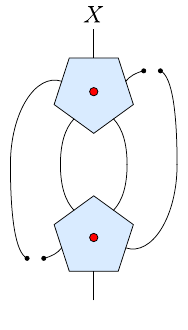}
    \end{gathered}
    =
    \begin{gathered}
        \includegraphics[width=0.12\textwidth]{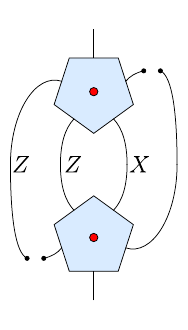}
    \end{gathered}
    =
    \begin{gathered}
        \includegraphics[width=0.12\textwidth]{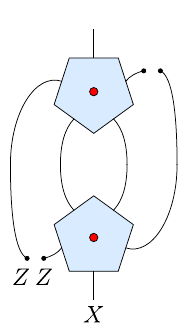}
    \end{gathered} \ ,
\end{align}
which reproduces the $ZZ$ ancilla insertion we showed in \eqref{EQ_IOTA_ZZ_INSERTION}.

Let us summarize the relationship between the three approaches discussed above. The second and third approach both amount to setting a unique operator-pushing flow that associates a unique boundary operator to a bulk operator $\cO_a \cO_{\chi_a} \cO_{\chi_{\g_A}}$ acting on a subregion $a$. Crucially, the resulting operator-pushing map $\iota$ is always unital, i.e.,
\begin{equation}
    \iota(\id_a \id_{\chi_a} \id_{\chi_{\g_A}}) = \id_A \ .
\end{equation}
Similarly, in the stabilizer code picture we consider boundary operators logically equivalent if they differ only by a product with an operator $\iota(\id_a \cO_{\chi_a} \id_{\chi_{\g_A}})$ where $\cO_{\chi_a} \ket{\chi_a} = \ket{\chi_a}$. However, changing the operator-pushing flow does not simply correspond to taking the product of every logical operator with a fixed stabilizer, which would not preserve $\id_A$. Instead, as we have seen above, the new operator pushing flow is equivalent to applying specific stabilizers on specific logical operators, which can be implemented as a tensor network map such as \eqref{EQ_PUSHING_FLOW_TRANSFORMATION}.
Incidentally, the role of the RT algebra $\cA_{\gamma_A}$ can be understood in a similar vein as the ancilla algebra $\cA_{\chi_a}$: By pushing a pair of conjugate operators, such as two copies of a Pauli operator, from the RT surface $\gamma_A$ to both $A$ and $A^c$, we obtain the stabilizers that map between logical representations in either region. The existence of such exact stabilizers is an algebraic way of identifying EPR-like entanglement between $A$ and $A^c$.

Throughout the rest of this paper, we consider a fixed configuration of opened bulk legs in the tensor network, leading to a unique operator map $\iota$. This simplifies the construction of the inductive limit, where we consider the image of all operators acting on the subregion, i.e., $\cH_a \otimes\cH_{\c_a}\otimes\cH_{\g_A}$. The algebra of operators acting on $\cH_{\c_a}$ then includes both the stabilizers and their conjugate error operators, i.e., those that anti-commute with at least one stabilizer. While this includes operators $O_{\c_a}$ that do not preserve the codespace, the choice of including or excluding such operators will not affect the algebra type in our setting of finitely-entangled bulk states. This is because we may think of the Hilbert space $\cH_{\c_A}$ simply as additional bulk qubits in our entanglement wedge $a$ that encode degrees of freedom beyond the codespace.

\begin{figure*}[ht]
    \centering
    \includegraphics[width=0.75\textwidth]{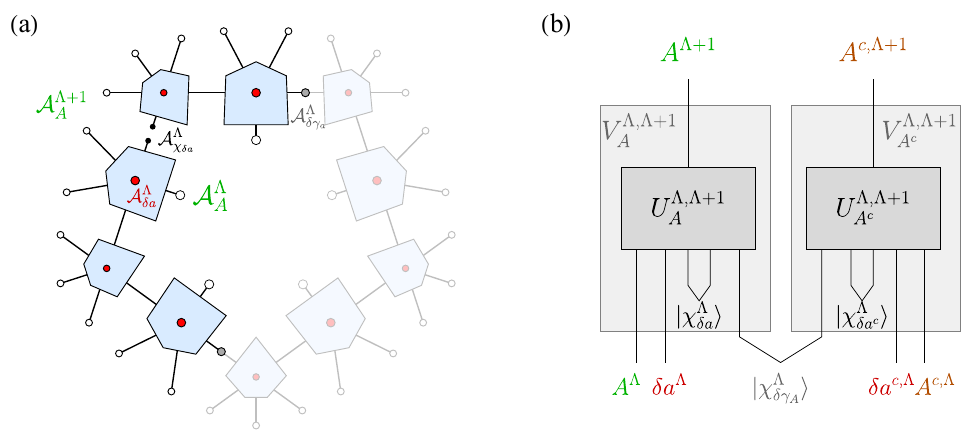}
    \caption{Subregion algebra mapping with one layer of the HaPPY model.
    (a) A single vertex inflation layer of the ``opened-up'' HaPPY code of Fig.\ \ref{fig:happy_to_unitary_algebras}, acting as a unitary map from the subregion algebra $\mathcal{A}_A^\Lambda$ at layer $\Lambda$ and the algebras of the degrees of freedom of the new layer, the bulk algebra $\mathcal{A}_{\d a}^\Lambda$ (red), wedge ancilla algebra $\mathcal{A}_{\chi_{\d a}}^\Lambda$ (black), and Ryu-Takayanagi algebra $\mathcal{A}_{\d \gamma_A}^\Lambda$ (gray) to the subregion algebra $\mathcal{A}_A^{\Lambda+1}$ on the next layer.
    (b) The generic form of a layer of the HaPPY code with ancillas, written as a circuit diagram with the two unitary subregion maps $U_A^{\Lambda,\Lambda+1}$ (highlighted in (a)) and $U_{A^c}^{\Lambda,\Lambda+1}$.
    }
    \label{fig:happy_to_unitary_algebras_layer}
\end{figure*}

\subsection{Mapping algebras layer by layer}
In Sec.\ \ref{sec:happy_fixed_layer} we studied the HaPPY code at a fixed number of layers $\L$.
We found that after splitting the boundary into subsystems $A,A^c$, we can
further divide it into three subsystems associated to the entanglement wedge
algebra $A_a$, the $RT$ algebra $A_{\g_A}$ and the wedge ancilla algebra
$A_{\c_a}$. This division is implemented by an unitary $U_A^{\dagger}$. Now we
can grow the HaPPY code that consists of $\L$ layers by contracting it with
another layer of the tensor network. Here we will repeat the procedure and push
operators of layer $\L$ through to layer $\L+1$. We will from now on indicate the layer at which a given object lives by the superscript $\L$.
We will also refer to the process of mapping operators at layer $\L$ to layer
$\L+1$ as operator pushing, but one should be aware that this layer-to-layer
pushing, indicated by the map $\p^{\L,\L+1}$, is different from the
bulk-to-boundary pushing $\iota$ as it maps the whole boundary of a given layer
to the next, not just the bulk operator to the boundary.
We can represent the tensor network of the
outermost layer again as an unitary map 
\begin{equation}
U^{\L,\L+1}_{A} : \cH_A^{\L} \otimes \cH_{\d a}^{\L+1} \otimes \cH^{\L+1}_{ \c_{\d a}}
\otimes \cH^{\L+1}_{\d \g_A} \rightarrow \cH_A^{\L+1}   
\end{equation}
that takes the boundary $\cH_A^{\L}$ at layer $\L$, the logical information of
the next layer $\cH_{\d a}^{\L+1}$ and further auxiliary and RT degrees of freedom
$\cH^{\L+1}_{\c_{\d a}} \otimes \cH^{\L+1}_{\d \g_A}$ and embeds them into the next layer. This induces the isometric map $\g^{\L,\L+1}$ we discussed in the context of inductive systems. Following the discussion
of the previous section, we could also consider $\g^{\L,\L+1}$ to be made from the isometry $V^{\L,\L+1} =
U^{\L,\L+1}_A \ket{\c_{\d a}}$ as again arising from the contraction of some
auxiliary entangled pairs into the unitary but keeping the unitary picture is
convenient for the following discussion as it provides an direct decomposition
of the boundary algebra of $A$ at any layer.  We will comment on this more in Sec.\ \ref{sec:inductive_limits}. We remind the reader that the boundary subregion $A$
is chosen such that its successive mappings under operator pushing have a support that satisfies complementary recovery.
Because of this, we will not
distinguish between $A$, the subregion of the boundary we considered at layer
$\L$ and its lightcone $J^+(A)$ at later layers and collectively denote the
subregion as $A$, where the respective lightcone is implicit by the
layer-label $\L$. Since we rephrased the growing of the network as a unitary embedding of the
old network with additional degrees of freedom corresponding to the added layer, it is
clear that we have an embedding of algebras 
\begin{equation}
     U^{\L,\L+1}_A \cA_A^{\L} U^{\L,\L+1 \dagger}_A  \subset
     \cA^{\L+1}_A. 
\end{equation}
Now we have to decide how the state grows. We will suppress the layer index $\L$ for degrees of freedom that are added by the new layer. For the bulk, we choose that the bulk qubits are, at each layer, put in reference states where there is no entanglement between the bulk entanglement wedges $a$ and $a^c$, so that at each layer the "new" bulk qubits come in product states $\ket{  i_{\d a}},\ket{ j_{\d a^c}}$. We note that this is an arbitrary choice we made and one could also consider states with bulk entanglement. Then one would have to be more careful about the structure of the entanglement to compute the type. In addition, the network representing the bulk-to-boundary isometry at layer $\L+1$ is obtained by projecting additional Bell pairs $\ket{\c_{\d a}}\ket{\c_{\d a^c}}$ onto the open legs of $U^{\L,\L+1}_A U^{\L,\L+1}_{A^c}$ that were opened up to generate the unitaries, as well as additional Bell pairs $\ket{\c_{\d \g_A}}$ into the legs that extend the RT surface from layer $\L$ by a new bond.
Therefore, the state on $\cH^{\L+1}$ is 
\begin{equation}
    \ket{\y}_{\L+1} = U^{\L,\L+1}_A U^{\L,\L+1}_{A^c} \ket{\y}_{\L}
    \ket{i_{\d a}}\ket{j_{\d a^c}}\ket{\c_{\d a}}\ket{\c_{\d a^c}}
    \ket{\c_{\d \g_A}} 
\end{equation}
and defines the same state on the image of $\cB(\cH^{\L})$ as the state of the
previous layer, i.e.,  for
\begin{equation}\label{eq:embedding_local_operators}
    O^{\L+1} = U^{\L,\L+1}_A O^{\L} U_A^{\L,\L+1 \dagger}, O^{\L} \in
    \cB(\cH_A^{\L})
\end{equation}
we have
\begin{equation}\label{eq:commutativity_operator_state_pushing}
    \bra{\y}_{\L+1} O^{\L+1}\ket{\y}_{\L+1} = \bra{\y}_{\L} O^{\L}\ket{\y}_{\L}.
\end{equation}
This only works because $U^{\L,\L+1}_A$ is a unitary. Note that, because of this preservation of the state of the previous layer, as well as having a unitary embedding, the algebra of the previous layer is preserved into the next, i.e. 
\begin{equation}\label{eq:algebra_layer_decomp}
    \cA_A^{\L+1} \cong \cA^\L_A \otimes \d \cA_A^{\L+1},
\end{equation}
so the algebra of the new layer decomposes into the algebra of the previous layer together with addititional degrees of freedom $\d \cA_A^{\L+1}$. 
In summary, we obtain a sequence of algebras $\{\cA_A^{\L}\}_{\L \in \mathbb{N}}$
and a sequence of states on each algebra $\{\ket{\y}_\L\}_{\L \in \mathbb{N}}$,
where we have an embedding $\phi^{\L,\L+1}(\cA^{\L}_A) \subset \cA^{\L+1}_{A}$ implemented
by the unitary transformation \eqref{eq:embedding_local_operators}.
We see that, due to the unitary nature of the embedding, we can at each layer
decompose $\cA_A^{\L}$ into the tensor product 
\begin{equation}
    \cA_A^{\L} \sim \cA_a^{\L} \otimes \cA_{\c_a}^{\L} \otimes \cA^{\L}_{\g_A}
\otimes \id_{A^c}
\end{equation}
with its commutant 
\begin{equation}
    \cA_{A^c}^{\L} \sim \id_A \otimes \cA^{\L}_{\g_{A^c}} \otimes \cA_{a^c}^{\L} \otimes
 \cA_{\c_{a^c}}^{\L}
\end{equation}
on which the tensor network state takes the form
\begin{equation} \label{eq:boundary_state_decomposition}
    \ket{\Y^{\L}} \sim \ket{i}_a \otimes \ket{\c_a} \otimes \ket{\c_\g} \otimes
    \ket{j} \otimes \ket{\c_{a^c}},
\end{equation}
where $i(j)$ is the bulk logical state in $a(a^c)$, $\c_{a(a^c)}$ is the state of the internal
auxiliary degrees of freedom in the entanglement wedge $a(a^c)$ and $\c_\g$ is the
maximal entangled state that comes from the contraction of the two sides of the
tensor network along the RT surface $\g$. This decomposition is preserved between layers via Eq. \eqref{eq:algebra_layer_decomp}. We note that all the entanglement entropy
of the two boundary sides $A, A^c$ comes from $\c$, if the bulk state
$\ket{ij}_{aa^c}$ is pure. Because we rewrote the construction of the network through opening
the legs associated to $\c_a, \c_{a^c}$, this split of the boundary state
is obvious. 

\subsection{Limit algebras}

We can now take the inductive limit of the procedure defined above. It is clear that, as we grow the network, due to the unitary
nature of the embedding, the boundary state always decomposes as in
\ref{eq:boundary_state_decomposition} at every layer and one has essentially
the same setup as in the Araki-Woods-Powers factors described in Appendix \ref{sec:type_classification} just with tensor
products between the different kinds of spin chains and that the decomposition
is manifestly true only after the application of local unitaries
$U_A,U_{A^c}$. Note that one has unentangled
logical and auxiliary degrees of freedom $ \ket{ij},\ket{\c_a},\ket{\c_{a^c}}$
corresponding to the type I case and a maximally entangled state for the
algebra $\cA_{\g}$ as in the type II$_1$ case. We therefore expect that, as we
increase the number of layers to infinity, given a pure bulk-input state
$\ket{ij}$, the algebra $\cA_{A}$ in the direct limit Hilbert space
becomes type II$_{\infty}$, because it reduces to the algebra of the form $B(\cH) \otimes \text{II}_1$ for $\cH$ encoding the Hilbert space built out of auxiliary and wedge degrees of freedom in $a$ and the II$_1$ factor acting on the RT surface $\gamma$. 

Let us now make this statement precise. We can first consider the inductive-limit algebra $\mathcal{A}_{\gamma_A}$. Let $A^\Lambda,B^\Lambda\in\mathcal{A}_{\gamma_A}^\Lambda.$ The state $\chi_A$ is maximally mixed on $\mathcal{A}_{\gamma_A}^\Lambda$, so that \begin{align}\bra{\chi_A}[A^\Lambda,B^\Lambda]\ket{\chi_A}=0.\end{align} Since \begin{align}\hat{\mathcal{A}}_{\gamma_A}=\varinjlim \mathcal{A}_{\gamma_A}^\Lambda,\end{align} we deduce by continuity that $\ket{[\chi_{\gamma_A}^\Lambda]}$ induces a tracial state on $\hat{\mathcal{A}}_{\gamma_A}$. Now $\ket{[\Psi_{\gamma_A}^\Lambda]}$, which restricts to $\ket{[\chi_{\gamma_A}^\Lambda]}$ on $\hat{\mathcal{A}}_{\gamma_A}$, is a state on the inductive-limit Hilbert space, so it is normal on $\mathcal{A}_{\gamma_A}=\pi(\hat{A}_{\gamma_A})^{\prime\prime}$. Moreover, it is tracial on a weak-operator dense subalgebra of $\mathcal{A}_{\gamma_A}$, so it extends by continuity for the weak operator topology to a tracial normal state on $\mathcal{A}_{\gamma_A}$. From this we deduce:
\begin{theorem}
    The inductive-limit RT von Neumann algebra $\mathcal{A}_{\gamma_A}$ has type II$_1$.
\end{theorem}

By an exactly similar reasoning, since the states $\ket{[\chi_a]}$ and $\ket{[i]}$ are pure on $\mathcal{A}_{\chi_a}$ and $\mathcal{A}_{a}$, we deduce  
\begin{theorem}
    The inductive-limit ancilla and bulk von Neumann algebras $\mathcal{A}_{\chi_a}$ and $\mathcal{A}_{a}$ have type $I$.
\end{theorem}

We can then deduce the type of full boundary algebra from the following observation (see for example \cite{van2024convergence}): the algebra $\hat{\mathcal{A}}_A$ can be decomposed as \begin{align}\hat{\mathcal{A}}_A=\hat{\mathcal{A}}_{a,\chi_a}\otimes\hat{\mathcal{A}}_{\gamma_A},\end{align} where the tensor product of $C^\ast$-algebras is unambiguously defined because all considered algebras are nuclear. We then have \begin{align}\mathcal{A}_A=\mathcal{A}_{a,\chi_a}\overline{\otimes}\mathcal{A}_{\gamma_A}.\end{align} Since the first tensor factor has type I and the second tensor factor has type II$_1$, we deduce \begin{theorem}The boundary subregion algebra $\mathcal{A}_A$ has type II$_\infty$.\end{theorem}
A useful way of seeing this result in view of the next section is that we have decomposed $\mathcal{A}_A$ (using unitary equivalence) into an infinite tensor product of finite-dimensional factors, where the tensor network state is pure on some of them (corresponding to bulk inputs and ancillas), whereas it is maximally entangled on others (the RT degrees of freedom). The Araki--Woods classification of infinite tensor products then tells us that the algebra $\mathcal{A}_A$ has type II$_\infty$.
Note that the above results will also hold if the bulk state $\ket{ij}$
carries an $\cO(1)$ amount of entanglement, where the counting parameter is the
number of layers $\L$. If the bulk state carries a divergent amount of
entanglement, it will be able to change the type of the resulting bulk algebra, depending on its
entanglement structure. We note that the geometrical entanglement of the RT
surface will always lead to a type II factor $\cA_{\g}$ associated to the
RT surface due to maximal entanglement in the state that glues the two wedges
$a$ and $a^c$.

In the next section we will explain how our discussion extends to a more general class of tensor networks with a layered structure that satisfy complementary recovery.
\section{An abstract perspective}
\label{sec:inductive_limits}
In the previous discussion, we focused on the HaPPY code that we could open up to write the operator pushing from bulk to boundary and between layers as an explicit conjugation by a unitary. In this section, we identify the mathematical barebones of our construction. In the first section, we explain, following \cite{Faulkner:2020hzi,Gesteau:2022hss,Faulkner:2022ada,Gesteau:2023hbq} that the structure of a holographic code with complementary recovery can be formalized as a code subspace-preserving conditional expectation. We then show that this conditional expectation structure can be leveraged to define a general notion of inductive system of codes, for which results akin to the ones derived in the previous section hold.

\subsection{Codes and conditional expectations}

Holographic tensor networks truncated at a finite layer number form holographic codes with complementary recovery. Throughout this section, we will denote the bulk
``code'' Hilbert space by $\mathcal{H}$. Specifying the local algebra and Hilbert space associated with a bulk subregion $a$ (at a finite cutoff) will be done by the labels $\mathcal{H}_a$ and $\mathcal{A}_a$. We will label boundary ``physical'' Hilbert spaces with $\mathcal{K}$ and we will label the boundary algebra with $\mathcal{A}_\cK$. Choosing a boundary subregion $A$, we will label the local Hilbert space and algebra with $\mathcal{K}_A$ and $\mathcal{A}_A$. No subscripts will denote the full boundary or bulk objects.

The bulk-to-boundary isometry of a code will be labeled $V:\mathcal{H}
\rightarrow \mathcal{K}$. Given a subregion $A$ on the boundary, there exists a
region $a$ in the bulk, the entanglement wedge of $A$ such that $\mathcal{A}_a$,$\mathcal{A}^\prime_a$ 
are recoverable in $\mathcal{A}_A$ and $\mathcal{A}^\prime_A$, respectively. This supplies us with operator pushing
maps $\iota_a, \iota^\prime_a$, which are faithful unital $\star$-homomorphisms
$\iota_a:\mathcal{A}_a\rightarrow \mathcal{A}_A$, $\iota^\prime_a:\mathcal{A}^\prime_a\rightarrow \mathcal{A}^\prime_A$. Defining $\alpha:\mathcal{A}\rightarrow\mathcal{O}$ by $\alpha(x) = V^\dag x V$, the map $\iota_a\circ\alpha:\mathcal{A}_A \rightarrow \iota_a{\mathcal{A}_a}$ is a conditional expectation from $\mathcal{A}_A$ onto the image of $\iota_a$, i.e. a linear map with $E(\id) = \id$ and 
\begin{equation}\label{eq:cond_exp}
    E(abc) = a E(b) c,\forall a,c \in \cN, b \in \cM.
\end{equation}
Other work on the connection of error-correcting codes with conditional
expectations appeared in
\cite{Faulkner:2020hzi, Faulkner:2022ada, Furuya:2020tzv, Gesteau:2023hbq}. Similarly, the map $\iota^\prime_a\circ\alpha$ is a conditional expectation onto $\iota^\prime_a{\mathcal{A}^\prime_a}$.
Furthermore, we assume that all algebras involved are factors. The existence of a conditional expectation guarantees that $\mathcal{A}_A \cong
\iota_a(\mathcal{A}_a)\otimes \iota_a(\mathcal{A}_a)^c$ \cite{Stratila_2020}, where we use $\cong$ to denote unitary equivalence and where the second term denotes the relative commutant in
$\mathcal{A}_A$,
\begin{equation}
\iota_a(\mathcal{A}_a)^c := \iota(\cA_a)' \cap \cA_{A}.   
\end{equation}
The first term in $\cA_A$ is the set of those operators acting on the Hilbert space of logical states, and the second term is the operators acting on $\mathcal{H}_A$ which do not affect the logical degrees of freedom. 
Since, in the case of finite layers, all the above algebras are Type I factors, the existence of a conditional expectation guarantess, see Sec. 9.15 in \cite{Stratila_2020}, that 
\begin{equation}
\cK_A \cong \cH_a \otimes \cK_c,
\end{equation}
where $\cK_c$ denotes the space on which the relative commutant acts, i.e. $\iota_a(\cA_a)^c = \cB(\cK_c)$.
These unitary equivalences can be seen as an algebraic version of Harlow's theorem \ref{eq:Harlow's_thm} that appears if the operator pushing map is unital.
In the language of the previous section, this isomorphism is
implemented by the conjugation of operators in $\cA_A$ with $U_A$. As before, we now
assume that a particular isomorphism has been chosen at every layer.
We can write down the subsystem decomposition for the full physical space 
    \begin{align}\label{eq:phys_decomp} 
        \mathcal{K} &\cong
        \mathcal{H}_a\otimes\mathcal{H}_{\overline{a}}\otimes\mathcal{K}_c\otimes
        \mathcal{K}_{\overline{c}}\\
        \mathcal{A} &\cong \mathcal{A}_a\otimes\mathcal{A}_{\overline{a}}\otimes\mathcal{A}_c\otimes \mathcal{A}_{\overline{c}}
    \end{align}

The conditional expectations $\iota_a\circ\alpha$, $\iota^\prime_a\circ\alpha$ project $\mathcal{A}_A$ onto $\mathcal{A}_a$ and $\mathcal{A}^\prime_A$ onto $\mathcal{A}^\prime_a$, respectively. States in the code subspace are invariant under these conditional expectations, i.e. they can be written as $\ket{\psi}_{code}\otimes\ket{\chi}$, where $\ket{\psi}_{code}\in \mathcal{K}_a\otimes\mathcal{K}_{\overline{a}}$ and $\ket{\chi}$ is one, fixed reference state on $\mathcal{K}_c\otimes\mathcal{K}_{\bar{c}}$. The analog of $\ket{\chi}$ in the case of the HaPPY code at one fixed layer truncation is the tensor product of the RT state and the ancilla state.

We will now show how to study the growth of such a system once suitable inductive maps are defined. 

\subsection{Inductive systems of codes}
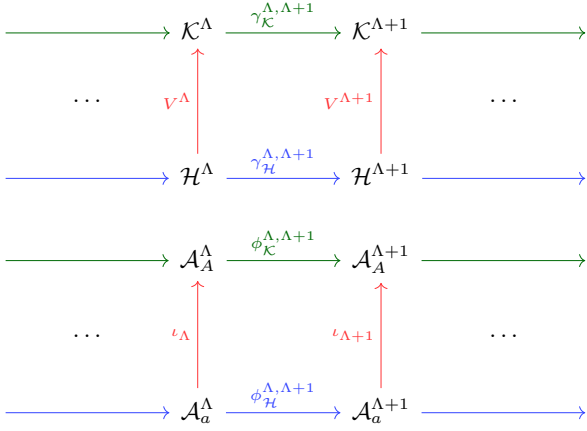
\begin{figure}[ht!]
\centering
\begin{tikzcd}
	{} && {\mathcal{K}^{\Lambda}} && {\mathcal{K}^{\Lambda+1}} && {} \\
	& \dots &&&& \dots \\
	{} && {\mathcal{H}^{\Lambda}} && {\mathcal{H}^{\Lambda+1}} && {} \\
	{} && {\mathcal{A}_A^\Lambda} && {\mathcal{A}_A^{\Lambda+1}} && {} \\
	& \dots &&&& \dots \\
	{} && {\mathcal{A}_a^{\Lambda}} && {\mathcal{A}_a^{\Lambda+1}} && {}
	\arrow[color={rgb,255:red,0;green,107;blue,7}, from=1-1, to=1-3]
	\arrow["{\gamma_\mathcal{K}^{\Lambda,\Lambda+1}}", color={rgb,255:red,0;green,107;blue,7}, from=1-3, to=1-5]
	\arrow[color={rgb,255:red,0;green,107;blue,7}, from=1-5, to=1-7]
	\arrow[color={rgb,255:red,51;green,68;blue,255}, from=3-1, to=3-3]
	\arrow["{V^\Lambda}", color={rgb,255:red,255;green,51;blue,54}, from=3-3, to=1-3]
	\arrow["{\gamma_{\mathcal{H}}^{\Lambda,\Lambda+1}}", color={rgb,255:red,51;green,68;blue,255}, from=3-3, to=3-5]
	\arrow["{V^{\Lambda+1}}", color={rgb,255:red,255;green,51;blue,54}, from=3-5, to=1-5]
	\arrow[color={rgb,255:red,51;green,68;blue,255}, from=3-5, to=3-7]
	\arrow[color={rgb,255:red,0;green,107;blue,7}, from=4-1, to=4-3]
	\arrow["{\phi_{\mathcal{K}}^{\Lambda,\Lambda+1}}", color={rgb,255:red,0;green,107;blue,7}, from=4-3, to=4-5]
	\arrow[color={rgb,255:red,0;green,107;blue,7}, from=4-5, to=4-7]
	\arrow[color={rgb,255:red,51;green,68;blue,255}, from=6-1, to=6-3]
	\arrow["{\iota_\Lambda}", color={rgb,255:red,255;green,51;blue,54}, from=6-3, to=4-3]
	\arrow["{\phi_\mathcal{H}^{\Lambda,\Lambda+1}}", color={rgb,255:red,51;green,68;blue,255}, from=6-3, to=6-5]
	\arrow["{\iota_{\Lambda+1}}", color={rgb,255:red,255;green,51;blue,54}, from=6-5, to=4-5]
	\arrow[color={rgb,255:red,51;green,68;blue,255}, from=6-5, to=6-7]
\end{tikzcd}
\caption{Commutative diagram summarizing the structure required for an inductive limit of codes. The sequence of logical Hilbert spaces and their isometries is shown on the top diagram, while the sequence of algebras and their operator pushing maps is shown on the bottom diagram. We ask that the arrows of the same color on the commutative diagram satisfy the compatibility conditions \eqref{eq:pushing_layer_commutativity} (for the red ones), and \eqref{eq:pushing_layer_commutativity2} (for the blue and green ones). A similar diagram to the bottom one must also hold for commutant algebras and maps.}
\label{fig:commutatived}
\end{figure}

We now explain how to take the inductive limit of a family of exact holographic codes. The data we want are: \begin{enumerate}
\item A sequence of logical Hilbert spaces $\mathcal{H}_\Lambda$, which should be seen as the Hilbert spaces of bulk logical legs for a network truncated at layer $\Lambda$.
\item A sequence of physical Hilbert spaces $\mathcal{K}_\Lambda$, which should be seen as the Hilbert spaces of boundary legs for a network truncated at layer $\Lambda$.
\item Bulk-to-boundary isometries $V_\Lambda$ for each truncation at layer $\Lambda$. They are the usual holographic maps defined by holographic tensor networks.
\item Bulk-to-bulk isometries $\gamma_{\mathcal{H}}^{\Lambda,\Lambda+1}$. They correspond to enlarging the bulk Hilbert spaces with more bulk qubits put in a (usually disentangled) reference state.
\item Boundary-to-boundary isometries $\gamma_{\mathcal{K}}^{\Lambda,\Lambda+1}$. They correspond to enlarging the boundary Hilbert spaces by acting with one layer of the tensor network, the bulk qubits being each put in the same reference state as the one chosen for the bulk-to-bulk maps.
\item A sequence of logical algebras $\mathcal{A}_a^\Lambda$, $\mathcal{A}_a^{\prime\Lambda}$ which should be seen as the operators acting on of bulk logical legs on either side of the RT surface for a network truncated at layer $\Lambda$.
\item A sequence of physical algebras $\mathcal{A}_A^\Lambda$, $\mathcal{A}_A^{\prime\Lambda}$, which should be seen as the operators acting on the boundary on either side of the RT surface for a network truncated at layer $\Lambda$.
\item Bulk-to-boundary $\star$-homomorphisms $\iota_\Lambda$ for each truncation at layer $\Lambda$. They are the usual operator pushing maps defined by holographic tensor networks.
\item Bulk-to-bulk $\star$-homomorphisms $\phi_{\mathcal{H}}^{\Lambda,\Lambda+1}$ and $\phi_{\mathcal{H}}^{\prime\Lambda,\Lambda+1}$ for each truncation at layer $\Lambda$. They correspond to tensoring bulk operators with identities on the next layer on either side of the RT surface.
\item Boundary-to-boundary  $\star$-homomorphisms $\phi_{\mathcal{K}}^{\Lambda,\Lambda+1}$ and $\phi_{\mathcal{K}}^{\prime\Lambda,\Lambda+1}$ for each truncation at layer $\Lambda$. They correspond to pushing boundary operators at layer $\Lambda$ through one layer of the tensor network on either side of the RT surface, the bulk qubits being each put in the same reference state as the one chosen for the bulk-to-bulk maps.
\end{enumerate} 
We also require compatibility of the operator reconstruction maps with the bulk-to-boundary isometries, and of the layer-to-layer isometries and operator pushing maps between each other, i.e., for $O\in\mathcal{A}_{\mathcal{H}}^\Lambda,O^\prime\in\mathcal{A}_{\mathcal{H}}^{\prime\Lambda}$,
\begin{equation}\label{eq:pushing_layer_commutativity}
   V_\L O= \iota_\L (O)V_\L,\quad
   V_\L O^\prime= \iota^\prime_\L (O^\prime)V_\L,
\end{equation}
and for $O\in\mathcal{A}_{\mathcal{H},\mathcal{K}}^\Lambda,O^\prime\in\mathcal{A}_{\mathcal{H},\mathcal{K}}^{\prime\Lambda}$,\footnote{Strictly speaking, this last equation is not required to have a well-defined inductive limit code, but it allows to keep track of the RT and auxiliary Bell pair degrees of freedom added at each step.}
\begin{equation}\label{eq:pushing_layer_commutativity2}
   \gamma_{\mathcal{H},\mathcal{K}}^{\L,\L+1} O= \phi_{\mathcal{H},\mathcal{K}}^{\L,\L+1} (O)V_\L,\quad \gamma_{\mathcal{H},\mathcal{K}}^{\L,\L+1} O^\prime= \phi_{\mathcal{H},\mathcal{K}}^{\prime\L,\L+1} (O^\prime)V_\L.
\end{equation}
The above structure is summarized by the commutative diagram in Figure \ref{fig:commutatived}.
All these operator equations are also assumed to hold on commutant algebras.
First, Equation \eqref{eq:pushing_layer_commutativity} implies that \begin{equation}
    V^{\L} \ket{\varphi} \cong \ket{\varphi} \otimes \ket{\Y^\L},
\end{equation}
where $\Y_\L$ is a reference state which is identified in the case of the HaPPY code with the RT and auxiliary Bell pair degrees of freedom at layer $\L$. Second, Equation \eqref{eq:pushing_layer_commutativity2} implies that the layer-to-layer maps \textit{also} implement error-correcting codes with complementary recovery, both at the level of the bulk and at the level of the boundary. We therefore deduce from the inherited conditional expectation structure that in the bulk, 
\begin{equation}
    \mathcal{H}^{\Lambda+1} \cong \mathcal{H}^\Lambda \otimes \d \mathcal{H}^\Lambda,
\end{equation}
and
\begin{equation}
    \gamma_\cH^{\Lambda,\Lambda+1}\ket{\psi} \cong
    \ket{\psi}\otimes\ket{\Omega^{\Lambda}} \in \mathcal{H}^{\Lambda+1},
\end{equation}
with $\ket{\Omega^{\Lambda}}$ fixed. Similarly on the boundary,
\begin{equation}
    \begin{aligned}
        \cK^{\L+1} &\cong \cK^{\L} \otimes \d \cK^{\L},\\
        \gamma_\cK^{\Lambda,\Lambda+1}\ket{\psi} &\cong
        \ket{\psi}\otimes\ket{\Q^{\Lambda}} \in \mathcal{K}^{\Lambda+1}.
    \end{aligned}
\end{equation}
This choice fits the structure of the HaPPY code, where $\ket{\Theta^\L}$ corresponds to the extra bulk qubits and $\ket{\Q^{\L}}$ corresponds to the extra RT- and auxiliary Bell pairs.
We can say more about this structure by recognizing that because of the
compatibility of operator pushing between layers and bulk-to boundary operator
pushing combined with the compatibility of operator pushing with the
layer-to-layer Hilbert space isometries
\eqref{eq:pushing_layer_commutativity},\eqref{eq:compatibility} 
we also have a conditional expectation that decomposes
\begin{equation}
    \d \cK^{\L} \cong \d \cH^{\L} \otimes \d \bar{\cK}^{\L} 
\end{equation}
such that 
\begin{equation}
    \ket{\Q^{\L}} \cong \ket{\W^{\L}} \otimes \ket{\Y^{\L}},
\end{equation}
where $\ket{\Y}$ are all ``new'' degrees of freedom that come from growing the
code, such as the extra RT-pairs and auxiliary degrees of freedom in the HaPPY code.

\subsection{Limit algebras}

With this structure, we define the inductive sequence of bulk and boundary Hilbert spaces
$\{\mathcal{H}^\Lambda, \gamma_\cH^{\Lambda,
\Lambda+1}\},\{\mathcal{K}^\Lambda, \gamma_\cK^{\Lambda, \Lambda+1}\}$, and algebras $\{\mathcal{A}_a^\Lambda, \phi_\cH^{\Lambda,
\Lambda+1}\},\{\mathcal{A}_A^\Lambda, \phi_\cK^{\Lambda, \Lambda+1}\},\{\mathcal{A}_a^{\prime\Lambda}, \phi_\cH^{\prime\Lambda,
\Lambda+1}\},\{\mathcal{A}_A^{\prime\Lambda}, \phi_\cK^{\prime\Lambda, \Lambda+1}\}$. The choice of how
to grow the inductive system is fully contained in the choice of
$\ket{\Omega^\Lambda},\ket{\Q^{\L}}$. For example, if we consider the HaPPY code and choose the
$\ket{\Omega^\Lambda}$ to be just the $\ket{0}$ state on the additional bulk legs,
one can think of the inductive limit of the bulk as the Hilbert space of bulk
states which are asymptotically in the $\ket{0}$ state.
To construct the inductive limit bulk and boundary Hilbert spaces and algebras from the
$\Lambda$-layer Hilbert spaces, we now take the direct limit accordingly: 
\begin{equation} \mathcal{H} \equiv \varinjlim \mathcal{H}^\Lambda,\qquad \mathcal{K} \equiv \varinjlim \mathcal{K}^\Lambda,\end{equation} and similarly for the algebras of observables. The above compatibility relations ensure that these algebras of observables have a valid representation on the inductive limit Hilbert space, so that one can take their bicommutant and construct an inductive limit von Neumann algebra. Moreover, the bulk-to-boundary maps $\iota_\Lambda$ and $\iota^\prime_\Lambda$ also extend to the inductive limit, and can be extended by continuity to unital normal $\star$-homomorphisms, so that the conditional expectation structure is preserved in the limit. If the states $\ket{\Theta^\Lambda}$ and $\ket{\Omega}^\Lambda$ have a similar structure to the HaPPY code, we can compute the types of the various algebras appearing in this section in a similar way, by essentially reducing the calculation of the type to an Araki--Woods-like situation, where the $\ket{\Omega^\Lambda}$ and $\ket{\Theta^\Lambda}$ lead to a decomposition into an infinite tensor product of finite density matrices. In the case of the HaPPY code, we found that the $\ket{\Omega^\Lambda}$ were pure, while the $\ket{\Theta^\Lambda}$ had a maximally entangled part, which led to the type II$_\infty$ structure. In more general cases, the entanglement properties of the $\ket{\Omega^\Lambda}$ and $\ket{\Theta^\Lambda}$ similarly lead to the type of the algebra through the Araki--Woods classification.
\section{Examples}\label{sec:examples}

    \subsection{HaPPY code from Majorana dimers}\label{sec:HaPPY_dimers}
        Here 
        we describe a specific instance of the HaPPY
        code in terms of Majorana dimers as described in \cite{Jahn:2019nmz}.
        Although it does not add any new conclusions to the type of boundary algebras in the HaPPY code, we develop techniques that can be applied to analyze the algebras of other networks that are not based on perfect tensors but arise from the contraction of dimer states. Additionally, this
        provides us with a graphical understanding of the codespace and local algebras on the boundary of the HaPPY code. In short, Majorana dimer states are states in qubit systems that have a graphical representation in terms of graphs, where each edge corresponds to a single qubit and has two nodes on it, which represent Majorana operators. Each node is connected to a different node by a \textit{dimer}, which indicates that a fermionic annihilation operator build of the Majorana operators associated to the two nodes annihilates the state. As we demonstrate below, one can associate a single qubit to such a pair of nodes connected by a dimer.
        A review of Majorana dimers is given in Appendix \ref{app:dimer_review}.
        We use the following encoded representation of logical states as dimer states
        
        \begin{align}
            \label{eq:HAPPY_ZERO}
                \ket{\bar{0}}_5\; = \quad
                \begin{gathered}
                \includegraphics[height=0.12\textheight]{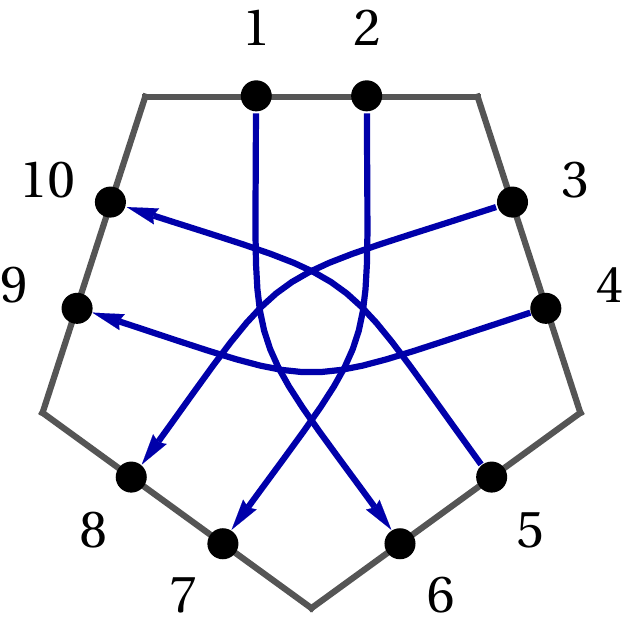}
                \end{gathered} \\
                \label{eq:HAPPY_ONE}
                \ket{\bar{1}}_5\; = \quad
                \begin{gathered}
                \includegraphics[height=0.12\textheight]{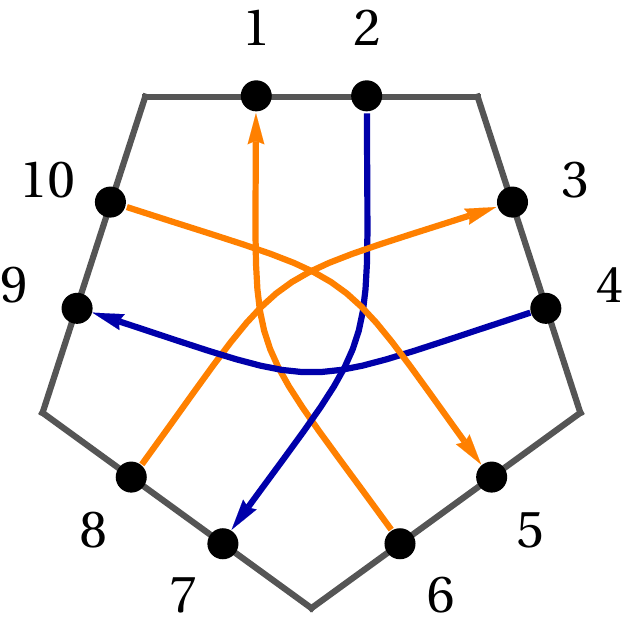}
                \end{gathered} 
        \end{align}

        In the following, we will describe how one can interpret the previous
        considerations explicitly in the dimer picture of the HaPPY code by giving a graphical interpretation of the unitaries $U_A,U_{A^c}$ of Harlow's theorem \eqref{eq:Harlow's_thm}. 
        
        \subsubsection{Disentangling the bulk}
        As a first step in the explicit construction of $U_A$ of equation
        \eqref{eq:boundary_unitary}, we need to identify how the bulk logical
        information is encoded in the boundary state.
        For this we note that each dimer originates originally from some bulk qubit. For each bulk qubit of $a$, we will have some dimers coming from the local tensor that pierce the RT surface and some dimers that stay in the subgregion, thus they begin and end in $A$. 
        We want to associate some particular dimer that stays in $A$ with the
        information carried by this bulk qubit. To do so, we note the following.

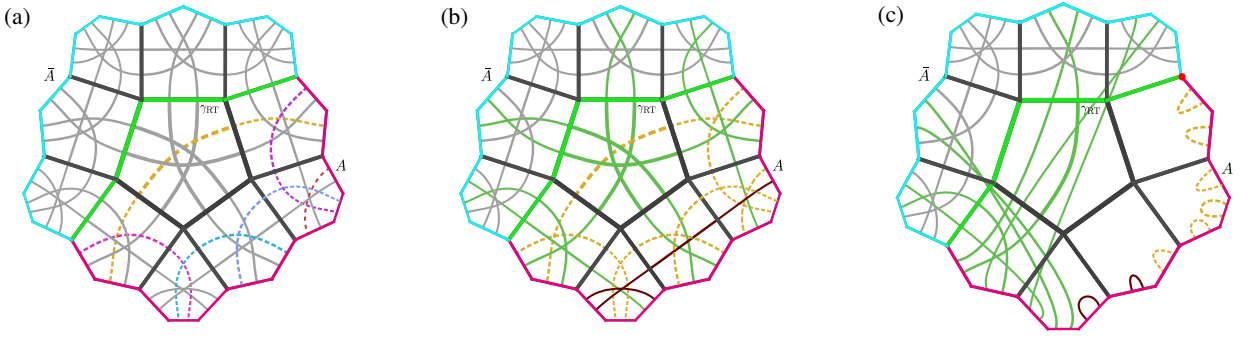
\begin{figure*}[ht]
\centering
\begin{tikzpicture}
  \node at (0,0) {\includegraphics[width=0.25\textwidth]{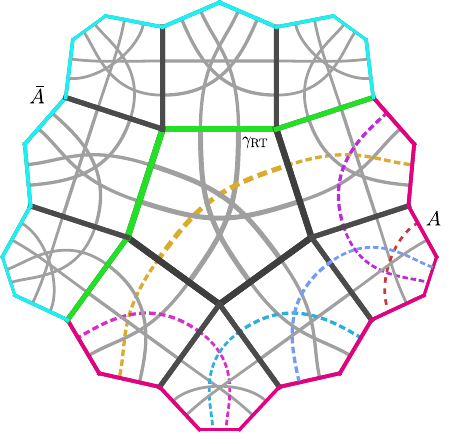}};
  \node at (-2.3,2) {(a)};
\end{tikzpicture}
\hspace{0.04\textwidth}
\begin{tikzpicture}
  \node at (0,0) {\includegraphics[width=0.25\textwidth]{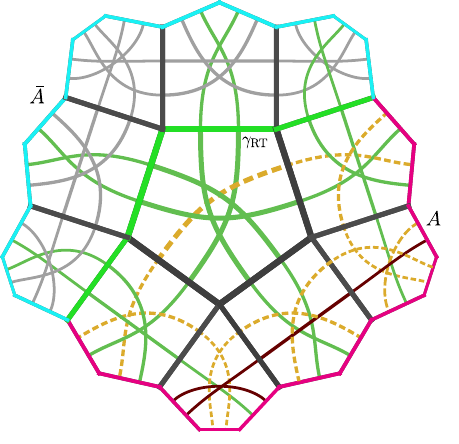}};
  \node at (-2.3,2) {(b)};
\end{tikzpicture}
\hspace{0.04\textwidth}
\begin{tikzpicture}
  \node at (0,0) {\includegraphics[width=0.25\textwidth]{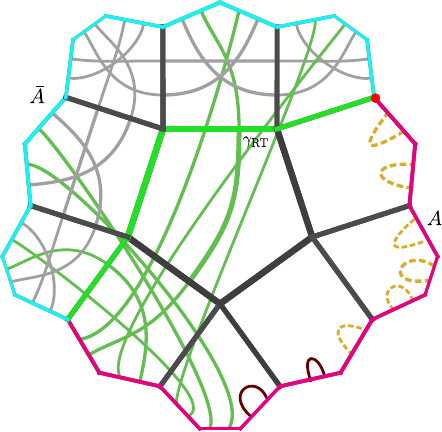}};
  \node at (-2.3,2) {(c)};
\end{tikzpicture}
        \caption{(a) Two boundary regions $A,A^c$ with their respective
            entanglement wedges and RT surface $\g_{\text{RT}}$ and dimers carrying logical information drawn with colored dashed lines. Note that there is one dimer for each bulk qubit.
            (b) HaPPY code with different dimers in region $A$ colored. Green are dimers belonging to $D_{\g}$, red are dimers that belong to $D^A_a$ and and dashed yellow are the logical dimers in $D^A_l$. The parity of the dimers was neglected in this figure.
            (c) Dimers in $A$ after disentangling logical and auxiliary
            dimers by applying local swap operations.}
\label{fig:HaPPY_logical_dimers}
\end{figure*}

        \begin{theorem}\label{Thm_1}
            In the HaPPY code represented by Majorana dimers on a $\{5,4\}$ tiling of
            the hyperbolic plane, there exists a collection of dimers beginning and ending in $A$ of which the parities are different between any two basis states of the codespace, that differ only in bulk qubits in the entanglement wedge of our subregion, independent of the state in the complementary wedge.
        \end{theorem}
        This theorem is proven in Appendix \ref{app:prf_thm_2}. Note that we do not mean that given just the parities of the logical dimers, the bulk logical state can be trivially read of, i.e. we do not mean that if a particular bulk is in state $\ket{1}$ that the associated logical dimer will have its parity inverted compared to the state $\ket{0}$ but that the collection of parities of logical dimers is in one-to-one correspondence with the logical state.
        This theorem states that there is a collection of dimers that stand in one-to-one correspondence with the logical state of the entanglement
        wedge, independent of what the state is in the complementary wedge, an example
of a collection of such logical dimers is shown in Fig.
        \ref{fig:HaPPY_logical_dimers} (a).

        Following Theorem \ref{Thm_1} we have a set of dimers which we can
        associate with information in the entanglement wedge. What we want to do
        now is to construct a unitary $C^A_i$ that distills this information from
        the codespace state $\ket{\bar{i j}}$, where $i$ is the logical state in
        $a$ and $j$ the logical state in $a^c$.
        To achieve this, we use the following corollary.
        
        \begin{lemma}
        Given the above setup, there exists a local unitary $\tilde{U}_A\otimes \id: \cH_A
        \otimes \cH_{A^c}
        \rightarrow  \cH_A
        \otimes \cH_{A^c} $ such that for any bulk input $\ket{j} \in \cH_{a^c}$: 
        \begin{equation}\label{eq:factorize_i}
            \tilde{U}_A \ket{\bar{ij}} = \ket{i} \ket{\c_a} \ket{\chi_{\g,j}},
        \end{equation}
        where $i$ is the logical state written out in a set of qubits that are
        associated to the logical dimers in theorem \ref{Thm_1},  $\chi_j$ is a $j$- dependent state that carries the entanglement that
         comes from the dimers associated to the RT surface and the complementary
         and $\ket{\c_a}$ is an arbitrary fixed state of an additional set of
         qubits that come from dimers that do not belong to either the logical or
         RT dimers. Furthermore, there exists a local unitary in $A^c$ that satisfies Eq. \eqref{eq:factorize_i} if one swaps $i$ with $j$ and $a$ with $a^c$ on the r.h.s. Their combined action satisfies Harlow's theorem \eqref{eq:Harlow's_thm}
         \begin{equation}
             \tilde{U}_{A^c}\tilde{U}_A \ket{\bar{ij}} = \ket{i}\ket{j} \ket{\c_a} \ket{c_{a^c}}\ket{\chi_{\g}},
         \end{equation}
         where now $\ket{j}$ is associated to logical dimers of the complementary region, $\ket{\c_{a^c}}  $ comes from auxilliary dimers in the complementary region and $\ket{\c_\g}$ is a maximally entangled state made up from all the dimers that cross the RT surface and connected region $A$ with $A^c$.
        \end{lemma}
        
\begin{figure}[ht]
            \includegraphics[scale=0.5]{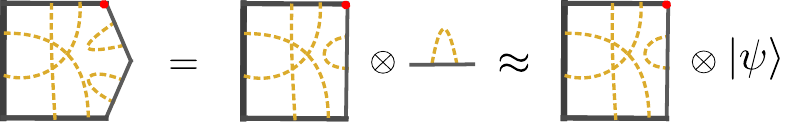}
            \caption{Illustration that a dimer state in which an edge is connected to
            itself next to the pivot can be factorized into a qubit system and the
        remaining dimer state.}
            \label{fig:detach_dimer}
        \end{figure}

        \textit{Sketch of Proof.} The full proof can be found in \ref{app:prf_cor_1}. We give a sketch of it since it conveys the conceptual action of the unitaries $U_A$. One begins by grouping all dimers ending in region $A$
        into groups $D_l^A$ of logical dimers, dimers $D_\g$ that cross the
        RT surface and the remaining dimers $D_a$, as illustrated in Fig.\ \ref{fig:HaPPY_logical_dimers}(b).
        Then, independently of the bulk input, one can perform swap operations $\cS_A$ that are local unitaries in $A$ and move the logical and auxiliary dimers so that after the swaps,
        each dimer will start and end at the same edge, as illustrated in Fig. \ref{fig:HaPPY_logical_dimers} (c).
        As shown in \cite{Jahn:2019nmz}, whenever a dimer starts and ends at the first edge after the pivot, the state factorizes as
illustrated in Fig. \ref{fig:detach_dimer}, where the parity of the dimer that factorizes can be directly translated into whether $\y$ is in the state $\ket{0}$ or $\ket{1}$.
        
        By aligning all the logical and auxiliary information on the edges
        directly following the pivot, one ends up with a state that has the form 
        
        \begin{equation}
        \cS_A \ket{\bar{ij}} = \ket{\y_{i}}_a \ket{\c_{i,a}}_{\c_a} \ket{\c_j}_{\g},
        \end{equation}
        
        where $\y_i,\c_{i,a}$ are states in qubit Hilbert spaces that depend only on the bulk state in $\a$ and $\ket{\c_j}_{\g}$ is the part of the state
        associated to the dimers in the complementary entanglement wedge and the
        RT dimers. Now we can apply state dependent unitaries $\cX^A_i$ made out of
        $X$-operators acting on the factorized qubits\footnote{At this stage we
        treat dimers and qubits in a hybrid setting, where we relabeled the
    Jordan-Wigner transformation in such a way that only the qubits that did not
    factorize in the previous steps, i.e., the dimers associated to the RT surface and the complementary
    wedge $a^c$, participate in the Jordan-Wigner transformation and the
    factorized ones are excluded and we treat them as regular qubits with Pauli
    operators acting on them. However, a single factorized dimer represents a qubit,
    so we can think of the factorized qubits graphically also in terms of dimers having each their own Jordan-Wigner transformation associated to them.} via 
        
        \begin{equation}
            \cX^A_i \ket{ \y_i}_a \ket{\c_{i,a}}_{\c_a} = \ket{i}\ket{\c_a},
        \end{equation}
        i.e., they extract the original bulk state $i$ in $a$ and safe it in the
        qubits associated to logical dimers and set the qubits associated to
        auxiliary dimers into a fixed reference $\c_a$. This state dependent
        unitaries can be combined into a fixed unitary $\tilde{U}_A$ that applies the
        correct transformation depending on the bulk input. A similar
        transformation can be applied in the complementary region, which can be
        made local by first shifting the pivot as described in the appendix to the
        leftmost edge of the complementary region $A^c$ and repeating the same
        logic of disentangling via swaps, factorization of logical and auxiliary
        dimers and subsequent state dependent flipping via a unitary
        $\cX^{A^c}_j$. This will also comprise a unitary $\tilde{U}_{A^c}$ that
        acts analogous as $\tilde{U}_A$. The combined action of $\tilde{U}_a
        \tilde{U}_{A^c}$ will result in a state 
        \begin{equation}
            \tilde{U}_A \tilde{U}_{A^c} \ket{\bar{ij}} = \ket{i} \ket{j}
        \ket{\c_a}\ket{\c_{a^c}}\ket{\c'},
        \end{equation}
        where $\c'$ is a fixed maximally entangled state between $A$ and $A^c$
        made from dimers that cross the RT surface. \qed

        In the last step $\c'$ is just a maximally entangled state and it is a priori unclear that the associated algebra of operators acting on it factorizes into a simple tensor product. However, all the operators acting
        on this state originate, as discussed above, from operators that can act on
        the edges of the tensor network that cross the RT surface and were mapped
        unitarily to the boundary.
        The preceding arguments can be summarized in 
        \begin{corollary}
            In the Majorana dimer version of the HaPPY code, there exist local
            unitaries $U_A, U_{A^c}$ such that the full boundary
            algebra $B(\cH_A) \otimes \id$ is mapped to 
            \begin{align}
                U_A U_{A^c} \big( B(\cH_A) \otimes \id_{A^c} \big) (U_A U_{A^c})^{\dagger} &=\nonumber\\ 
                B(\cH_a) \otimes B(\cH_{\c_a})& \otimes B(\cH_{a,\g}) \otimes
                \id_{A^c},
            \end{align}
            where $\cH_a$ is a Hilbert space made out of qubits that are formed
            from dimers that are the logical dimers from Thm. \ref{Thm_1}, $\cH_{\c_a}$
            are the auxiliary dimers and $\cH_{a,\g}$ are the qubits associated to
            Bell-pairs that originate from dimers that cross the RT surface and in
            particular each of the $B(\cH_i)$ can be written as a tensor product of
            algebras that act on tensor products of $\mathbb{C}^2$ or, as in the
            case of $B(\cH_{a,\g})$, it can be written as the tensor product of
            algebras that act on the tensor product of Bell-pairs that originate from dimers that cross the RT surface.
        \end{corollary}
        
        The preceding theorem makes it clear that in the case of the Majorana dimer
        version of the HaPPY code, all one has to keep track of, when growing the
        network to understand how the algebras are mapped between layers, is what
        happens to the logical, RT and auxiliary dimers. Furthermore the fact that
        growing the network corresponds to feeding the network at layer $\L$ into a
        unitary together with extra Bell-pairs now becomes evident from the fact
        that a logical dimer at layer $\L$ still is a logical dimer at layer
        $\L+1$ and the same holds for auxiliary and RT dimers. Accordingly the
        operators that correspond to operators acting on the logical, auxiliary
        and RT dimers stay of this kind, when embedding them in the next layer,
        because the unitaries $U_A^{\L+1}, U_{A^c}^{\L+1}$ will again disentangle them in such a way
        that their image acts on a qubit that is associated to the same dimer as in
        the previous layer. 
        \subsection{Dimerized networks}
We now consider general tensor networks built out of contraction of dimer
states that have a similar structure as the HaPPY code:
In the HaPPY code, we considered sequences of subregions $A, J^+(A),\hdots$ with
their associated algebra $\cA^{1}_A, \cA^{2}_{J^+(A)}, \hdots$ where the full
algebra $\cA_A$
was mapped completely to a subalgebra of $\cA_2$ when growing the network. This
is most evident from the circuit picture (c) \ref{fig:happy_to_unitary}, where
all that connects the two subsystems is the maximally entangled state
$\ket{\c_\g}$ of the RT surface whose one-sided algebra gets completely mapped
to $A$. In the
dimer picture, this was visible by the fact that all dimers that ended in $A$
at layer $\L$ were extended to $J^+(A)$ when growing the network and all that
changed was that additional dimers were added, i.e., no dimers left the region
$A$. We saw that we could then decompose the full algebra into operators that act on pairs of
dimers that connect $A$ with its complement $A^c$ and dimers that remain
completely in $A$ and that this decomposition is respected by the embedding
because under growing the network, dimers just get extended to the larger
region and do not leave it.
This is the dimer version of complementary recovery.
We can repeat the analysis of the previous section for any network generated by
contracting dimer states that has this property and
find that the total algebra decomposes into
\begin{equation}
    \cB(\cH^{\L}) = \cB(\cH^{\L}_{A}) \otimes (\cB(\cH^{\L}_{A,\g}) \otimes
    \cB(\cH^{\L}_{A^c,\g})) \otimes \cB(\cH^{\L}_{A^c}),
\end{equation}
where $\cH^{\L}_{A(A^c)}$ denotes the Hilbert space of dimers that begin
and end in $A (A^c)$ and
$\cH^{\L}_{A,\g}$ the Hilbert space of dimers that connect $A$ with $A^c$.
By an analogous derivation as in the previous section, we can then construct disentangling unitaries $U_A,
U^{\L}_{A^c}$ that decompose the total state $\ket{\Y^{\L}}$ of the network at
layer $\L$ into 
\begin{equation}
    U^{\L}_A U^{\L}_{A^c} \ket{\Y^{\L}} = \ket{\x_{A}}\ket{\c_\g}\ket{\x_{A^c}},
\end{equation}
where $\c_{\g}$ is maximally entangled and $\x_A,\x_{A^c}$ are the states
of the dimers building $\cH^{\L}_{A(A^c)}$.
Since we assume that the network does not make dimers leave the subregion under
growing of the network, each additional layer will again just add maximal 
entangled pairs to $\c_{\g}$ or dimers that remain in $A$, so that again,
growing a layer just amounts to the addition of Bell pairs to $\g$ or
unentangled reference states to $\x_A$. We can therefore conclude again that
the inductive limit of the algebra of $A$ decomposes into
\begin{equation}
    \cA_A = \cB(\cH_{\x}) \otimes \cA_{A,\g} \otimes \id_{A^c},
\end{equation}
where $\cA_{A,\g}$ is the hyperfinite type II$_1$ factor. Depending
on, whether $\cH_{\x}$ is finite or infinite dimensional, we see again that the
resulting algebra either has type II$_{1}$ or type II$_{\infty}$.

\subsection{Type II factors and the absence of magic}
Dimer states, as well as the image of bulk states $\ket{ij}$ in the boundary of the HaPPY code built out of perfect tensors, are called stabilizer states, which are states that can be obtained from a
unitary circuit $U$ applied to $\ket{0}$, where the circuit $U$ is made out of
Clifford gates. This is a subset of all unitary gates that can be efficiently simulated by a classical computer using the
Gottesman-Knill theorem \cite{gottesman1998}. Clifford gates are defined as the stabilizer of the
Pauli group, i.e., all unitaries that map Pauli operators to Pauli operators
under conjugation. States that are prepared by circuits containing gates that
do not belong to the Clifford group are said to have \textit{magic} and their
properties are tightly linked to the efficiency of quantum compared
to classical computation. We saw above several examples of inductive limits of stabilizer
states that lead to type II factors. We suspect that this is a general
feature of stabilizer circuits and subregions that with respect to the circuit
satisfy complementary recovery, that is, that the state prepared by a
layered Clifford
circuit will generate local algebras that are at most type II, but never type III.
An intuitive reason for that is that a generating set for Clifford circuits is given by the Pauli group together with the CNOT and Hadamard gate, of which the latter two satisfy 
\begin{equation}
    CNOT (H\otimes \id) \ket{0 0 } = \frac{1}{\sqrt{2}}(\ket{11} + \ket{00}),
\end{equation}
i.e., they naturally generate maximally entangled states. One can more generally
argue that the bipartite entanglement spectrum of stabilizer states only
contains inverse powers of 2, so that one can decompose the state into
maximally entangled pairs and unentangled states and one cannot achieve the
submaximal entanglement spectrum necessary for a type III factor. It is
tempting to conjecture that the requirement for complementary recovery causes the
possible circuits to be of the form \ref{fig:happy_to_unitary} so that again
all the entanglement comes from gluing the two halves together by maximal
entangled pairs.
In light of this, we formulate the heuristic \textit{Magic is necessary for type III algebras}
which appears to provide a different point of view on recent observations
concerning magic in quantum field theory \cite{White:2020zoz,Cao:2023mzo} as
local algebras in quantum field theory are type III$_1$ factors.
In a system with complementary recovery, the information of both
halves does not get mixed under the induction step, and one can hope to provide an analogous decomposition into
entangled pairs and unentangled local pieces as in the dimer picture. A more
precise form of our conjecture is 
\textit{
    The inductive limits of local algebras of tensor networks that are prepared by Clifford circuits and are associated with regions that satisfy complementary recovery are never type III.}
A detailed investigation of this claim is left for future work \cite{cliffordpaper:2025}.

\subsection{MERA}

\begin{figure}[t]
\centering
\includegraphics[width=0.48\textwidth]{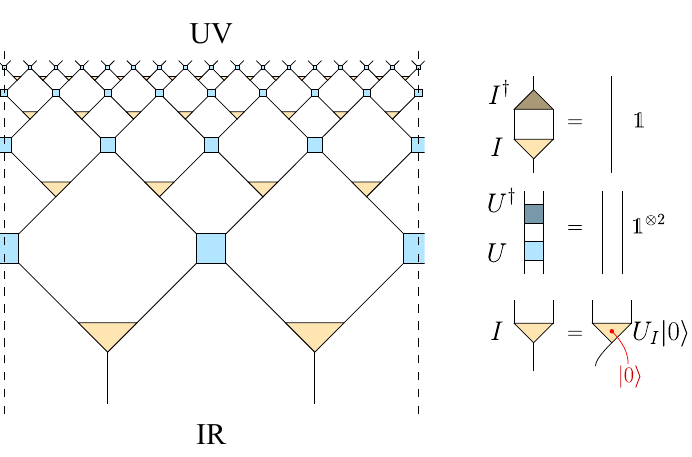}
\caption{The multi-scale entanglement renormalization ansatz (MERA) with four layers. The MERA is a tensor network that maps coarse-grained (IR) degrees of freedom to fine-grained (UV) ones, designed to describe the renormalization group flow of critical, gapless theories. It is constructed from isometries $I$ (triangles) and unitary \emph{disentanglers} $U$ (squares). As shown in the legend, $I$ can be rewritten as a unitary map $U_I$ postselected onto a reference state $\ket{0}$.
Here we show the MERA with periodic boundary conditions, denoted by dashed lines.
}
\label{fig:MERAC1}
\end{figure}

\begin{figure*}[ht]
\centering
\includegraphics[width=1.0\textwidth]{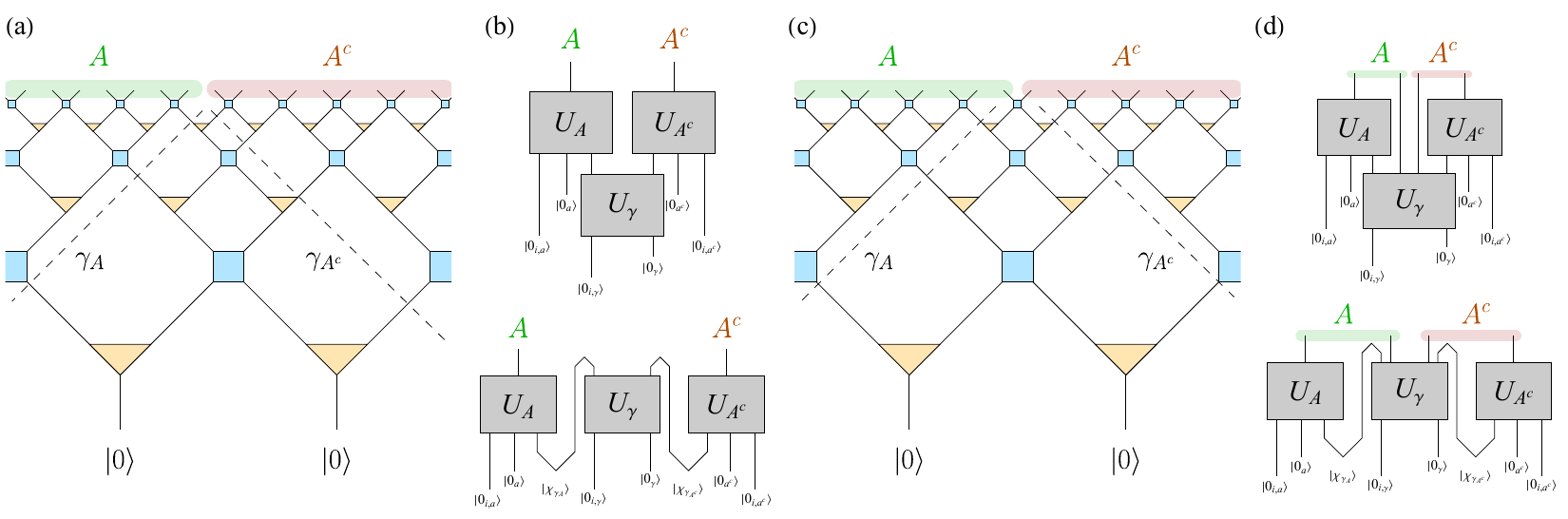}
    \caption{Bipartitions of MERA boundary sites. 
    (a) A choice of a half-infinite system $A$ that only contains both
        legs of any disentangler in its support. The boundary of its
        preimage is marked as $\gamma_A$. If one grows the network
        by an additional layer, this algebra is completely embedded into a
        set of bulk legs at the next layer but its image does have
        spatial overlap with the image of $A^c$, so that it breaks complementary recovery. The bulk separates into three regions $a$ (between $A$ and $\gamma_A$), $a^c$ (between $A^c$ and $\gamma_{A^c}$), and the ``thick'' RT region $\gamma$ (between $\gamma_A$ and $\gamma_{A^c}$).
    (b) A circuit representation of the previous setup. Here $\ket{0_{a}},          \ket{0_{a^c}},\ket{0_{\g}}$ are the auxiliary bulk states that are fed into each unitary defining an isometry tensor in each bulk region, and $\ket{0_{i,a}},\ket{0_{i,a^c}},\ket{0_{i,\g}}$ are the input states at the bottom layer.
    The bottom circuit shows a parallel implementation with maximally entangled input pairs $\ket{\chi_{\gamma_A}}$ and $\ket{\chi_{\gamma_{A^c}}}$, along with post-selection, highlighting the difference from the holographic encoding circuit in Fig.\ \ref{fig:happy_to_unitary}(c).
    (c) Another choice of bi-partitioning the MERA into half-infinite subsystems that splits the output qubits of a disentangler, also breaking complementary recovery.
    (d) Circuit representations of (c).
    }
    \label{fig:MERAC2}
\end{figure*}

The multi-scale entanglement renormalization ansatz (MERA) is a tensor network ansatz for the discretized renormalization group flow of critical, gapless theories. As shown in Fig.\ \ref{fig:MERAC1}, it can map operators and states between coarse-grained ``infrared'' (IR) and fine-grained ``ultraviolet'' (UV) degrees of freedom. This ansatz has been shown to include good approximations of grounds state correlation functions of simple critical theories and the spectrum of their primary operators \cite{Pfeifer:2009criticalMERA,Zou:2019dnc}. Its geometry also mimics the path integral of a conformal field theory \cite{Milsted:2018san,Zou:2019dnc}, which appear in the continuum limit of certain discrete critical models. Assuming that the MERA, given instances with suitable input parameters on their two types of tensors, can well-approximate states of a (conformal) field theory in its infinite scaling limit, it should then be expected that the local subregion algebra of such instances is described by a type III von Neumann algebra. Like the HaPPY code, the MERA has been proposed as a model of a holographic bulk-to-boundary map \cite{Swingle:2009bg}; if it leads to type III instead of type II algebras, then the MERA must possess qualitative features that deviate from our discussion of holographic codes so far.
As we will briefly show now, this is due to a breakdown of complementary recovery, which allows for nontrivial entanglement spectra between two sides of a bipartition to appear.
To understand the algebras of half-infinite systems associated with the MERA, it is convenient to extend the isometry tensors into unitaries by adding a bulk leg to each tensor, with a projection of this leg onto a reference state $\ket{0}$ recovering the initial isometry (see legend in Fig.\ \ref{fig:MERAC1}).
This extension is always possible and allows us to turn the entire tensor network into a unitary map and consider the pre-image of any boundary subalgebra.
Now we want to determine the type of the algebra associated to half-infinite systems in this network. For this we have essentially two choices: At a given layer, one can either split the systems between two disentanglers as in Fig.\ \ref{fig:MERAC2}(a), or at a disentangler as in Fig.\ \ref{fig:MERAC2}(c).
Both choices for the subregion $A$ have the property that adding a new layer to the network results in a nontrivial overlap of the lightcones of $A$ and $A^c$, thus breaking complementary recovery. 
For concreteness, let us now fix the choice to (a). The effect of the breaking of complementary recovery is that for operators in $\cA_{A}$ that are in the image of operators that act on the RT surface $\g_A$, the state prepared by the tensor network is not necessarily maximally entangled anymore.
This is visible in the circuit picture \ref{fig:MERAC2}(b) from the fact that the state $\ket{\c_{\g_A}}$ does not merely connect to the input legs of the unitaries $U_A$, $U_{A^c}$, and $U_\gamma$ but connects input and output legs. This gives rise to a ``thick'' Ryu-Takayanagi (RT) surface $\g$ in Fig.\ \ref{fig:MERAC2}(a) that consists of several tensors between the cuts $\g_A$ and $\g_{A^c}$ that bound the causal past of $A$ and $A^c$. 
While the algebras $\cA_a$ and $\cA_{a^c}$ of each bulk region above these cuts are unitarily mapped to $A$ and $A^c$, respectively, the algebra $\cA_\g$ of bulk operators in $\g$ (acting on the extended isometries) is shared between both boundary regions.
In contrast to the HaPPY code, where $\g_A$ and $\g_{A^c}$ typically coincide and $\cA_\g$ lives in the center of $\cA_A$ and $\cA_{A^c}$, complementary recovery is therefore not a feature of a general MERA. Even asymptotically, there is generally no complementary recovery, as each iterative step of growing the tensor network adds more tensors and algebra elements to $\gamma$.
The choice of subregions in Fig.\ \ref{fig:MERAC2}(a) leads to a further complication that the cut $\g_{A^c}$ is not stable under such an iteration, as tensors near the boundary that belonged to $a^c$ become part of $\gamma$ in the next iteration step. This can be ameliorated by the subregion choice of Fig.\ \ref{fig:MERAC2}(c), but does not solve the problem that operators with support entirely on $A$ or $A^c$ in one step can have overlapping support on both in the next.
In terms of entanglement between $A$ and $A^c$, the breakdown of complementary recovery then allows for complicated non-bipartite entanglement, potentially approximating the entanglement spectra of QFT subregions.
In the scaling limit, one may then recover the type III factors expected from such a QFT \cite{Fredenhagen:1984dc}. We note that we can still associate type $I$ sub-factors to each of the half-infinite systems that arise from the operators acting on bulk  and input legs for which MERA network just prepares the state $\ket{0_a}\otimes \ket{0_{i,a}}$ which has no intrinsic entanglement.

There is an additional issue in any one of the above algebras that makes them
unsuitable to fit into the structure of a quantum field theory. By the
Reeh-Schlieder Theorem \cite{Witten:2018zxz}, for any local algebra the
vacuum state $\ket{\Omega}$ is cyclic and separating, where cyclic means that every state in the Hilbert space can be approximated by states of the form  $O\ket{\Omega}$, where $O$ is an operator in the local algebra and separating means that $\ket{\Omega}$ is not annihilated by any operator in the local algebras that is not identically
zero. That the state prepared by MERA is not separating for the local algebra can be seen as follows: One can push the operator $\ket{1}\bra{1}$ that
acts on any of the input states $\ket{0}$ through the network and obtain
an operator that is localized in the region $A$ or $A^c$ and thus belongs
to the local algebra, but this operator by construction annihilates the
state of the MERA. We see that each of the local algebras has a large set
of operators that annihilates the state represented by the MERA, breaking
the separability. This feature is also true in the HaPPY code, where bulk projections pushed to the boundary can annihilate the boundary state but here it is essential to deal with it, if one wants to recover a quantum field theory. It thus seems evident that to recover a quantum field theory, one has to make further restrictions on the allowed operators.

\section{Discussion}\label{sec:discussion}

Our exploration of infinite, layered tensor networks has revealed a link between features of quantum error-correcting codes implemented by such tensor networks, in particular the property of complementary recovery, and the classification of hyperfinite factors as introduced by Araki-Woods and Powers \cite{Araki1968ACO,Powers1967RepresentationsOU}. 
A particular focus and practical example of our studies has been the HaPPY code \cite{Pastawski:2015qua}, a tensor network model of holographic quantum error correction in which complementary recovery naturally appears.
Generalizing to layered tensor network codes, we showed that these lead to the emergence of von Neumann algebras associated to boundary subregions, with complementary recovery further restricting these algebras to type II factors.

\subsection{Summary}

Employing the theory of inductive limits of Hilbert spaces and $C^\star$-algebras, we developed a framework that enables us to identify local algebras in infinitely large tensor networks. This binds together tensor network states and operator algebras, allowing for a systematic examination of emergent features. A major finding is that the subregion algebras in the HaPPY code form the unique hyperfinite type II$_\infty$ factor, a conclusion that extends to any exact quantum error-correcting tensor network with complementary recovery between layers and between the Ryu-Takayanagi (RT) surfaces that glue the complementary regions of the network together. This stems from the unitary equivalence between the network state, the Araki-Woods-Powers construction, and the subregion algebra. The type II nature is inherited from maximal entanglement across the bulk RT surface.

In addition to the HaPPY code, we studied networks under a spin-to-fermion equivalence via the Jordan-Wigner transformation, finding that those built from a contraction of Majorana dimer tensors \cite{Jahn:2019nmz} also generically lead to type II factors, provided the network satisfies complementary recovery. We commented on similar observations for algebras for subsystems of Clifford circuits that appear to be fixed to type II or type I factors. We discussed how the MERA network \cite{Vidal:2008zz}, due to the appearance of a ``thick'' RT surface, breaks complementary recovery, thus allowing for more complicated entanglement patterns than in the simple Araki-Woods-Powers factors. We conjectured that the type III algebras that are associated with half-infinite subsystems in quantum field theory, which is a widely expected limit of the MERA for a suitable choice of tensors, might arise from this thick RT surface. We also argued that it appears reasonable that our usage of all operators is too relaxed and that one should employ the possibility in the MERA to define an operator pushing map using the superoperator. This would allow for the restriction to operators made of primary fields as already suggested in \cite{Pfeifer:2009criticalMERA}. Based on this, the approach of \cite{Morinelli:2020uea} of using the GNS construction and focus on more complicated inductive maps seems like a better avenue to study the limit of the MERA.
The presumably most useful observation stemming from our work concerns the connection between the notion of complementary recovery and the Araki-Woods-Powers factors, which allows for strong control of the entanglement structure of the state of the network as it is grown layer by layer.

\subsection{Outlook}

Multiple open directions emerge. First, our analysis centers upon networks with maximal entanglement in link states, intrinsically favoring type II factors. Examining networks with submaximal entanglement \cite{Cheng:2022ori} may result in type III algebras. In particular, random tensor networks \cite{Hayden:2016cfa} in the limit of large bond dimension are made out of approximately perfect tensors. So, an extension of our method to random networks might be a fruitful next step to understand operator algebras in holographic systems, at least in a probabilistic or average sense. The absence of complementary recovery in MERA networks suggests that its ``RT algebra'' allows for more complicated entanglement patterns that might lead to type III algebras. A better understanding of the thick RT surfaces could deepen our understanding of the role of tensor networks in approximating CFTs. 
Furthermore, our conjecture on the appearance of type II factors from systems prepared by Clifford circuits provides a new perspective on the interplay of operator algebras with quantum computation. Understanding the relation of this line of thought to recent work \cite{vanLuijk:2024nnx,vanLuijk:2025ufz} on von Neumann algebras in many-body quantum systems and the operational perspective on the type classification that the authors provide seems like a natural extension of our work.
From the perspective of holography, it seems interesting that the type II nature of HaPPY code subregions arises from maximal entanglement across the RT surface, as observed in fixed-area states in quantum gravity \cite{Dong:2018seb}. 
We want to mention a direct parallel to the construction by Soni \cite{Soni:2023fke} which considered a similar system based on the notion of holographic codes or gauge-invariant Hilbert spaces in lattice gauge theories that also carry maximal entanglement across the subsystem boundary \cite{Donnelly:2011hn,Donnelly:2014gva,Aoki:2015bsa}. However, in contrast to our choice of fixing the bulk of the HaPPY code to finitely-entangled states so that the associated algebras are type I, Soni considered a limit where the bulk logical algebras are type III and the resulting limit mimics the gravitational crossed product that was used in \cite{Witten:2021unn, Chandrasekaran:2022cip} in the context of perturbative quantum gravity to obtain local algebras of type II, which has been studied extensively in recent years \cite{Chandrasekaran:2022eqq, Jensen:2023yxy, Klinger:2023tgi, Kudler-Flam:2023qfl,Chen:2024rpx,Gomez:2022eui,Kudler-Flam:2023hkl,DeVuyst:2024uvd, AliAhmad:2024saq} motivated by the discovery of type transitions in holographic dualities \cite{Leutheusser:2021frk}l. A more concrete realization of this limit within our framework might be useful to gain intuition on the appearance of the crossed product. It would be interesting to understand whether this entanglement-induced structure of the algebra underlies type II factors in gravity, as could be made possible by thinking of the RT degrees of freedom as gravitational edge modes that contribute to the area operator, which in turn contributes to the entropy. The idea to connect edge modes with black hole entropy can be found at several places in the literature \cite{Harlow:2016vwg, Klinger:2023tgi, Haco:2018ske}. 
In \cite{Furuya:2020tzv} a framework for connecting error-correcting codes and conditional expectations to the real-space version of the renormalization group were made and it would appear as a natural consideration to apply our results in this context. By exploring these questions, future work can further clarify the interplay between tensor networks, operator algebras and their role in quantum field theory and quantum gravity.

\section*{Acknowledgement} 
We thank Charles Cao and Zolt\'{a}n Zimbor\'{a}s for numerous helpful initial insights and comments. We also thank Shadi Ali Ahmad, Charlie Cummings, Jens Eisert, and Alexander Stottmeister, for illuminating discussions and comments on the draft. LS thanks Lorenzo Leone, Lennart Bittel and Dimitris Saraidaris for discussions in the course of related work. The authors would like to express their sincere gratitude for Wayne Weng's early-stage support and commitment to this project, which was essential to its development. AJ and LS are grateful for support from the Einstein Foundation Berlin and the Einstein Research Unit ``Perspectives of a quantum digital transformation''. EG is supported by the Office of High Energy Physics of U.S.\ Department of Energy under grant Contract Number DE-SC0012567 and DE-SC0020360 (MIT contract number 578218). DM is supported by the Office of High Energy Physics of the U.S. Department of Energy under grant Contract Number DE-SC0018407.

    \newpage
    
    \appendix
    
\section{The type classification and entanglement}\label{sec:type_classification}
    In this section, we summarize the classification of types of von Neumann algebras through the lens of spin systems, drawing on the description by Witten
    \cite{Witten:2018zxz}. This presents a physicist's perspective on the factors
introduced by Powers \cite{Powers1967RepresentationsOU} and later expanded upon by Araki and
    Woods \cite{Araki1968ACO}. Consider a system consisting of an
    infinite collection of pairs of qubits; formally,  we have the infinite
    tensor product \cite{vonNeumann1939}
    \begin{equation}
        \hat{\cH} = \bigotimes_{i = 1}^{\infty} \mathbb{C}^2 \otimes
        \mathbb{C}^2. 
    \end{equation}
    This space is non-separable, and care needs to be taken to determine which vectors
    actually belong to it. To compute inner products, one must consider infinite products of numbers, which introduces complications. As a result,
    this space falls outside the typical scope
    of physics \cite{vonNeumann1939}. To obtain a separable Hilbert space, i.e., 
    a space with a countable basis, we have to choose a vacuum or reference
    state within this large
    Hilbert space and consider finite excitations placed on top of it. 
    This will provide a toy example of layered tensor networks.
    We consider the sequence of Hilbert spaces
    \begin{equation}
        \mathbb{C}^2 \otimes \mathbb{C}^2 \rightarrow (
        \mathbb{C}^2 \otimes \mathbb{C}^2)^{\otimes 2}\rightarrow (
        \mathbb{C}^2 \otimes \mathbb{C}^2)^{\otimes 3}\rightarrow \hdots,
    \end{equation}
    where each step adds a pair of qubits.
    The layered tensor network is replaced by a sequence of states  
    \begin{equation}
        \ket{\y^{(0)}} \rightarrow \ket{\y^{(0)}} \otimes \ket{\y^{(1)}} \rightarrow
        \hdots,
    \end{equation}
    where, at each level, a new state of a qubit-pair is tensored in. In the language of the main body of the text, this is equivalent
    to a layered tensor network with the isometry
    \begin{equation}\label{eq:isometry_spin_chain}
        \ket{\Y^{\L+1}} = \g^{\L,\L+1}\ket{\Y^{\L}} = \ket{\Y^{\L}} \otimes \ket{\y^{(\L+1)}}.
    \end{equation}
    Similarly, the operator pushing map between layers is then
    \begin{equation}
        \p^{\L,\L+1}(O) = O \otimes \id
    \end{equation}
    so that the operator becomes the identity on the additional qubits.
    Following the inductive limit procedure described in Sec. \ref{sec:inductive_limits_intro}, we consider the sequence of states $\Y^{\L}$ and Hilbert
    spaces with the isometries $\g^{\L,\L+1}$ as an inductive system with an limiting Hilbert space $\cH_\Y$ that contains a reference vacuum state $\ket{\Y}$ that represents the state
    \begin{equation}
        \ket{\Y} = \bigotimes_{\L=1}^{\infty} \ket{\y^\L}.
    \end{equation}
    As a
    subregion, we consider the algebra of operators that act only on one qubit of each qubit
    pair \footnote{Because the locality here is ambiguous, we just make an
        arbitrary choice which of the two qubits we consider at each layer,
        since due to the embedding \eqref{eq:isometry_spin_chain} both choices
    are valid to consider as lightcones.}.
    These algebras again form an inductive system with inductive limit $\cA$ that can be represented on
    the limiting Hilbert space $\cH_{\Y}$ induced by the sequence
    $\ket{\Y^{\L}}$ and define a von Neumann algebra $\cA_{\Y}$. Note that the Hilbert space isometries and the operator pushing maps are compatible in the sense of eq. \eqref{eq:compatibility}.
    Now consider the specific states
    \begin{equation}
    \ket{\y^{(i)}_{\l}} := \ket{\y_{\l}} =\frac{1}{\sqrt{1+\l^2}} (\ket{\uparrow \uparrow} + \l
        \ket{\downarrow \downarrow}).
    \end{equation}
    so that the chain of states is homogeneous between the levels. We denote
    the resulting one-sided von Neumann algebras by $\cA_{\l}$ and the Hilbert
    spaces by $\cH_{\l}$.
    With this setup one has 
    \begin{itemize}
        \item[(I)] If $\l = 0$, the underlying state is unentangled, and one can
            show that the resulting Hilbert space takes the form 
            \begin{equation}
                \cH_{0} = \cH_A \otimes \cH_B,
            \end{equation}
            where $\cH_A$ and $\cH_B$ are separable, infinite-dimensional Hilbert spaces.
            In particular, one has 
            \begin{equation}
                \cA_{0} = \cB(\cH_A) \otimes \id.
            \end{equation}
            This is a particular case of a more general situation: a von Neumann algebra that can be represented as the bounded operators of a
            Hilbert space is called a \emph{type I$_\infty$ factor}, with the
            subscript $\infty$
            denoting the infinite dimensionality of the Hilbert space. If the
            Hilbert space is finite dimensional they are called \emph{type
            I$_n$ factors}.
        \item[(II)] If $\l = 1,$ the state $\ket{\y_1}$ is a maximally entangled
            Bell pair. As a result, the reference state $\ket{\Y} \in \cH_\Y$
            satisfies
            \begin{equation}\label{eq:tracial_state}
                \braket{\Y|ab|\Y} = 
                \braket{\Y|ba|\Y},\ \forall a,b \in \cA_{1}.
            \end{equation}
            This is the defining property of a tracial state. In particular, since
            equivalence classes $[a]$ in the inductive limit algebra that come
            from finite-level operators act only on finitely many qubit-pairs,
            the number \ref{eq:tracial_state} has a finite value on each $[a]$, and by extension, on each
            element of $\cA_{1}$. Therefore, $\cA_{1}$ allows for a normal\footnote{Normal means that the state behaves well with limits of sequences of operators.} state
            that is tracial on each element of $\cA_{1}$. This is the defining property of a 
            von Neumann algebra of \emph{type II$_1$}. \\
        \item[(III)] For $0 < \l < 1,$ one can show that none of the above holds, so $\cA_{\l}$ can not be represented as $\cB(\cH)$ for some
            Hilbert space, and it does not allow for a finite trace. Specifically,
            every function that could be cyclic on $\cA_{\l}$ must take the
            values $0$ or $\infty$, a property usually referred to as
            $\cA_{\l}$ being a properly infinite factor\footnote{In contrast to case (II)
            which is a finite factor.}. Such an algebra is said to have \emph{type
            III$_{\l}$}.
        \item[(IV)] As a generalization, one can consider, instead of a fixed
            $\Y_{\l},$ an alternating sequence 
            \begin{equation}
                \ket{\Y_{\l_1}} \rightarrow
                \ket{\Y_{\l_1}}\ket{\Y_{\l_2}} \rightarrow
                \ket{\Y_{\l_1}}\ket{\Y_{\l_2}}\ket{\Y_{\l_1}},
            \end{equation}
            where $\frac{\l_1}{\l_2}$ is not a rational number. This leads to
            an entanglement spectrum between the two sides which becomes
            continuous in the limit of infinitely many levels. The one-sided
            von Neumann algebra $\cA$ then has \emph{type III$_1$} and neither admits 
            representations as $\cB(\cH)$ nor a tracial state. These algebras
            describe causally complete subregions in quantum field theory
            \cite{Fredenhagen:1984dc}.
    \end{itemize}
    All of the von Neumann algebras listed above are referred to as \textit{hyperfinite}
    factors, meaning that they are the weak operator closure of an increasing union of finite-dimensional
    subalgebras. Another type of algebra that is relevant to us in the following is the
    hyperfinite factor of \emph{type II$_{\infty},$} which can be shown to be
    isomorphic to an algebra acting on a Hilbert space of the form
    \begin{equation}
        \cH = \cH_A \otimes \cH_B,
    \end{equation}
    and the algebra being of the form
    
    \begin{equation}
        \cA = \cB(\cH_A) \otimes \text{II}_{1},
    \end{equation}
    where the notation indicates that it is a II$_{1}$ factor on $\cH_B$ multiplied
    with $\cB(\cH_A)$, where $\cH_A$ is a separable, infinite dimensional
    Hilbert space. These algebras do allow for a trace but are not finite, as
    the trace of $\cH_1$ maps the identity to $\infty$.
    \section{Majorana dimers}\label{app:dimer_review}

    Here we introduce a basic description of states represented by Majorana
    dimers, which are heavily used in Sec.\ \ref{sec:HaPPY_dimers}.
    This can be applied whenever one has a Hilbert space of the form 
    \begin{equation}
        \cH = \bigotimes_{i=1}^N \mathbb{C}^{2}.
    \end{equation}
    Instead of considering the Hilbert space in this ``spin picture'' as a tensor
    product of local qubit Hilbert spaces, one performs a Jordan-Wigner
    transformation \cite{jordan1928} to obtain a ``Majorana picture'' of
    the Hilbert space and operators. In particular, we define 
    \begin{align}\label{eq:Jordan_Wigner_1}
        \g_{2k-1} &= Z_1 Z_2 \hdots Z_{k-1}X_k \ ,\\
        \g_{2k} &= Z_1 Z_2 \hdots Z_{k-1}Y_k \ , 
    \end{align}
    where $k \in \{1,\hdots N\}$, $N$ being the number of spins, and $Z_i,X_i$ denoting the respective Pauli operator acting on the $i$-th spin. These represent Majorana
    operators that satisfy $\{\g_l,\g_m\} = 2\d_{l,m}$. We can then define fermionic creation and annihilation operators via
    \begin{equation}
        f_k^\dagger := \frac{1}{2}(\g_{2k-1}- \ii \g_{2k}).
    \end{equation}
    Given any state 
    \begin{equation}
        \ket{\y} = \sum_{i_1,\hdots,i_N = 0}^{1} T_{i_1 \hdots i_N} \ket{i_1
        \hdots i_N},
    \end{equation}
    we can associate a fermionic representation of $\psi$ via 
    \begin{equation}
        \ket{\psi}_f = \sum_{i_1,\hdots,i_L=0}^1 T_{i_1\hdots i_N}
        (f_1^\dagger)^{i_1} \hdots (f_N^\dagger)^{i_N}\ket{\W},
    \end{equation}
    where $\ket{\W}$ is the fermionic vacuum, which coincides with the
    $\ket{0}^{\otimes N}$ state in the spin/qubit picture.
    The fermionic vacuum satisfies
    \begin{equation}
        f_i \ket{\W} = \frac{1}{2}(\g_{2i-1}+\ii\g_{2i}) \ket{\W} = 0, i \in \{1,\hdots,N\}.
    \end{equation}
    This provides $N$ conditions of the form 
    \begin{equation}\label{eq:dimer_def}
        (\g_k + \ii p_{k,l} \g_l)\ket{\psi} = 0,
    \end{equation} 
    where for the vacuum state  $\psi = \W, p_{k,l} = 1, l= k+1, k = 2m, m \in
    \{1,\hdots,N\}$. A state satisfying \eqref{eq:dimer_def} for $L$ pairs that are mutually exclusive is called a \textit{Majorana dimer state} and the numbers $p_{k,l}$ are the \textit{dimer parities}.
    The contraction of two states $\psi, \phi$ along the say third and fourth index is then a new state in a bigger system, i.e., we 
    define 
    \begin{align}
    \ket{C(\psi,3,\phi,4)} := 
    &\sum_{i_k,j_m}(T^\psi_{i_1 i_2 0 i_3 i_4 i_5} T^\phi_{j_1 j_2 j_3 0 j_5}\\
    &+ T^\psi_{i_1 i_2 1 i_3 i_4 i_5} T^\phi_{j_1 j_2 j_3 1 j_5}) \ket{i_1 i_2 i_3 i_5 j_1 j_2 j_3 j_5}.
    \end{align}
    It was shown in \cite{Jahn:2019nmz} that if the individual tensors $T^\phi,T^\psi$ give a dimer state in the dimer representation, the contracted tensor does so as well. Since the logical basis states $\bar{0}$ and $\bar{1}$ of the five-qubit Laflamme code are themselves represented by dimer states, the encoding of a bulk basis state via the full HaPPY code is itself a dimer state on the boundary. 
    
    \subsection{Dimer calculus}
\begin{figure}[ht]
        \includegraphics[scale=0.4]{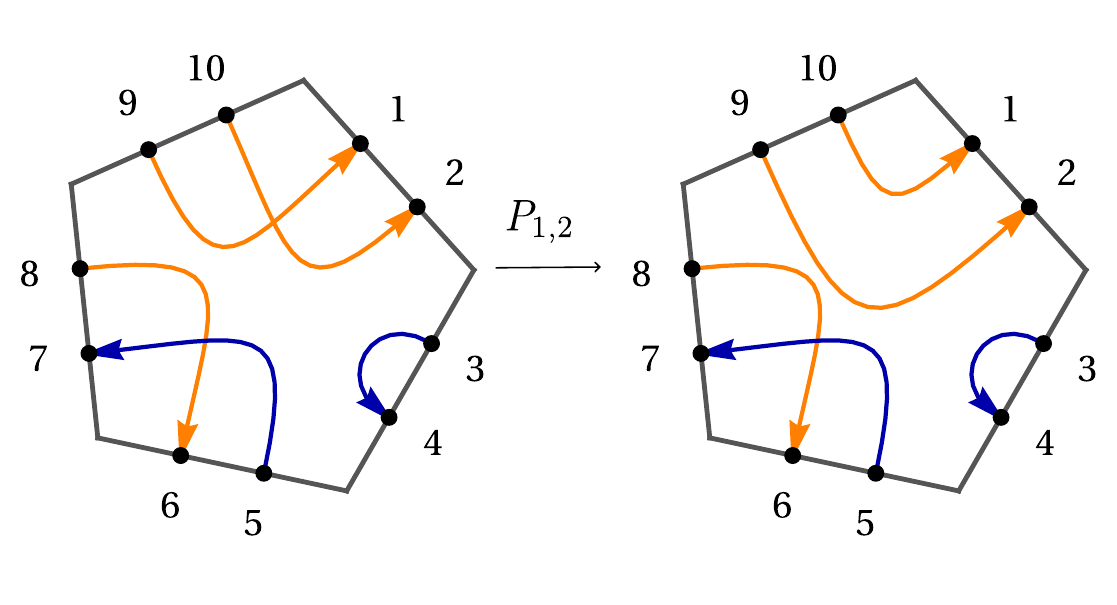}
        \caption{Representation of swap operation between node 1 and 2.}
        \label{fig:pentagon_swap}
    \end{figure}

    Given a dimer state, we can represent it using a simple graphical
    representation, as for the following two logical basis states of the five-qubit code (each qubit represented as the edge of a pentagon):
    \begin{align}
        \ket{\bar{0}}_5\; = \quad
        \begin{gathered}
        \includegraphics[height=0.12\textheight]{pentagon_net_happy_p.pdf}
        \end{gathered} \\
        \ket{\bar{1}}_5\; = \quad
        \begin{gathered}
        \includegraphics[height=0.12\textheight]{pentagon_net_happy_m.pdf}
        \end{gathered} 
    \end{align}
    Here a blue arrow from e.g.\ $1$ to $6$ in $\ket{\bar{0}}$ indicates that the state is annihilated by the operator $\g_1 + \ii \g_6$. The orange arrow from $1$ to $6$ in $\ket{\bar{1}}$ indicates that the corresponding state is annihilated by $\g_1 - \ii \g_6$ and so on. 
    One can check that application of a Majorana operator $\g_i$ has the effect
    of flipping the parity of the dimer associated to $\g_i$, i.e., if
    \begin{equation}
        \g_j + \ii  \g_i \ket{\y} = 0 \ , 
    \end{equation}
    then 
    \begin{align}
        (\g_j - \ii  \g_i) \g_i \ket{\y} 
        &= \g_i(-\g_j-\ii  \g_i)\ket{\y} \nonumber\\
        &= -\g_i(\g_j+\ii \g_i)\ket{\y} \nonumber\\
        &= 0 \ .
    \end{align}
    Therefore, we can think of the application of a Majorana operator on a dimer
    state in the pictorial representation as a flip of the respective color of
    the dimer.
    Also we can swap dimers by applying the swap operator
    \begin{equation}\label{eq:swap}
        P_{j,k} = \frac{\cP_{\text{tot}}}{\sqrt{2}}(\g_j-\g_k),
    \end{equation}
    the effect of which is just the exchange of the dimer connected to the
    point $j$ to become connected to the point $k$ and vice-versa. 
    Here $\cP_{\text{tot}}$ is the total parity operator
    \begin{equation}
    \cP_{\text{tot}} = \prod_{i=1} Z_i.
    \end{equation}
    An example is given in Fig.\ \ref{fig:pentagon_swap}.
    Given two such states, we can contract the corresponding tensors to a new
    state as in the following picture
    \begin{equation}
    \label{EQ_CONTR_EX1}
    \begin{gathered}
    \includegraphics[height=0.12\textheight]{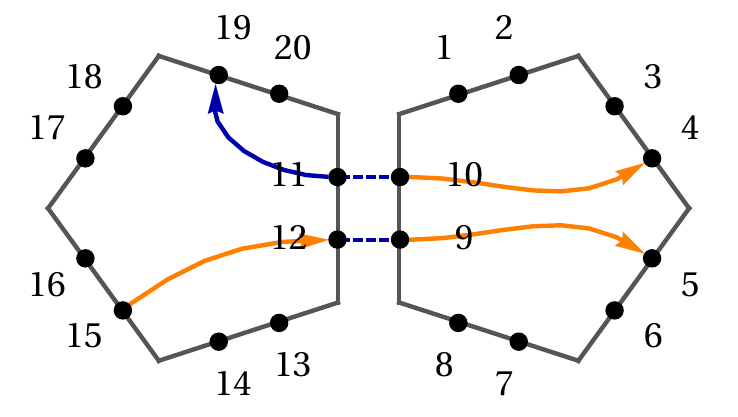}\\
    \scalebox{1.2}{$\parallel$}\\
    \includegraphics[height=0.12\textheight]{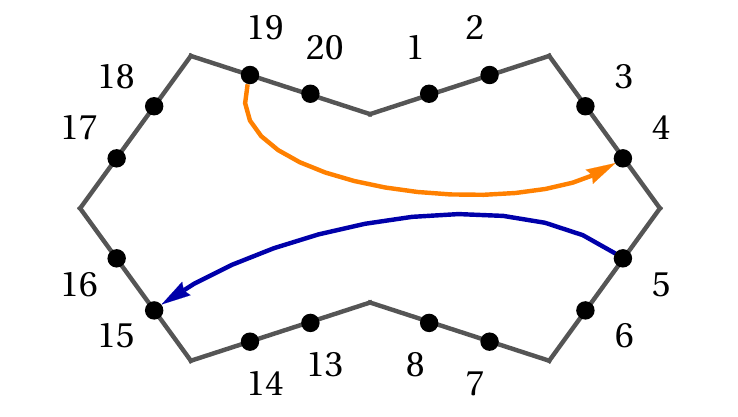}
    \end{gathered}
    \end{equation}
    i.e, if neighboring edges (here edge 5 and 6) are contracted, the neighboring dimers are extended and the resulting dimer parities are the product of the dimer parities of the dimers that were contracted. Here it is important that one contracts neighboring edges. In particular if one wants to contract the HaPPY code, one has to first choose an orientation of the local pentagons. This comes with choosing a \textit{pivot}, i.e.,  an edge at which one starts counting the dimers and that implicitly defines the starting point of the Jordan-Wigner transformation. 
    To be able to contract neighboring edges with non-consecutive indices, one first has to rotate the local dimers so that their pivots align upon contraction. As explained in \cite{Jahn:2019nmz}, for parity-even states\footnote{Parity-even dimer states are those where the number of parity-odd dimers times $(-1)^{N_c}$, $N_c$ being the number of dimer crossing points, is even} a rotation does not do anything except changing the order. For parity-odd states, the rotation of a dimer amounts to the insertion of a $Z$ string along the path of the pivot, which has the effect of flipping all dimer parities that are traversed by the pivot when moving from the old position to the new one. These $Z$ strings have to be taken into account when performing the full contraction.
    
    \subsection{Z strings}\label{app:Z_string}
\begin{figure}[ht]
        \includegraphics[scale=0.3]{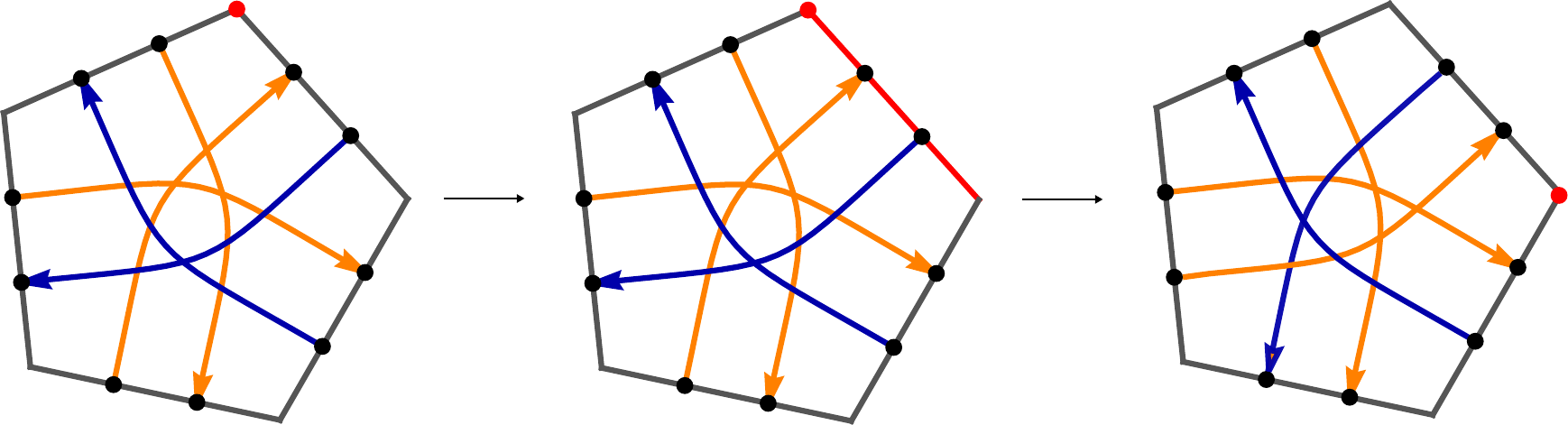}
        \caption{Demonstration of the generation of a $Z$ string as an effect of
        rotating the Pivot by one edge.}
        \label{fig:Z_string_1}
    \end{figure}

\begin{figure}[ht]
        \begin{tikzpicture}
          \node at (0,0) {};
          \node at (-4.5,0.8) {(a)};
          \node at (-2.4,0.8) {(b)};
          \node at (-0.3,0.8) {(c)};
          \node at (1.75,0.8) {(d)};
          \node at (-4.5,-0.6) {$\g_A$};
          \node at (-0.4,0) {\includegraphics[width=0.95\linewidth]{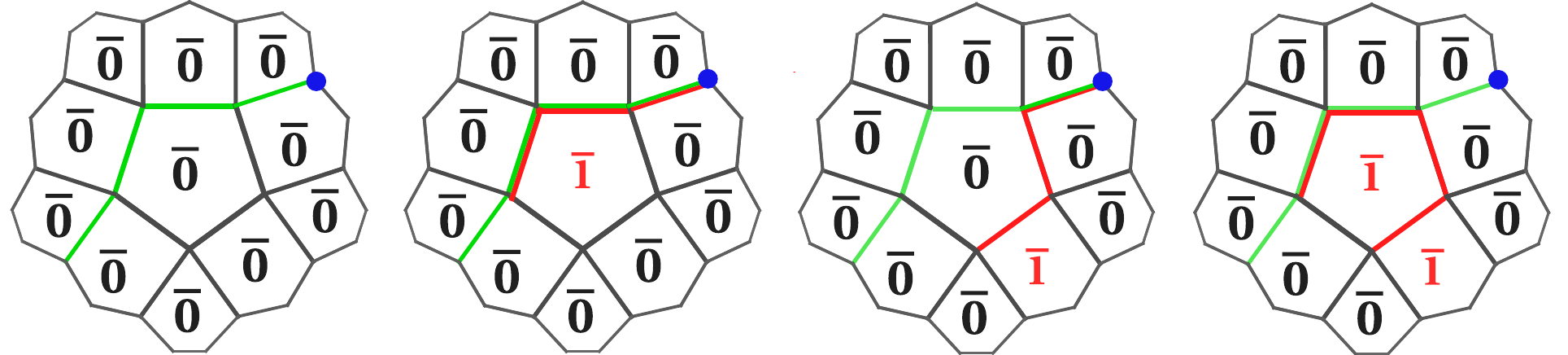}};
        \end{tikzpicture}
        \caption{Illustration of $Z$ strings that appear when exciting bulk qubits from $\ket{0}$ to $\ket{1}$. (a) The RT surface $\g_A$ is indicated by green edges and the pivot at the rightmost edge of the RT surface by a blue dot. (b) Only the central qubit gets excited and a $Z$ string, indicated by a red edge, stretches from its pivot to the global pivot located on the boundary. (c) A qubit in the second layer gets excited and the
        corresponding $Z$ string goes along the edge that connects the first with the
        second layer. (d) A qubit in the first and second layer get excited. Both $Z$
        strings from (b) and (c) appear such that a edge is contained in two individual
        $Z$ strings that cancels out, leading to a $Z$ string connecting the pivots of the
        individual pentagons.
    }
    \label{fig:z_strings_example}
\end{figure}
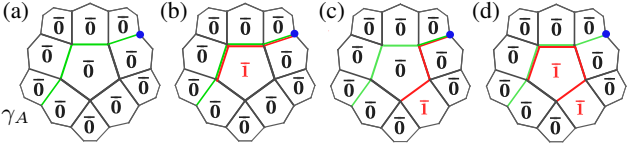

    Here we describe the use of $Z$ strings in the proof of theorem \ref{Thm_1}. The $Z$ strings appear when one changes the Jordan-Wigner transformation by a cyclic permutation of the spin indices when performing the transformation \eqref{eq:Jordan_Wigner_1}, which visually corresponds to ``rotating the pivot''. This is used when one wants to align the edges during the contraction of two dimer states such that the edges contracted are consecutive edges $k,k+1$, for which the rules for contracting dimer states from the previous section apply. 
    As described in \cite{Jahn:2019nmz}, if the dimer state has even total parity, a cyclic permutation does not generate any change in the individual dimer parities, but if the state has odd total parity, then the parity of every dimer whose endpoint is passed by the pivot is flipped (if the pivot passed over both endpoints, the dimer parity is preserved). 
    An example of the logical state $\bar{1}$ for a single pentagon is given in Fig. \ref{fig:Z_string_1} where the pivot is rotated over one edge. 
    When contracting the whole network for the HaPPY code the appearance of Z
    strings can be thought of as follows:
    If one starts out with the bulk vacuum $\ket{\bar{0}\,\bar{0}\hdots\bar{0}}$ no $Z$ strings appear during the contraction since the total state is parity-even and all
    dimers in the resulting state have positive parity. If a single bulk qubit is instead in the $\bar{1}$ state, the corresponding pentagon has to be rotated during contraction, and this rotation will produce a $Z$ string. As described in \cite{Jahn:2019nmz}, the $Z$ string will stretch from the initial pivot of the pentagon to the pivot of the complete state representing the fully contracted network, as  demonstrated in Fig.\ \ref{fig:z_strings_example} (b) and (c). In the following, we place the pivot of the contracted network at the end of the rightmost edge of the RT surface of a boundary subregion, as indicated in Fig.\ \ref{fig:z_strings_example}.(a). In this situation, any $Z$ string that arises when flipping a bulk qubit in layer $n$ can be located only on the edges that connect the layer $n$ with the layer $n-1$ until it hits the RT surface as demonstrated in Fig.\ \ref{fig:z_strings_example}(c). Thinking of $Z$ strings in this way makes it clear that the effect of flipping any qubit in layer $n$ will flip only dimers that state $\bar{1}$ from state $\bar{0}$ or dimers that connect layer $n$ with layer $n-1$ or with the complementary entanglement wedge via the RT surface. For this reason, in the proof of Theorem \ref{Thm_1} we can focus on dimers that connect the layer $n$ to itself, since all other dimers were considered in the previous layers.

 \subsection{Proof of Theorem 4}\label{app:prf_thm_2}
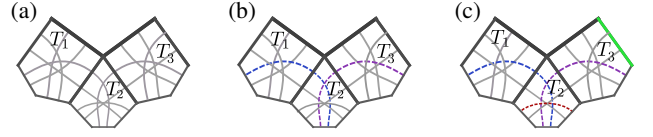
\begin{figure}[ht]
        \begin{tikzpicture}
          \node at (0,0) {\includegraphics[width=0.95\linewidth]{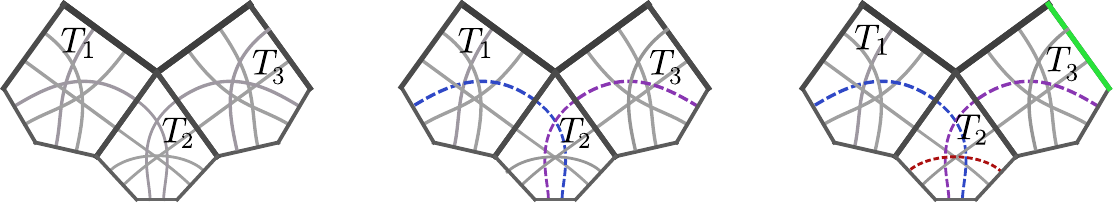}};
          \node at (-4.0,0.8) {(a)};
          \node at (-1.1,0.8) {(b)};
          \node at (1.9,0.8) {(c)};
        \end{tikzpicture}
        \caption{(a) A corner piece which illustrates all situations that can occur in
        the $\{5,4\}$ tiling of the hyperbolic plane. The lower three pentagons are
    representing layer $n$. Every layer consists of pentagons contracted as the
    bottom three pentagon. (b) Logical dimers of the two leftmost pentagons in
            dashed lines if the rightmost pentagon does not
        border the RT surface. (c) Logical dimers of $T_1, T_2, T_3$ in
            dashed lines if the $T_3$ does border the RT surface indicated by a green edge.}
        \label{fig:logical_proof}
    \end{figure}

    \textit{Proof.} We will prove the theorem for every layer individually, where
    we count the layers according to how the tensor network is grown, i.e., some
    pentagons directly at the RT surface are the deepest in the bulk and come from
    the same layer of growing the tensor network, such as the central pentagon in Fig.
    \ref{fig:HaPPY_logical_dimers}. These pentagons comprise the layer $n=1$. The
    pentagons immediately surrounding the $n=1$ pentagons form the layer $n=2$ and so
    on. We will now show that for any $n$ we can make a consistent choice of
    logical dimers that does not depend on the choice of the previous layer.
    In our setup, we assume that the pivot is located at the rightmost edge of
    our subregion, so the
    node nearest to the RT surface is node 1.
    If we now perform a logical operation in the layer $n$ on a single bulk qubit, we
    will flip 3 dimers that distinguish the logical $0$ from the logical $1$ state.
    Additionally, as explained in Appendix \ref{app:Z_string}, there will be
    a $Z$ string stretching from the pivot of the bulk qubit to the pivot
    of the subregion $A$. This $Z$ string only passes edges that lie between layer
    $n$ and $n+1$, therefore only flipping dimers that existed at layer $n$.
    Therefore, if we can make a consistent choice for logical dimers in layer $n$ that do not arise from
    the previous layer, we can make this selection at each layer separately. For
    the code on the $\{5,4\}$ tiling, one can see a consistent choice considering a corner,
    as illustrated in Fig.\ \ref{fig:logical_proof}(a). Each layer consists of a chain
    of tensors like $T_1, T_2$ aligned along a chain.

Regardless of whether $T_1$ borders the RT surface, as long as $T_3$ does not border the RT surface, logical dimers can be chosen as in Fig.\ \ref{fig:logical_proof}(b) for the pentagons $T_1,T_2$. This choice can be repeated as one goes to the right through the layer. Note that these dimers are flipped independently of where the pivot of
    $T_1$ or $T_2$ is located, as can be seen from the representation of the logical
    states in dimer form \eqref{eq:HAPPY_ONE}.

    If $T_3$ borders the RT surface, then logical dimers can be chosen as in Fig.\ \ref{fig:logical_proof}(c).
    The only difference between these two configurations is that in the case that
    the RT surface borders the rightmost pentagon, one cannot choose the dimers as
    in Fig.\ \ref{fig:logical_proof}(b) because this would make the logical dimers
    cross the RT surface. If however, the RT surface borders the corner Pentagon
    that is not connected with the previous layer, then one can go for the logical
    dimers as in \ref{fig:logical_proof}(c) by not coloring the logical dimer of the
    neighboring pentagon that lies in the complementary entanglement wedge. \qed
    
    \subsection{Proof of Lemma 1}\label{app:prf_cor_1}
\begin{figure}[ht]
         \includegraphics[scale = 0.3]{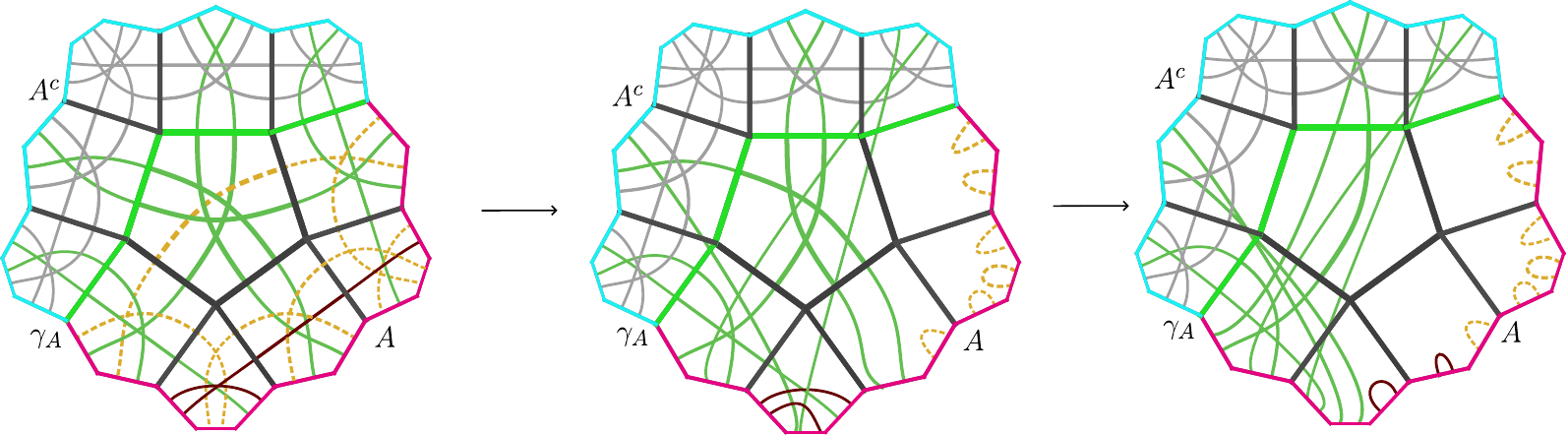}
         \caption{Illustration of a sequence of swap operations that disentangles the
         logical from the RT and the auxiliary degrees of freedom. Between each
     step a set of swap operations was applied. Note that the parity of
    dimers was neglected. The colors are used to show the logical affiliation
    of each dimer, dashed and yellow being in $D^A_l$, red being in $D_a$ and
green dimers are in $D_{\g}$. The boundary region $A$ is indicated in pink and its complement $A^c$ in blue. The RT surface $\g_A$ is also drawn in green to fit the respective dimers.}
         \label{fig:HaPPY_disentangled}
     \end{figure}

    \textit{Proof.} As established previously, we divide the dimers that end in
    region $A$ into three sets: Logical dimers $D_l^A$ as established by Thm.
    \ref{Thm_1}, dimers that cross the RT surface $D_\g$ and the remaining
    dimers that are not logical but start and end in region A as the auxilliary
    dimers $D_a$. We think of the state $\ket{\bar{ij}}$ as a state where all the
    bulk qubits which are in the $\ket{1}$ state as the result of applying the
    respective logical $X$ operators on the bulk state in which all qubits are in
    the $\ket{0}$ state. Furthermore, we locate the pivot on the rightmost edge
    of the RT surface as indicated in Fig. \ref{fig:z_strings_example}. We will
    construct $U_A$ by defining a \textit{state-dependent} unitary $C_i^A$ for
    each logical state $\ket{i}$ in the
    entanglement wedge $a$ that implements the action of $U_A$ just on this
    state and then combine each of the $C_i^A$ together to the unitary
    $U_A$.
    For a fixed state $\ket{i}$ in the entanglement wedge $a$, $C^A_i$ will have the action
    \begin{equation}\label{eq:cia}
        C^A_i \ket{\bar{ij}} = \ket{i} \ket{\c_a} \ket{\c_{\g,j}},
    \end{equation}
    so that $\ket{\c_a}$ is a fixed reference state of the auxilliary dimers
    $D_{a}$ and $\ket{\c_{\g,j}}$ is a
    state such that if $\ket{j} = \ket{0}$, the RT dimers have a fixed parity.
    Given the state $\ket{\bar{ij}}$ we explicitly construct $C^A_i$ as follows: First, we recall \cite{Jahn:2019nmz} that a state that has a dimer connecting the same edge factorizes with the rest, as demonstrated in Fig.\ \ref{fig:detach_dimer}. 
    If the dimer has positive parity $\ket{\y} = \ket{0}$, if it has
    negative parity $\ket{\y} = \ket{1}$. The first action that we want to
    implement is to achieve the factorization of \ref{eq:cia} by applying swaps on the logical
    and auxilliary dimers, so that the resulting state factorizes between the
    respective qubits. A selection of logical, RT and auxilliary dimers is illustrated in Fig.\
    \ref{fig:HaPPY_logical_dimers}(b).
    We will now apply swap operations (For more information on swaps, see Appendix \ref{app:dimer_review}).
     \begin{equation}
        P_{j,k} = \frac{\cP_{tot}}{\sqrt{2}}(\g_j - \g_k),
     \end{equation}
     to the dimers in $D^A_l, D_{a},D_{\g}$ to put them in a configuration that can be used with the rule
     illustrated in Fig.\ \ref{fig:detach_dimer} to obtain a logical and auxiliary
     system that factorizes with the rest. The swap $P_{j,k}$ has the effect of
     swapping the dimers beginning or ending at $j,k$ to now begin or end at $k,j$.
     We produce the desired state by swapping the position of the endpoint of any
     dimer in $D^A_l, D_{a}$ so that it is located on the same edge as it begins, thus
     providing a factorizing state. This disentangling procedure is illustrated in Fig.\ \ref{fig:HaPPY_disentangled}.
     
\begin{figure}[ht]
\centering
\includegraphics[width=0.25\textwidth]{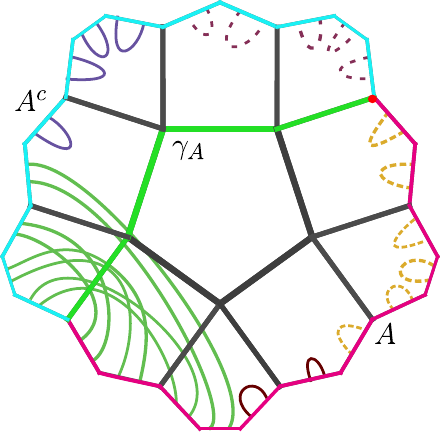}
        \caption{Dimers after full disentangling. All RT dimers begin and end in pairs at the same edge to form maximally entangled pairs.}
\label{fig:HaPPY_disentangled_Abar}
\end{figure}

     In addition, we will swap the RT dimers so that dimers that cross the RT
     surface at the same edge also end at the same edge in $A$.
     After we have done this, we will end up
     only with RT dimers connecting $A$ to the edges of $A^c$ and a piece that
     factorizes with the rest. In total, we get a system that, after applying the
     factorization rule of Fig.\ \ref{fig:detach_dimer}, has the form $\cH =
     \cH_a \otimes \cH_{\c_a} \otimes \cH_{\g,A^c}$ as in Eq.
     \eqref{eq:factorize_i}, so we have a product state between the three Hilbert
     spaces that has the form  
     \begin{equation}\label{eq:postswap}
         \cS_A\ket{\bar{ij}} = \ket{\psi_i}\ket{\phi_{i,a}} \ket{\chi_{ij}}
     \end{equation}
    where $\c_{ij}$ is the state of the dimers in $D_\g$, $\ket{\y_i}$ the
    state of the logical dimers and $\ket{\p_{i,a}}$ the state of the
    auxilliary dimers.
    Here, we have denoted the sequence of swap operations by $\cS_A$. This map is
    independent of $\ket{i}$ because the dimers of the logical $0$ and $1$ state
    differ only in parity, not in connectivity as provided by Thm.\
    \ref{Thm_1}. Therefore, to get to the factorized form we have to apply the
    same swaps for any bulk input. This swap can be made unitary because one
    has $P_{i,j}^\dagger = P_{i,j}^{-1}=-P_{i,j}$. 
    So if we needed to apply an odd number of swaps to get to the factorizing state, we can just apply an additional swap on an edge carrying a logical or auxiliary dimer which will only change its parity. We end up with an even number of swaps $S_A$ that is a local unitary operation.
    We will now abuse notation and go into a hybrid between dimer and qubit
    language, where we will talk about operators that act on $\cH_{\g,A^c}$ in
    dimer language and we will talk about operators that act on the, now
    disentangled dimers, in qubits language because the set of disentangled
    dimers is just a system of qubits, where each dimer represents a single
    qubit. Another way of thinking about it is that we define a Jordan-Wigner transformation only on the Hilbert space $\cH_{\g,A^c}$ without
    involving the disentangled dimers.\\
    In Eq. \eqref{eq:postswap} the state of the dimers of $D_a, D_\g$ is still
    $i$-dependent. To remove this dependence, we define a local unitary
    $\cX_i^A$ that will remove the $i$-dependence. First, the state $\p_{i,a}$
    is a state described by zeros and ones because it comes from disentangled
    dimers in $D_a$. We can apply $X$ operators to put them into the $\ket{0}$
    state. This sequence of $X$'s forms a unitary $\cX_{i,a}$. Furthermore, if
    the state in the complementary wedge is $\ket{j} = \ket{0}$, the parity of
    the RT dimers only depends on the state $i$. We will now apply a product of
    majorana operators $\g$ that act on these RT dimers and set all their
    parities to be positive, if $\ket{j} = \ket{0}$. 
    This comprises a local unitary $\cX_{i,\g}$.
    At last, we can apply $X$ operators on the logical dimers to transform the
    sequence of zeros and once in $\y_i$ into the state $\ket{i}$ where the
    logical dimer associated to each bulk qubit in $a$ is in the state the
    corresponding bulk qubit is in. This last step is not necessary, we could
    just continue to work with $\y_i$ but for concreteness sake we will also
    perform such an respective application of $X$'s via a unitary $\cX_{i,l}$.
    We then define
    \begin{equation}
        \cX_{i}^A =\cX_{i,l} \cX_{i,a} \cX_{i,\g}.
    \end{equation}
    We now define
    \begin{equation}
        C^A_i = \cX^A_i \cS_A.
    \end{equation} 
    This unitary will satisfy \eqref{eq:cia} by construction where $\ket{\p_a} =
    \ket{0}$.
    We can repeat the same construction for the complementary region to obtain
    a unitary $C_{j}^{A^c}$. \footnote{Note that we have to move the pivot to the rightmost edge of the
    complementary region $A^c$ so that the swap operators one has to apply
are local unitaries.} The connectivity of the state $C_i^{A}C_j^{A^c} \ket{\bar{ij}}$ is
represented graphically in
    Fig.\ \ref{fig:HaPPY_disentangled_Abar}. Note that we constructed
    $\cS_A, \cS_{A^c}$ such that pairs of RT dimers that crossed the same edge
    on the RT surface also begin and end on the same edge. We can also choose
    them such that the dimers associated to one RT edge do not cross each
    other. As was shown in \cite{Jahn:2019nmz}, such dimer pairs give maximally entangled
    states. We defined $\cX_{i,\g}$ such that if
    $\ket{j} = \ket{0}$ all the RT dimers have positive parity in $C_{i}^A
    \ket{\bar{i0}}$. Because of this, all the
    maximally entangled pairs associated to each edge will be in the same
    state with the same parities for any bulk state $\ket{ij}$, because if
    going from $\ket{\bar{00}}$ to $\ket{i 0}$ a pair of RT dimers flipped
    their parities, then $\cX_{i,\g}$ will reverse this parity change. The same
    parity-flip reversal takes place from $\ket{\cX_{j,\g}}$ when the dimers
    change their parity when going from $\ket{\bar{00}}$ to $\ket{\bar{0j}}$,
    so that the combined action $\cX_{j,\g}\cX_{i,\g}$ will flip the parity of
    these RT dimers twice, so that their parity is the same as in the
    $\ket{\bar{00}}$ state, namely positive. 

\begin{figure}[ht]
        \includegraphics[scale=0.4]{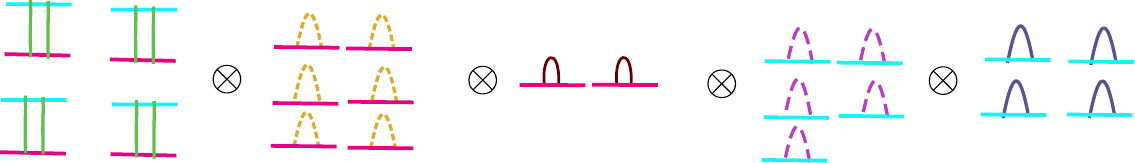}
        \caption{Fully disentangled codestate without indication of parities.}
        \label{fig:HaPPY_disentangled_full}
    \end{figure}
    
    By construction we then have
    \begin{equation}
        C^A_i C^{A^c}_j \ket{\bar{ij}} = \ket{i}\ket{j}\ket{\c_a}\ket{\c_{a^c}}
        \ket{\chi},
    \end{equation}
    where $\ket{\c_a} = \ket{0}, \ket{\c_{a^c}} = \ket{0}$ and $\ket{\c}$ is a
    collection of maximally entangled pairs between $A$ and $A^c$, one for each
    edge in the RT surface and that is
    independent of the bulk logical state $\ket{\bar{ij}}$
    Now that we can disentangle each state $\ket{\bar{ij}}$ individually, we construct the
    unitaries $U_A,U_{A^c}$.
    We can now define projectors 
    \begin{equation}
    \begin{aligned}
        P^A_{i} &= (C^A_i)^\dagger \ket{i}\bra{i} C^A_i\\
        P^{A^c}_{j} &= (C^{A^c}_j)^\dagger \ket{j}\bra{j}C^{A^c}_j,
    \end{aligned}
    \end{equation}
    where $\ket{i}\bra{i} = \ket{i}\bra{i} \otimes \id_{\cH_{\c_a}} \otimes
    \id_{\cH_{\g,A}} \otimes \id_{A^c}$, so that the identities act on the
    space of auxilliary, RT dimers and the complementary region $A^c$. In the following we will omit these identities.
    The projction $P^A_i$ projects on the dimer states in which the logical
    dimers of region $A$ have the same parity as they have in the codestate
    $\ket{i}$ but independent of what parities the other dimers have. 
    These are mutually orthogonal because in their product
    \begin{equation}\label{eq:P_i_orthogonality}
        P^A_i P^A_j = (C^A_i)^\dagger \ket{i}\bra{i} C^A_i(C^A_j)^\dagger \ket{j}\bra{j} C^A_j,
    \end{equation}
    we have 
    \begin{equation}
        C_i^A (C^A_j)^\dagger = \cX^A_i \cS_a \cS_a^\dagger (\cX^A_j)^\dagger = \cX^A_i (\cX^A_j)^{\dagger}
    \end{equation}
    in \eqref{eq:P_i_orthogonality} we will thus end up with $\cX_j^\dagger$
    flipping the spins in $\ket{j}$ to the parities that the corresponding
    dimers have in the original dimer picture and a analogous operation will be
    done by $\bra{i}\cX^A_i$. Here it is that Thm.\ \ref{Thm_1} is important, namely it forces the operator $\ket{i}\bra{i} \cX^A_i$ to multiply to zero with  $\cX_j^\dagger(\ket{j}\bra{j} $ 
    if $i \neq j$ because the application of $\cX^A_i$ on $\ket{i}\bra{i}$ will flip the spins in such a way that the parities are the 
    same as one had in the initial HaPPY state for the dimers that built up the logical $\ket{i}$. These parities (or effectively this collection of ones and zeros) will differ at least in 
    one spin from the state one gets by computing $\cX_j^\dagger (\ket{j}\bra{j} \otimes \id)$ exactly because this is the 
    content of proposition \ref{Thm_1} - the parities of logical dimers differ between any two distinct codestates. 
    Therefore we have
    \begin{equation}
    \begin{aligned}
        &P_i^A P_j^A\\
        & =  \delta_{ij} (C^A_i)^\dagger \ket{i}\bra{i} \cX^A_i (\cX^A_i)^\dagger  \ket{i}\bra{i} C^A_i\\
        &= \delta_{ij} (C^A_i)^\dagger \ket{i}\bra{i}C^A_i=P_i \delta_{ij}
    \end{aligned}
    \end{equation}
    Furthermore, $P_i^A$ is by construction local in $A$ and thus commutes with all $P_i^{A^c}$.\\
    We now define the operators 
    \begin{equation}
    \begin{aligned}
        U_A &:= \sum_i C^A_i P^A_i = \sum_i \ket{i}\bra{i}\cX^A_i \cS_A\\
         U_{A^c} &:= \sum_j C^{A^c}_j P^{A^c}_j = \sum_i \ket{i}\bra{i}\cX^{A^c}_i \cS_{A^c}
    \end{aligned} 
    \end{equation}
 Lastly, we want to show that these are unitaries. We compute
    \begin{equation}
    \begin{aligned}
        U_A U_A^\dagger &= \sum_{ij} \ket{i}\bra{i} \cX^A_i \cS_A \cS_A^\dagger \cX_j^\dagger (\ket{j}\bra{j} \\
        &= \sum_{ij} \ket{i}\bra{i} \cX^A_i (\cX^A_j)^\dagger (\ket{j}\bra{j} \\
        &= \sum_{i} \ket{i}\bra{i} \cX^A_i (\cX^A_i)^\dagger \ket{i}\bra{i} \\
        &= \sum_{i} \ket{i}\bra{i}  = \id.
    \end{aligned}
    \end{equation}
    Here we used the unitarity of $\cS_A$ in the second line, the argument of
    orthogonality following from proposition \ref{Thm_1} as in the proof of the pairwise orthogonality of $P_i^A$ and in the end that $\ket{i}\bra{i}$ is an orthonormal basis for the logical Hilbert space.
    Similarly, we have 
    \begin{equation}
    \begin{aligned}
        U_A^\dagger U_A &= \sum_{ij} \cS_A^\dagger (\cX^A_j)^\dagger \ket{j}\bra{j} \ket{i}\bra{i} \cX^A_i \cS_A \\
        &= \sum_{i} \cS_A^\dagger (\cX^A_i)^\dagger \ket{i}\bra{i} \cX^A_i \cS_A = \sum_i P_i = \id.
    \end{aligned}
    \end{equation}
    Here, we used the fact that $P_i$ is summing to the identity, which follows from a dimension-counting argument. The $P_i$'s are mutually orthogonal and there are as
    many of them as we have bulk qubit states. Each of them projects onto a
    subspace that has the dimensionality $\text{Dim}(\cH) / 2^{n_{bulk}}$,
    where $n_{bulk}$ is the number of bulk qubits and $\cH$ is the full
    boundary Hilbert space. Since we have $2^{n_{bulk}}$ of those orthogonal
    subspaces, the sum of them must be the total Hilbert space, so the
    projections must sum to the identity.\qed

\bibliography{references}

\begin{thebibliography}{82}%
\makeatletter
\providecommand \@ifxundefined [1]{%
 \@ifx{#1\undefined}
}%
\providecommand \@ifnum [1]{%
 \ifnum #1\expandafter \@firstoftwo
 \else \expandafter \@secondoftwo
 \fi
}%
\providecommand \@ifx [1]{%
 \ifx #1\expandafter \@firstoftwo
 \else \expandafter \@secondoftwo
 \fi
}%
\providecommand \natexlab [1]{#1}%
\providecommand \enquote  [1]{``#1''}%
\providecommand \bibnamefont  [1]{#1}%
\providecommand \bibfnamefont [1]{#1}%
\providecommand \citenamefont [1]{#1}%
\providecommand \href@noop [0]{\@secondoftwo}%
\providecommand \href [0]{\begingroup \@sanitize@url \@href}%
\providecommand \@href[1]{\@@startlink{#1}\@@href}%
\providecommand \@@href[1]{\endgroup#1\@@endlink}%
\providecommand \@sanitize@url [0]{\catcode `\\12\catcode `\$12\catcode `\&12\catcode `\#12\catcode `\^12\catcode `\_12\catcode `\%12\relax}%
\providecommand \@@startlink[1]{}%
\providecommand \@@endlink[0]{}%
\providecommand \url  [0]{\begingroup\@sanitize@url \@url }%
\providecommand \@url [1]{\endgroup\@href {#1}{\urlprefix }}%
\providecommand \urlprefix  [0]{URL }%
\providecommand \Eprint [0]{\href }%
\providecommand \doibase [0]{https://doi.org/}%
\providecommand \selectlanguage [0]{\@gobble}%
\providecommand \bibinfo  [0]{\@secondoftwo}%
\providecommand \bibfield  [0]{\@secondoftwo}%
\providecommand \translation [1]{[#1]}%
\providecommand \BibitemOpen [0]{}%
\providecommand \bibitemStop [0]{}%
\providecommand \bibitemNoStop [0]{.\EOS\space}%
\providecommand \EOS [0]{\spacefactor3000\relax}%
\providecommand \BibitemShut  [1]{\csname bibitem#1\endcsname}%
\let\auto@bib@innerbib\@empty
\bibitem [{\citenamefont {Bridgeman}\ and\ \citenamefont {Chubb}(2017)}]{Bridgeman:2016dhh}%
  \BibitemOpen
  \bibfield  {author} {\bibinfo {author} {\bibfnamefont {J.~C.}\ \bibnamefont {Bridgeman}}\ and\ \bibinfo {author} {\bibfnamefont {C.~T.}\ \bibnamefont {Chubb}},\ }\bibfield  {title} {\bibinfo {title} {{Hand-waving and Interpretive Dance: An Introductory Course on Tensor Networks}},\ }\href {https://doi.org/10.1088/1751-8121/aa6dc3} {\bibfield  {journal} {\bibinfo  {journal} {J. Phys. A}\ }\textbf {\bibinfo {volume} {50}},\ \bibinfo {pages} {223001} (\bibinfo {year} {2017})},\ \Eprint {https://arxiv.org/abs/1603.03039} {arXiv:1603.03039 [quant-ph]} \BibitemShut {NoStop}%
\bibitem [{\citenamefont {Cirac}\ and\ \citenamefont {Verstraete}(2009)}]{cirac2009renormalization}%
  \BibitemOpen
  \bibfield  {author} {\bibinfo {author} {\bibfnamefont {J.~I.}\ \bibnamefont {Cirac}}\ and\ \bibinfo {author} {\bibfnamefont {F.}~\bibnamefont {Verstraete}},\ }\bibfield  {title} {\bibinfo {title} {Renormalization and tensor product states in spin chains and lattices},\ }\href@noop {} {\bibfield  {journal} {\bibinfo  {journal} {Journal of physics a: mathematical and theoretical}\ }\textbf {\bibinfo {volume} {42}},\ \bibinfo {pages} {504004} (\bibinfo {year} {2009})}\BibitemShut {NoStop}%
\bibitem [{\citenamefont {Swingle}(2012)}]{Swingle:2009bg}%
  \BibitemOpen
  \bibfield  {author} {\bibinfo {author} {\bibfnamefont {B.}~\bibnamefont {Swingle}},\ }\bibfield  {title} {\bibinfo {title} {{Entanglement Renormalization and Holography}},\ }\href {https://doi.org/10.1103/PhysRevD.86.065007} {\bibfield  {journal} {\bibinfo  {journal} {Phys. Rev. D}\ }\textbf {\bibinfo {volume} {86}},\ \bibinfo {pages} {065007} (\bibinfo {year} {2012})},\ \Eprint {https://arxiv.org/abs/0905.1317} {arXiv:0905.1317 [cond-mat.str-el]} \BibitemShut {NoStop}%
\bibitem [{\citenamefont {Bao}\ \emph {et~al.}(2017)\citenamefont {Bao}, \citenamefont {Cao}, \citenamefont {Carroll},\ and\ \citenamefont {Chatwin-Davies}}]{bao2017sitter}%
  \BibitemOpen
  \bibfield  {author} {\bibinfo {author} {\bibfnamefont {N.}~\bibnamefont {Bao}}, \bibinfo {author} {\bibfnamefont {C.}~\bibnamefont {Cao}}, \bibinfo {author} {\bibfnamefont {S.~M.}\ \bibnamefont {Carroll}},\ and\ \bibinfo {author} {\bibfnamefont {A.}~\bibnamefont {Chatwin-Davies}},\ }\bibfield  {title} {\bibinfo {title} {De sitter space as a tensor network: cosmic no-hair, complementarity, and complexity},\ }\href@noop {} {\bibfield  {journal} {\bibinfo  {journal} {Physical Review D}\ }\textbf {\bibinfo {volume} {96}},\ \bibinfo {pages} {123536} (\bibinfo {year} {2017})}\BibitemShut {NoStop}%
\bibitem [{\citenamefont {Bao}\ \emph {et~al.}(2019)\citenamefont {Bao}, \citenamefont {Penington}, \citenamefont {Sorce},\ and\ \citenamefont {Wall}}]{Bao:2019fpq}%
  \BibitemOpen
  \bibfield  {author} {\bibinfo {author} {\bibfnamefont {N.}~\bibnamefont {Bao}}, \bibinfo {author} {\bibfnamefont {G.}~\bibnamefont {Penington}}, \bibinfo {author} {\bibfnamefont {J.}~\bibnamefont {Sorce}},\ and\ \bibinfo {author} {\bibfnamefont {A.~C.}\ \bibnamefont {Wall}},\ }\href@noop {} {\bibinfo {title} {{Holographic Tensor Networks in Full AdS/CFT}}} (\bibinfo {year} {2019}),\ \Eprint {https://arxiv.org/abs/1902.10157} {arXiv:1902.10157 [hep-th]} \BibitemShut {NoStop}%
\bibitem [{\citenamefont {Bao}\ \emph {et~al.}(2015)\citenamefont {Bao}, \citenamefont {Cao}, \citenamefont {Carroll}, \citenamefont {Chatwin-Davies}, \citenamefont {Hunter-Jones}, \citenamefont {Pollack},\ and\ \citenamefont {Remmen}}]{Bao:2015uaa}%
  \BibitemOpen
  \bibfield  {author} {\bibinfo {author} {\bibfnamefont {N.}~\bibnamefont {Bao}}, \bibinfo {author} {\bibfnamefont {C.}~\bibnamefont {Cao}}, \bibinfo {author} {\bibfnamefont {S.~M.}\ \bibnamefont {Carroll}}, \bibinfo {author} {\bibfnamefont {A.}~\bibnamefont {Chatwin-Davies}}, \bibinfo {author} {\bibfnamefont {N.}~\bibnamefont {Hunter-Jones}}, \bibinfo {author} {\bibfnamefont {J.}~\bibnamefont {Pollack}},\ and\ \bibinfo {author} {\bibfnamefont {G.~N.}\ \bibnamefont {Remmen}},\ }\bibfield  {title} {\bibinfo {title} {{Consistency conditions for an AdS multiscale entanglement renormalization ansatz correspondence}},\ }\href {https://doi.org/10.1103/PhysRevD.91.125036} {\bibfield  {journal} {\bibinfo  {journal} {Phys. Rev. D}\ }\textbf {\bibinfo {volume} {91}},\ \bibinfo {pages} {125036} (\bibinfo {year} {2015})},\ \Eprint {https://arxiv.org/abs/1504.06632} {arXiv:1504.06632 [hep-th]} \BibitemShut {NoStop}%
\bibitem [{\citenamefont {Jahn}\ \emph {et~al.}(2022)\citenamefont {Jahn}, \citenamefont {Zimbor\'as},\ and\ \citenamefont {Eisert}}]{Jahn:2020ukq}%
  \BibitemOpen
  \bibfield  {author} {\bibinfo {author} {\bibfnamefont {A.}~\bibnamefont {Jahn}}, \bibinfo {author} {\bibfnamefont {Z.}~\bibnamefont {Zimbor\'as}},\ and\ \bibinfo {author} {\bibfnamefont {J.}~\bibnamefont {Eisert}},\ }\bibfield  {title} {\bibinfo {title} {{Tensor network models of AdS/qCFT}},\ }\href {https://doi.org/10.22331/q-2022-02-03-643} {\bibfield  {journal} {\bibinfo  {journal} {Quantum}\ }\textbf {\bibinfo {volume} {6}},\ \bibinfo {pages} {643} (\bibinfo {year} {2022})},\ \Eprint {https://arxiv.org/abs/2004.04173} {arXiv:2004.04173 [quant-ph]} \BibitemShut {NoStop}%
\bibitem [{\citenamefont {Miyaji}\ \emph {et~al.}(2017)\citenamefont {Miyaji}, \citenamefont {Takayanagi},\ and\ \citenamefont {Watanabe}}]{Miyaji:2016mxg}%
  \BibitemOpen
  \bibfield  {author} {\bibinfo {author} {\bibfnamefont {M.}~\bibnamefont {Miyaji}}, \bibinfo {author} {\bibfnamefont {T.}~\bibnamefont {Takayanagi}},\ and\ \bibinfo {author} {\bibfnamefont {K.}~\bibnamefont {Watanabe}},\ }\bibfield  {title} {\bibinfo {title} {{From path integrals to tensor networks for the AdS/CFT correspondence}},\ }\href {https://doi.org/10.1103/PhysRevD.95.066004} {\bibfield  {journal} {\bibinfo  {journal} {Phys. Rev. D}\ }\textbf {\bibinfo {volume} {95}},\ \bibinfo {pages} {066004} (\bibinfo {year} {2017})},\ \Eprint {https://arxiv.org/abs/1609.04645} {arXiv:1609.04645 [hep-th]} \BibitemShut {NoStop}%
\bibitem [{\citenamefont {Brown}\ \emph {et~al.}(2020)\citenamefont {Brown}, \citenamefont {Gharibyan}, \citenamefont {Penington},\ and\ \citenamefont {Susskind}}]{Brown:2019rox}%
  \BibitemOpen
  \bibfield  {author} {\bibinfo {author} {\bibfnamefont {A.~R.}\ \bibnamefont {Brown}}, \bibinfo {author} {\bibfnamefont {H.}~\bibnamefont {Gharibyan}}, \bibinfo {author} {\bibfnamefont {G.}~\bibnamefont {Penington}},\ and\ \bibinfo {author} {\bibfnamefont {L.}~\bibnamefont {Susskind}},\ }\bibfield  {title} {\bibinfo {title} {{The Python\textquoteright{}s Lunch: geometric obstructions to decoding Hawking radiation}},\ }\href {https://doi.org/10.1007/JHEP08(2020)121} {\bibfield  {journal} {\bibinfo  {journal} {JHEP}\ }\textbf {\bibinfo {volume} {08}}\bibfield  {number} {\bibinfo  {number} { (121)}},\ }\Eprint {https://arxiv.org/abs/1912.00228} {arXiv:1912.00228 [hep-th]} \BibitemShut {NoStop}%
\bibitem [{\citenamefont {Pfeifer}\ \emph {et~al.}(2009{\natexlab{a}})\citenamefont {Pfeifer}, \citenamefont {Evenbly},\ and\ \citenamefont {Vidal}}]{Pfeifer:2008jt}%
  \BibitemOpen
  \bibfield  {author} {\bibinfo {author} {\bibfnamefont {R.~N.~C.}\ \bibnamefont {Pfeifer}}, \bibinfo {author} {\bibfnamefont {G.}~\bibnamefont {Evenbly}},\ and\ \bibinfo {author} {\bibfnamefont {G.}~\bibnamefont {Vidal}},\ }\bibfield  {title} {\bibinfo {title} {{Entanglement renormalization, scale invariance, and quantum criticality}},\ }\href {https://doi.org/10.1103/PhysRevA.79.040301} {\bibfield  {journal} {\bibinfo  {journal} {Phys. Rev. A}\ }\textbf {\bibinfo {volume} {79}},\ \bibinfo {pages} {040301} (\bibinfo {year} {2009}{\natexlab{a}})},\ \Eprint {https://arxiv.org/abs/0810.0580} {arXiv:0810.0580 [cond-mat.str-el]} \BibitemShut {NoStop}%
\bibitem [{\citenamefont {Verstraete}\ and\ \citenamefont {Cirac}(2010)}]{Verstraete:2010ft}%
  \BibitemOpen
  \bibfield  {author} {\bibinfo {author} {\bibfnamefont {F.}~\bibnamefont {Verstraete}}\ and\ \bibinfo {author} {\bibfnamefont {J.~I.}\ \bibnamefont {Cirac}},\ }\bibfield  {title} {\bibinfo {title} {{Continuous Matrix Product States for Quantum Fields}},\ }\href {https://doi.org/10.1103/PhysRevLett.104.190405} {\bibfield  {journal} {\bibinfo  {journal} {Phys. Rev. Lett.}\ }\textbf {\bibinfo {volume} {104}},\ \bibinfo {pages} {190405} (\bibinfo {year} {2010})},\ \Eprint {https://arxiv.org/abs/1002.1824} {arXiv:1002.1824 [cond-mat.str-el]} \BibitemShut {NoStop}%
\bibitem [{\citenamefont {Haegeman}\ \emph {et~al.}(2013)\citenamefont {Haegeman}, \citenamefont {Osborne}, \citenamefont {Verschelde},\ and\ \citenamefont {Verstraete}}]{Haegeman:2011uy}%
  \BibitemOpen
  \bibfield  {author} {\bibinfo {author} {\bibfnamefont {J.}~\bibnamefont {Haegeman}}, \bibinfo {author} {\bibfnamefont {T.~J.}\ \bibnamefont {Osborne}}, \bibinfo {author} {\bibfnamefont {H.}~\bibnamefont {Verschelde}},\ and\ \bibinfo {author} {\bibfnamefont {F.}~\bibnamefont {Verstraete}},\ }\bibfield  {title} {\bibinfo {title} {{Entanglement Renormalization for Quantum Fields in Real Space}},\ }\href {https://doi.org/10.1103/PhysRevLett.110.100402} {\bibfield  {journal} {\bibinfo  {journal} {Phys. Rev. Lett.}\ }\textbf {\bibinfo {volume} {110}},\ \bibinfo {pages} {100402} (\bibinfo {year} {2013})},\ \Eprint {https://arxiv.org/abs/1102.5524} {arXiv:1102.5524 [hep-th]} \BibitemShut {NoStop}%
\bibitem [{\citenamefont {Witteveen}\ \emph {et~al.}(2022)\citenamefont {Witteveen}, \citenamefont {Scholz}, \citenamefont {Swingle},\ and\ \citenamefont {Walter}}]{Witteveen:2019lsk}%
  \BibitemOpen
  \bibfield  {author} {\bibinfo {author} {\bibfnamefont {F.}~\bibnamefont {Witteveen}}, \bibinfo {author} {\bibfnamefont {V.}~\bibnamefont {Scholz}}, \bibinfo {author} {\bibfnamefont {B.}~\bibnamefont {Swingle}},\ and\ \bibinfo {author} {\bibfnamefont {M.}~\bibnamefont {Walter}},\ }\bibfield  {title} {\bibinfo {title} {{Quantum Circuit Approximations and Entanglement Renormalization for the Dirac Field in 1+1 Dimensions}},\ }\href {https://doi.org/10.1007/s00220-021-04274-w} {\bibfield  {journal} {\bibinfo  {journal} {Commun. Math. Phys.}\ }\textbf {\bibinfo {volume} {389}},\ \bibinfo {pages} {75} (\bibinfo {year} {2022})},\ \Eprint {https://arxiv.org/abs/1905.08821} {arXiv:1905.08821 [quant-ph]} \BibitemShut {NoStop}%
\bibitem [{\citenamefont {Stottmeister}\ \emph {et~al.}(2021)\citenamefont {Stottmeister}, \citenamefont {Morinelli}, \citenamefont {Morsella},\ and\ \citenamefont {Tanimoto}}]{Stottmeister:2020ezd}%
  \BibitemOpen
  \bibfield  {author} {\bibinfo {author} {\bibfnamefont {A.}~\bibnamefont {Stottmeister}}, \bibinfo {author} {\bibfnamefont {V.}~\bibnamefont {Morinelli}}, \bibinfo {author} {\bibfnamefont {G.}~\bibnamefont {Morsella}},\ and\ \bibinfo {author} {\bibfnamefont {Y.}~\bibnamefont {Tanimoto}},\ }\bibfield  {title} {\bibinfo {title} {{Operator-Algebraic Renormalization and Wavelets}},\ }\href {https://doi.org/10.1103/PhysRevLett.127.230601} {\bibfield  {journal} {\bibinfo  {journal} {Phys. Rev. Lett.}\ }\textbf {\bibinfo {volume} {127}},\ \bibinfo {pages} {230601} (\bibinfo {year} {2021})},\ \Eprint {https://arxiv.org/abs/2002.01442} {arXiv:2002.01442 [math-ph]} \BibitemShut {NoStop}%
\bibitem [{\citenamefont {Osborne}\ and\ \citenamefont {Stottmeister}(2023)}]{Osborne:2021ppp}%
  \BibitemOpen
  \bibfield  {author} {\bibinfo {author} {\bibfnamefont {T.~J.}\ \bibnamefont {Osborne}}\ and\ \bibinfo {author} {\bibfnamefont {A.}~\bibnamefont {Stottmeister}},\ }\bibfield  {title} {\bibinfo {title} {{Conformal Field Theory from Lattice Fermions}},\ }\href {https://doi.org/10.1007/s00220-022-04521-8} {\bibfield  {journal} {\bibinfo  {journal} {Commun. Math. Phys.}\ }\textbf {\bibinfo {volume} {398}},\ \bibinfo {pages} {219} (\bibinfo {year} {2023})},\ \Eprint {https://arxiv.org/abs/2107.13834} {arXiv:2107.13834 [math-ph]} \BibitemShut {NoStop}%
\bibitem [{\citenamefont {Haag}(1996)}]{Haag:1996hvx}%
  \BibitemOpen
  \bibfield  {author} {\bibinfo {author} {\bibfnamefont {R.}~\bibnamefont {Haag}},\ }\href {https://doi.org/10.1007/978-3-642-61458-3} {\emph {\bibinfo {title} {{Local Quantum Physics}}}},\ Theoretical and Mathematical Physics\ (\bibinfo  {publisher} {Springer},\ \bibinfo {address} {Berlin},\ \bibinfo {year} {1996})\BibitemShut {NoStop}%
\bibitem [{\citenamefont {van Luijk}\ \emph {et~al.}(2024{\natexlab{a}})\citenamefont {van Luijk}, \citenamefont {Stottmeister},\ and\ \citenamefont {Werner}}]{van2024convergence}%
  \BibitemOpen
  \bibfield  {author} {\bibinfo {author} {\bibfnamefont {L.}~\bibnamefont {van Luijk}}, \bibinfo {author} {\bibfnamefont {A.}~\bibnamefont {Stottmeister}},\ and\ \bibinfo {author} {\bibfnamefont {R.~F.}\ \bibnamefont {Werner}},\ }\bibfield  {title} {\bibinfo {title} {Convergence of dynamics on inductive systems of banach spaces},\ }in\ \href@noop {} {\emph {\bibinfo {booktitle} {Annales Henri Poincar{\'e}}}}\ (\bibinfo {organization} {Springer},\ \bibinfo {year} {2024})\ pp.\ \bibinfo {pages} {1--56}\BibitemShut {NoStop}%
\bibitem [{\citenamefont {Osborne}\ and\ \citenamefont {Stiegemann}(2020)}]{Osborne:2017woa}%
  \BibitemOpen
  \bibfield  {author} {\bibinfo {author} {\bibfnamefont {T.~J.}\ \bibnamefont {Osborne}}\ and\ \bibinfo {author} {\bibfnamefont {D.~E.}\ \bibnamefont {Stiegemann}},\ }\bibfield  {title} {\bibinfo {title} {{Dynamics for holographic codes}},\ }\href {https://doi.org/10.1007/JHEP04(2020)154} {\bibfield  {journal} {\bibinfo  {journal} {JHEP}\ }\textbf {\bibinfo {volume} {04}}\bibfield  {number} {\bibinfo  {number} { (154)}},\ }\Eprint {https://arxiv.org/abs/1706.08823} {arXiv:1706.08823 [quant-ph]} \BibitemShut {NoStop}%
\bibitem [{\citenamefont {Kang}\ and\ \citenamefont {Kolchmeyer}(2021)}]{Kang:2019dfi}%
  \BibitemOpen
  \bibfield  {author} {\bibinfo {author} {\bibfnamefont {M.~J.}\ \bibnamefont {Kang}}\ and\ \bibinfo {author} {\bibfnamefont {D.~K.}\ \bibnamefont {Kolchmeyer}},\ }\bibfield  {title} {\bibinfo {title} {{Entanglement wedge reconstruction of infinite-dimensional von Neumann algebras using tensor networks}},\ }\href {https://doi.org/10.1103/PhysRevD.103.126018} {\bibfield  {journal} {\bibinfo  {journal} {Phys. Rev. D}\ }\textbf {\bibinfo {volume} {103}},\ \bibinfo {pages} {126018} (\bibinfo {year} {2021})},\ \Eprint {https://arxiv.org/abs/1910.06328} {arXiv:1910.06328 [hep-th]} \BibitemShut {NoStop}%
\bibitem [{\citenamefont {Gesteau}\ and\ \citenamefont {Kang}(2020{\natexlab{a}})}]{Gesteau:2020hoz}%
  \BibitemOpen
  \bibfield  {author} {\bibinfo {author} {\bibfnamefont {E.}~\bibnamefont {Gesteau}}\ and\ \bibinfo {author} {\bibfnamefont {M.~J.}\ \bibnamefont {Kang}},\ }\href@noop {} {\bibinfo {title} {{The infinite-dimensional HaPPY code: entanglement wedge reconstruction and dynamics}}} (\bibinfo {year} {2020}{\natexlab{a}}),\ \Eprint {https://arxiv.org/abs/2005.05971} {arXiv:2005.05971 [hep-th]} \BibitemShut {NoStop}%
\bibitem [{\citenamefont {Gesteau}\ \emph {et~al.}(2022)\citenamefont {Gesteau}, \citenamefont {Marcolli},\ and\ \citenamefont {Parikh}}]{Gesteau:2022hss}%
  \BibitemOpen
  \bibfield  {author} {\bibinfo {author} {\bibfnamefont {E.}~\bibnamefont {Gesteau}}, \bibinfo {author} {\bibfnamefont {M.}~\bibnamefont {Marcolli}},\ and\ \bibinfo {author} {\bibfnamefont {S.}~\bibnamefont {Parikh}},\ }\bibfield  {title} {\bibinfo {title} {{Holographic tensor networks from hyperbolic buildings}},\ }\href {https://doi.org/10.1007/JHEP10(2022)169} {\bibfield  {journal} {\bibinfo  {journal} {JHEP}\ }\textbf {\bibinfo {volume} {10}}\bibfield  {number} {\bibinfo  {number} { (169)}},\ }\Eprint {https://arxiv.org/abs/2202.01788} {arXiv:2202.01788 [hep-th]} \BibitemShut {NoStop}%
\bibitem [{\citenamefont {Gesteau}\ and\ \citenamefont {Kang}(2020{\natexlab{b}})}]{Gesteau:2020rtg}%
  \BibitemOpen
  \bibfield  {author} {\bibinfo {author} {\bibfnamefont {E.}~\bibnamefont {Gesteau}}\ and\ \bibinfo {author} {\bibfnamefont {M.~J.}\ \bibnamefont {Kang}},\ }\href@noop {} {\bibinfo {title} {{Thermal states are vital: Entanglement Wedge Reconstruction from Operator-Pushing}}} (\bibinfo {year} {2020}{\natexlab{b}}),\ \Eprint {https://arxiv.org/abs/2005.07189} {arXiv:2005.07189 [hep-th]} \BibitemShut {NoStop}%
\bibitem [{\citenamefont {Morinelli}\ \emph {et~al.}(2021)\citenamefont {Morinelli}, \citenamefont {Morsella}, \citenamefont {Stottmeister},\ and\ \citenamefont {Tanimoto}}]{Morinelli:2020uea}%
  \BibitemOpen
  \bibfield  {author} {\bibinfo {author} {\bibfnamefont {V.}~\bibnamefont {Morinelli}}, \bibinfo {author} {\bibfnamefont {G.}~\bibnamefont {Morsella}}, \bibinfo {author} {\bibfnamefont {A.}~\bibnamefont {Stottmeister}},\ and\ \bibinfo {author} {\bibfnamefont {Y.}~\bibnamefont {Tanimoto}},\ }\bibfield  {title} {\bibinfo {title} {{Scaling Limits of Lattice Quantum Fields by Wavelets}},\ }\href {https://doi.org/10.1007/s00220-021-04152-5} {\bibfield  {journal} {\bibinfo  {journal} {Commun. Math. Phys.}\ }\textbf {\bibinfo {volume} {387}},\ \bibinfo {pages} {299} (\bibinfo {year} {2021})},\ \Eprint {https://arxiv.org/abs/2010.11121} {arXiv:2010.11121 [math-ph]} \BibitemShut {NoStop}%
\bibitem [{\citenamefont {Stottmeister}(2022)}]{Stottmeister:2022ptp}%
  \BibitemOpen
  \bibfield  {author} {\bibinfo {author} {\bibfnamefont {A.}~\bibnamefont {Stottmeister}},\ }\href@noop {} {\bibinfo {title} {{Anyon braiding and the renormalization group}}} (\bibinfo {year} {2022}),\ \Eprint {https://arxiv.org/abs/2201.11562} {arXiv:2201.11562 [quant-ph]} \BibitemShut {NoStop}%
\bibitem [{\citenamefont {Almheiri}\ \emph {et~al.}(2015)\citenamefont {Almheiri}, \citenamefont {Dong},\ and\ \citenamefont {Harlow}}]{almheiri2015bulk}%
  \BibitemOpen
  \bibfield  {author} {\bibinfo {author} {\bibfnamefont {A.}~\bibnamefont {Almheiri}}, \bibinfo {author} {\bibfnamefont {X.}~\bibnamefont {Dong}},\ and\ \bibinfo {author} {\bibfnamefont {D.}~\bibnamefont {Harlow}},\ }\bibfield  {title} {\bibinfo {title} {Bulk locality and quantum error correction in ads/cft},\ }\href@noop {} {\bibfield  {journal} {\bibinfo  {journal} {Journal of High Energy Physics}\ }\textbf {\bibinfo {volume} {2015}},\ \bibinfo {pages} {1} (\bibinfo {year} {2015})}\BibitemShut {NoStop}%
\bibitem [{\citenamefont {Pastawski}\ \emph {et~al.}(2015)\citenamefont {Pastawski}, \citenamefont {Yoshida}, \citenamefont {Harlow},\ and\ \citenamefont {Preskill}}]{Pastawski:2015qua}%
  \BibitemOpen
  \bibfield  {author} {\bibinfo {author} {\bibfnamefont {F.}~\bibnamefont {Pastawski}}, \bibinfo {author} {\bibfnamefont {B.}~\bibnamefont {Yoshida}}, \bibinfo {author} {\bibfnamefont {D.}~\bibnamefont {Harlow}},\ and\ \bibinfo {author} {\bibfnamefont {J.}~\bibnamefont {Preskill}},\ }\bibfield  {title} {\bibinfo {title} {{Holographic quantum error-correcting codes: Toy models for the bulk/boundary correspondence}},\ }\href {https://doi.org/10.1007/JHEP06(2015)149} {\bibfield  {journal} {\bibinfo  {journal} {JHEP}\ }\textbf {\bibinfo {volume} {06}}\bibfield  {number} {\bibinfo  {number} { (149)}},\ }\Eprint {https://arxiv.org/abs/1503.06237} {arXiv:1503.06237 [hep-th]} \BibitemShut {NoStop}%
\bibitem [{\citenamefont {Araki}\ and\ \citenamefont {Woods}(1968)}]{Araki1968ACO}%
  \BibitemOpen
  \bibfield  {author} {\bibinfo {author} {\bibfnamefont {H.}~\bibnamefont {Araki}}\ and\ \bibinfo {author} {\bibfnamefont {E.~J.}\ \bibnamefont {Woods}},\ }\bibfield  {title} {\bibinfo {title} {A classification of factors},\ }\href {https://api.semanticscholar.org/CorpusID:54930318} {\bibfield  {journal} {\bibinfo  {journal} {Publications of The Research Institute for Mathematical Sciences}\ }\textbf {\bibinfo {volume} {4}},\ \bibinfo {pages} {51} (\bibinfo {year} {1968})}\BibitemShut {NoStop}%
\bibitem [{\citenamefont {Powers}(1967)}]{Powers1967RepresentationsOU}%
  \BibitemOpen
  \bibfield  {author} {\bibinfo {author} {\bibfnamefont {R.~T.}\ \bibnamefont {Powers}},\ }\bibfield  {title} {\bibinfo {title} {{Representations of Uniformly Hyperfinite Algebras and Their Associated von Neumann Rings}},\ }\href {https://api.semanticscholar.org/CorpusID:124896803} {\bibfield  {journal} {\bibinfo  {journal} {Annals of Mathematics}\ }\textbf {\bibinfo {volume} {86}},\ \bibinfo {pages} {138} (\bibinfo {year} {1967})}\BibitemShut {NoStop}%
\bibitem [{\citenamefont {Jahn}\ and\ \citenamefont {Eisert}(2021)}]{Jahn:2021uqr}%
  \BibitemOpen
  \bibfield  {author} {\bibinfo {author} {\bibfnamefont {A.}~\bibnamefont {Jahn}}\ and\ \bibinfo {author} {\bibfnamefont {J.}~\bibnamefont {Eisert}},\ }\bibfield  {title} {\bibinfo {title} {{Holographic tensor network models and quantum error correction: a topical review}},\ }\href {https://doi.org/10.1088/2058-9565/ac0293} {\bibfield  {journal} {\bibinfo  {journal} {Quantum Sci. Technol.}\ }\textbf {\bibinfo {volume} {6}},\ \bibinfo {pages} {033002} (\bibinfo {year} {2021})},\ \Eprint {https://arxiv.org/abs/2102.02619} {arXiv:2102.02619 [quant-ph]} \BibitemShut {NoStop}%
\bibitem [{\citenamefont {Pollack}\ \emph {et~al.}(2022)\citenamefont {Pollack}, \citenamefont {Rall},\ and\ \citenamefont {Rocchetto}}]{Pollack:2021yij}%
  \BibitemOpen
  \bibfield  {author} {\bibinfo {author} {\bibfnamefont {J.}~\bibnamefont {Pollack}}, \bibinfo {author} {\bibfnamefont {P.}~\bibnamefont {Rall}},\ and\ \bibinfo {author} {\bibfnamefont {A.}~\bibnamefont {Rocchetto}},\ }\bibfield  {title} {\bibinfo {title} {{Understanding holographic error correction via unique algebras and atomic examples}},\ }\href {https://doi.org/10.1007/JHEP06(2022)056} {\bibfield  {journal} {\bibinfo  {journal} {JHEP}\ }\textbf {\bibinfo {volume} {06}}\bibfield  {number} {\bibinfo  {number} { (056)}},\ }\Eprint {https://arxiv.org/abs/2110.14691} {arXiv:2110.14691 [quant-ph]} \BibitemShut {NoStop}%
\bibitem [{\citenamefont {Ryu}\ and\ \citenamefont {Takayanagi}(2006)}]{Ryu:2006bv}%
  \BibitemOpen
  \bibfield  {author} {\bibinfo {author} {\bibfnamefont {S.}~\bibnamefont {Ryu}}\ and\ \bibinfo {author} {\bibfnamefont {T.}~\bibnamefont {Takayanagi}},\ }\bibfield  {title} {\bibinfo {title} {{Holographic derivation of entanglement entropy from AdS/CFT}},\ }\href {https://doi.org/10.1103/PhysRevLett.96.181602} {\bibfield  {journal} {\bibinfo  {journal} {Phys. Rev. Lett.}\ }\textbf {\bibinfo {volume} {96}},\ \bibinfo {pages} {181602} (\bibinfo {year} {2006})},\ \Eprint {https://arxiv.org/abs/hep-th/0603001} {arXiv:hep-th/0603001} \BibitemShut {NoStop}%
\bibitem [{\citenamefont {Harlow}(2017)}]{Harlow:2016vwg}%
  \BibitemOpen
  \bibfield  {author} {\bibinfo {author} {\bibfnamefont {D.}~\bibnamefont {Harlow}},\ }\bibfield  {title} {\bibinfo {title} {{The Ryu\textendash{}Takayanagi Formula from Quantum Error Correction}},\ }\href {https://doi.org/10.1007/s00220-017-2904-z} {\bibfield  {journal} {\bibinfo  {journal} {Commun. Math. Phys.}\ }\textbf {\bibinfo {volume} {354}},\ \bibinfo {pages} {865} (\bibinfo {year} {2017})},\ \Eprint {https://arxiv.org/abs/1607.03901} {arXiv:1607.03901 [hep-th]} \BibitemShut {NoStop}%
\bibitem [{\citenamefont {Faulkner}(2020)}]{Faulkner:2020hzi}%
  \BibitemOpen
  \bibfield  {author} {\bibinfo {author} {\bibfnamefont {T.}~\bibnamefont {Faulkner}},\ }\href@noop {} {\bibinfo {title} {{The holographic map as a conditional expectation}}} (\bibinfo {year} {2020}),\ \Eprint {https://arxiv.org/abs/2008.04810} {arXiv:2008.04810 [hep-th]} \BibitemShut {NoStop}%
\bibitem [{\citenamefont {Faulkner}\ and\ \citenamefont {Li}(2022)}]{Faulkner:2022ada}%
  \BibitemOpen
  \bibfield  {author} {\bibinfo {author} {\bibfnamefont {T.}~\bibnamefont {Faulkner}}\ and\ \bibinfo {author} {\bibfnamefont {M.}~\bibnamefont {Li}},\ }\href@noop {} {\bibinfo {title} {{Asymptotically isometric codes for holography}}} (\bibinfo {year} {2022}),\ \Eprint {https://arxiv.org/abs/2211.12439} {arXiv:2211.12439 [hep-th]} \BibitemShut {NoStop}%
\bibitem [{\citenamefont {Gesteau}(2025)}]{Gesteau:2023hbq}%
  \BibitemOpen
  \bibfield  {author} {\bibinfo {author} {\bibfnamefont {E.}~\bibnamefont {Gesteau}},\ }\bibfield  {title} {\bibinfo {title} {{Large N von Neumann Algebras and the Renormalization of Newton\textquoteright{}s Constant}},\ }\href {https://doi.org/10.1007/s00220-024-05192-3} {\bibfield  {journal} {\bibinfo  {journal} {Commun. Math. Phys.}\ }\textbf {\bibinfo {volume} {406}},\ \bibinfo {pages} {40} (\bibinfo {year} {2025})},\ \Eprint {https://arxiv.org/abs/2302.01938} {arXiv:2302.01938 [hep-th]} \BibitemShut {NoStop}%
\bibitem [{\citenamefont {{Vidal}}(2010)}]{Vidal.ERintro}%
  \BibitemOpen
  \bibfield  {author} {\bibinfo {author} {\bibfnamefont {G.}~\bibnamefont {{Vidal}}},\ }\href@noop {} {\bibinfo {title} {{Entanglement renormalization: An introduction}}} (\bibinfo {year} {2010})\BibitemShut {NoStop}%
\bibitem [{\citenamefont {Takesaki}(1979)}]{takesaki1979theory}%
  \BibitemOpen
  \bibfield  {author} {\bibinfo {author} {\bibfnamefont {M.}~\bibnamefont {Takesaki}},\ }\href {https://doi.org/10.1007/978-1-4612-6188-9} {\emph {\bibinfo {title} {Theory of Operator Algebras I}}},\ \bibinfo {edition} {1st}\ ed.\ (\bibinfo  {publisher} {Springer New York, NY},\ \bibinfo {address} {New York, NY},\ \bibinfo {year} {1979})\ pp.\ \bibinfo {pages} {VIII, 418},\ \bibinfo {note} {published: 05 November 2011 (softcover), 06 December 2012 (eBook). No further volume published. Part of Springer Book Archive}\BibitemShut {NoStop}%
\bibitem [{\citenamefont {Sunder}(1987)}]{sunder_1987}%
  \BibitemOpen
  \bibfield  {author} {\bibinfo {author} {\bibfnamefont {V.~S.}\ \bibnamefont {Sunder}},\ }\href {https://doi.org/10.1007/978-1-4613-8669-8} {\emph {\bibinfo {title} {An {Invitation} to von {Neumann} {Algebras}}}},\ edited by\ \bibinfo {editor} {\bibfnamefont {F.~W.}\ \bibnamefont {Gehring}}\ and\ \bibinfo {editor} {\bibfnamefont {P.~R.}\ \bibnamefont {Halmos}},\ Universitext\ (\bibinfo  {publisher} {Springer},\ \bibinfo {address} {New York, NY},\ \bibinfo {year} {1987})\BibitemShut {NoStop}%
\bibitem [{\citenamefont {Steinberg}\ \emph {et~al.}(2023)\citenamefont {Steinberg}, \citenamefont {Feld},\ and\ \citenamefont {Jahn}}]{Steinberg:2023wll}%
  \BibitemOpen
  \bibfield  {author} {\bibinfo {author} {\bibfnamefont {M.}~\bibnamefont {Steinberg}}, \bibinfo {author} {\bibfnamefont {S.}~\bibnamefont {Feld}},\ and\ \bibinfo {author} {\bibfnamefont {A.}~\bibnamefont {Jahn}},\ }\bibfield  {title} {\bibinfo {title} {{Holographic codes from hyperinvariant tensor networks}},\ }\href {https://doi.org/10.1038/s41467-023-42743-z} {\bibfield  {journal} {\bibinfo  {journal} {Nature Commun.}\ }\textbf {\bibinfo {volume} {14}},\ \bibinfo {pages} {7314} (\bibinfo {year} {2023})},\ \Eprint {https://arxiv.org/abs/2304.02732} {arXiv:2304.02732 [quant-ph]} \BibitemShut {NoStop}%
\bibitem [{\citenamefont {Blackadar}(2006)}]{blackadar2006operator}%
  \BibitemOpen
  \bibfield  {author} {\bibinfo {author} {\bibfnamefont {B.}~\bibnamefont {Blackadar}},\ }\href@noop {} {\emph {\bibinfo {title} {Operator algebras: theory of C*-algebras and von Neumann algebras}}},\ Vol.\ \bibinfo {volume} {122}\ (\bibinfo  {publisher} {Springer Science \& Business Media},\ \bibinfo {year} {2006})\BibitemShut {NoStop}%
\bibitem [{\citenamefont {Kadison}\ and\ \citenamefont {Ringrose}(1986)}]{kadison1986fundamentals}%
  \BibitemOpen
  \bibfield  {author} {\bibinfo {author} {\bibfnamefont {R.~V.}\ \bibnamefont {Kadison}}\ and\ \bibinfo {author} {\bibfnamefont {J.~R.}\ \bibnamefont {Ringrose}},\ }\href@noop {} {\emph {\bibinfo {title} {Fundamentals of the theory of operator algebras. Volume II: Advanced theory}}}\ (\bibinfo  {publisher} {Academic press New York},\ \bibinfo {year} {1986})\BibitemShut {NoStop}%
\bibitem [{\citenamefont {M.~Rørdam}(2000)}]{Roerdam}%
  \BibitemOpen
  \bibfield  {author} {\bibinfo {author} {\bibfnamefont {N.~L.~.}\ \bibnamefont {M.~Rørdam}, \bibfnamefont {F.~Larsen}},\ }\href@noop {} {\emph {\bibinfo {title} {{An Introduction to K-Theory for C*-Algebras}}}}\ (\bibinfo  {publisher} {Cambridge University Press},\ \bibinfo {year} {2000})\BibitemShut {NoStop}%
\bibitem [{\citenamefont {Brown}\ and\ \citenamefont {Ozawa}(2008)}]{brown_2008}%
  \BibitemOpen
  \bibfield  {author} {\bibinfo {author} {\bibfnamefont {N.}~\bibnamefont {Brown}}\ and\ \bibinfo {author} {\bibfnamefont {N.}~\bibnamefont {Ozawa}},\ }\href {https://doi.org/10.1090/gsm/088} {\emph {\bibinfo {title} {$C^*$-{Algebras} and {Finite}-{Dimensional} {Approximations}}}},\ \bibinfo {series} {Graduate {Studies} in {Mathematics}}, Vol.~\bibinfo {volume} {88}\ (\bibinfo  {publisher} {American Mathematical Society},\ \bibinfo {address} {Providence, Rhode Island},\ \bibinfo {year} {2008})\BibitemShut {NoStop}%
\bibitem [{\citenamefont {Laflamme}\ \emph {et~al.}(1996)\citenamefont {Laflamme}, \citenamefont {Miquel}, \citenamefont {Paz},\ and\ \citenamefont {Zurek}}]{Laflamme:1996iw}%
  \BibitemOpen
  \bibfield  {author} {\bibinfo {author} {\bibfnamefont {R.}~\bibnamefont {Laflamme}}, \bibinfo {author} {\bibfnamefont {C.}~\bibnamefont {Miquel}}, \bibinfo {author} {\bibfnamefont {J.~P.}\ \bibnamefont {Paz}},\ and\ \bibinfo {author} {\bibfnamefont {W.~H.}\ \bibnamefont {Zurek}},\ }\bibfield  {title} {\bibinfo {title} {{Perfect Quantum Error Correcting Code}},\ }\href {https://doi.org/10.1103/PhysRevLett.77.198} {\bibfield  {journal} {\bibinfo  {journal} {Phys. Rev. Lett.}\ }\textbf {\bibinfo {volume} {77}},\ \bibinfo {pages} {198} (\bibinfo {year} {1996})},\ \Eprint {https://arxiv.org/abs/quant-ph/9602019} {arXiv:quant-ph/9602019} \BibitemShut {NoStop}%
\bibitem [{\citenamefont {Gottesman}(1997)}]{Gottesman:1997zz}%
  \BibitemOpen
  \bibfield  {author} {\bibinfo {author} {\bibfnamefont {D.}~\bibnamefont {Gottesman}},\ }\href@noop {} {\bibinfo {title} {{Stabilizer codes and quantum error correction}}} (\bibinfo {year} {1997}),\ \Eprint {https://arxiv.org/abs/quant-ph/9705052} {arXiv:quant-ph/9705052} \BibitemShut {NoStop}%
\bibitem [{\citenamefont {Furuya}\ \emph {et~al.}(2022)\citenamefont {Furuya}, \citenamefont {Lashkari},\ and\ \citenamefont {Ouseph}}]{Furuya:2020tzv}%
  \BibitemOpen
  \bibfield  {author} {\bibinfo {author} {\bibfnamefont {K.}~\bibnamefont {Furuya}}, \bibinfo {author} {\bibfnamefont {N.}~\bibnamefont {Lashkari}},\ and\ \bibinfo {author} {\bibfnamefont {S.}~\bibnamefont {Ouseph}},\ }\bibfield  {title} {\bibinfo {title} {{Real-space RG, error correction and Petz map}},\ }\href {https://doi.org/10.1007/JHEP01(2022)170} {\bibfield  {journal} {\bibinfo  {journal} {JHEP}\ }\textbf {\bibinfo {volume} {01}}\bibfield  {number} {\bibinfo  {number} { (170)}},\ }\Eprint {https://arxiv.org/abs/2012.14001} {arXiv:2012.14001 [hep-th]} \BibitemShut {NoStop}%
\bibitem [{\citenamefont {Str{\v{a}}til{\v{a}}}(2020)}]{Stratila_2020}%
  \BibitemOpen
  \bibfield  {author} {\bibinfo {author} {\bibfnamefont {S.~V.}\ \bibnamefont {Str{\v{a}}til{\v{a}}}},\ }\href@noop {} {\emph {\bibinfo {title} {Modular Theory in Operator Algebras}}},\ Cambridge IISc Series\ (\bibinfo  {publisher} {Cambridge University Press},\ \bibinfo {year} {2020})\BibitemShut {NoStop}%
\bibitem [{\citenamefont {Jahn}\ \emph {et~al.}(2019)\citenamefont {Jahn}, \citenamefont {Gluza}, \citenamefont {Pastawski},\ and\ \citenamefont {Eisert}}]{Jahn:2019nmz}%
  \BibitemOpen
  \bibfield  {author} {\bibinfo {author} {\bibfnamefont {A.}~\bibnamefont {Jahn}}, \bibinfo {author} {\bibfnamefont {M.}~\bibnamefont {Gluza}}, \bibinfo {author} {\bibfnamefont {F.}~\bibnamefont {Pastawski}},\ and\ \bibinfo {author} {\bibfnamefont {J.}~\bibnamefont {Eisert}},\ }\bibfield  {title} {\bibinfo {title} {{Majorana dimers and holographic quantum error-correcting codes}},\ }\href {https://doi.org/10.1103/PhysRevResearch.1.033079} {\bibfield  {journal} {\bibinfo  {journal} {Phys. Rev. Research.}\ }\textbf {\bibinfo {volume} {1}},\ \bibinfo {pages} {033079} (\bibinfo {year} {2019})},\ \Eprint {https://arxiv.org/abs/1905.03268} {arXiv:1905.03268 [hep-th]} \BibitemShut {NoStop}%
\bibitem [{\citenamefont {Gottesman}(1998)}]{gottesman1998}%
  \BibitemOpen
  \bibfield  {author} {\bibinfo {author} {\bibfnamefont {D.}~\bibnamefont {Gottesman}},\ }\href {https://arxiv.org/abs/quant-ph/9807006} {\bibinfo {title} {The heisenberg representation of quantum computers}} (\bibinfo {year} {1998}),\ \Eprint {https://arxiv.org/abs/quant-ph/9807006} {arXiv:quant-ph/9807006 [quant-ph]} \BibitemShut {NoStop}%
\bibitem [{\citenamefont {White}\ \emph {et~al.}(2021)\citenamefont {White}, \citenamefont {Cao},\ and\ \citenamefont {Swingle}}]{White:2020zoz}%
  \BibitemOpen
  \bibfield  {author} {\bibinfo {author} {\bibfnamefont {C.~D.}\ \bibnamefont {White}}, \bibinfo {author} {\bibfnamefont {C.}~\bibnamefont {Cao}},\ and\ \bibinfo {author} {\bibfnamefont {B.}~\bibnamefont {Swingle}},\ }\bibfield  {title} {\bibinfo {title} {{Conformal field theories are magical}},\ }\href {https://doi.org/10.1103/PhysRevB.103.075145} {\bibfield  {journal} {\bibinfo  {journal} {Phys. Rev. B}\ }\textbf {\bibinfo {volume} {103}},\ \bibinfo {pages} {075145} (\bibinfo {year} {2021})},\ \Eprint {https://arxiv.org/abs/2007.01303} {arXiv:2007.01303 [quant-ph]} \BibitemShut {NoStop}%
\bibitem [{\citenamefont {Cao}(2024)}]{Cao:2023mzo}%
  \BibitemOpen
  \bibfield  {author} {\bibinfo {author} {\bibfnamefont {C.}~\bibnamefont {Cao}},\ }\href {https://doi.org/10.1007/JHEP11(2024)105} {\bibinfo {title} {{Non-trivial area operators require non-local magic}}} (\bibinfo {year} {2024}),\ \Eprint {https://arxiv.org/abs/2306.14996} {arXiv:2306.14996 [hep-th]} \BibitemShut {NoStop}%
\bibitem [{\citenamefont {Bittel}\ \emph {et~al.}(2025)\citenamefont {Bittel}, \citenamefont {Leone}, \citenamefont {Saraidaris},\ and\ \citenamefont {Shaposhnik}}]{cliffordpaper:2025}%
  \BibitemOpen
  \bibfield  {author} {\bibinfo {author} {\bibfnamefont {L.}~\bibnamefont {Bittel}}, \bibinfo {author} {\bibfnamefont {L.}~\bibnamefont {Leone}}, \bibinfo {author} {\bibfnamefont {D.}~\bibnamefont {Saraidaris}},\ and\ \bibinfo {author} {\bibfnamefont {L.}~\bibnamefont {Shaposhnik}},\ }\bibfield  {title} {\bibinfo {title} {{On the absence of magic in type II factors}}} (\bibinfo {year} {2025}),\ \bibinfo {note} {in preparation}\BibitemShut {NoStop}%
\bibitem [{\citenamefont {Pfeifer}\ \emph {et~al.}(2009{\natexlab{b}})\citenamefont {Pfeifer}, \citenamefont {Evenbly},\ and\ \citenamefont {Vidal}}]{Pfeifer:2009criticalMERA}%
  \BibitemOpen
  \bibfield  {author} {\bibinfo {author} {\bibfnamefont {R.~N.~C.}\ \bibnamefont {Pfeifer}}, \bibinfo {author} {\bibfnamefont {G.}~\bibnamefont {Evenbly}},\ and\ \bibinfo {author} {\bibfnamefont {G.}~\bibnamefont {Vidal}},\ }\bibfield  {title} {\bibinfo {title} {Entanglement renormalization, scale invariance, and quantum criticality},\ }\href {https://doi.org/10.1103/PhysRevA.79.040301} {\bibfield  {journal} {\bibinfo  {journal} {Phys. Rev. A}\ }\textbf {\bibinfo {volume} {79}},\ \bibinfo {pages} {040301(R)} (\bibinfo {year} {2009}{\natexlab{b}})},\ \Eprint {https://arxiv.org/abs/0810.0580} {arXiv:0810.0580 [cond-mat.str-el]} \BibitemShut {NoStop}%
\bibitem [{\citenamefont {Zou}\ \emph {et~al.}(2020)\citenamefont {Zou}, \citenamefont {Milsted},\ and\ \citenamefont {Vidal}}]{Zou:2019dnc}%
  \BibitemOpen
  \bibfield  {author} {\bibinfo {author} {\bibfnamefont {Y.}~\bibnamefont {Zou}}, \bibinfo {author} {\bibfnamefont {A.}~\bibnamefont {Milsted}},\ and\ \bibinfo {author} {\bibfnamefont {G.}~\bibnamefont {Vidal}},\ }\bibfield  {title} {\bibinfo {title} {{Conformal fields and operator product expansion in critical quantum spin chains}},\ }\href {https://doi.org/10.1103/PhysRevLett.124.040604} {\bibfield  {journal} {\bibinfo  {journal} {Phys. Rev. Lett.}\ }\textbf {\bibinfo {volume} {124}},\ \bibinfo {pages} {040604} (\bibinfo {year} {2020})},\ \Eprint {https://arxiv.org/abs/1901.06439} {arXiv:1901.06439 [cond-mat.str-el]} \BibitemShut {NoStop}%
\bibitem [{\citenamefont {Milsted}\ and\ \citenamefont {Vidal}(2018)}]{Milsted:2018san}%
  \BibitemOpen
  \bibfield  {author} {\bibinfo {author} {\bibfnamefont {A.}~\bibnamefont {Milsted}}\ and\ \bibinfo {author} {\bibfnamefont {G.}~\bibnamefont {Vidal}},\ }\href@noop {} {\bibinfo {title} {{Geometric interpretation of the multi-scale entanglement renormalization ansatz}}} (\bibinfo {year} {2018}),\ \Eprint {https://arxiv.org/abs/1812.00529} {arXiv:1812.00529 [hep-th]} \BibitemShut {NoStop}%
\bibitem [{\citenamefont {Fredenhagen}(1985)}]{Fredenhagen:1984dc}%
  \BibitemOpen
  \bibfield  {author} {\bibinfo {author} {\bibfnamefont {K.}~\bibnamefont {Fredenhagen}},\ }\bibfield  {title} {\bibinfo {title} {{On the Modular Structure of Local Algebras of Observables}},\ }\href {https://doi.org/10.1007/BF01206179} {\bibfield  {journal} {\bibinfo  {journal} {Commun. Math. Phys.}\ }\textbf {\bibinfo {volume} {97}},\ \bibinfo {pages} {79} (\bibinfo {year} {1985})}\BibitemShut {NoStop}%
\bibitem [{\citenamefont {Witten}(2018)}]{Witten:2018zxz}%
  \BibitemOpen
  \bibfield  {author} {\bibinfo {author} {\bibfnamefont {E.}~\bibnamefont {Witten}},\ }\bibfield  {title} {\bibinfo {title} {{APS Medal for Exceptional Achievement in Research: Invited article on entanglement properties of quantum field theory}},\ }\href {https://doi.org/10.1103/RevModPhys.90.045003} {\bibfield  {journal} {\bibinfo  {journal} {Rev. Mod. Phys.}\ }\textbf {\bibinfo {volume} {90}},\ \bibinfo {pages} {045003} (\bibinfo {year} {2018})},\ \Eprint {https://arxiv.org/abs/1803.04993} {arXiv:1803.04993 [hep-th]} \BibitemShut {NoStop}%
\bibitem [{\citenamefont {Vidal}(2008)}]{Vidal:2008zz}%
  \BibitemOpen
  \bibfield  {author} {\bibinfo {author} {\bibfnamefont {G.}~\bibnamefont {Vidal}},\ }\bibfield  {title} {\bibinfo {title} {{Class of Quantum Many-Body States That Can Be Efficiently Simulated}},\ }\href {https://doi.org/10.1103/PhysRevLett.101.110501} {\bibfield  {journal} {\bibinfo  {journal} {Phys. Rev. Lett.}\ }\textbf {\bibinfo {volume} {101}},\ \bibinfo {pages} {110501} (\bibinfo {year} {2008})},\ \Eprint {https://arxiv.org/abs/quant-ph/0610099} {arXiv:quant-ph/0610099} \BibitemShut {NoStop}%
\bibitem [{\citenamefont {Cheng}\ \emph {et~al.}(2024)\citenamefont {Cheng}, \citenamefont {Lancien}, \citenamefont {Penington}, \citenamefont {Walter},\ and\ \citenamefont {Witteveen}}]{Cheng:2022ori}%
  \BibitemOpen
  \bibfield  {author} {\bibinfo {author} {\bibfnamefont {N.}~\bibnamefont {Cheng}}, \bibinfo {author} {\bibfnamefont {C.}~\bibnamefont {Lancien}}, \bibinfo {author} {\bibfnamefont {G.}~\bibnamefont {Penington}}, \bibinfo {author} {\bibfnamefont {M.}~\bibnamefont {Walter}},\ and\ \bibinfo {author} {\bibfnamefont {F.}~\bibnamefont {Witteveen}},\ }\bibfield  {title} {\bibinfo {title} {{Random Tensor Networks with Non-trivial Links}},\ }\href {https://doi.org/10.1007/s00023-023-01358-2} {\bibfield  {journal} {\bibinfo  {journal} {Annales Henri Poincare}\ }\textbf {\bibinfo {volume} {25}},\ \bibinfo {pages} {2107} (\bibinfo {year} {2024})},\ \Eprint {https://arxiv.org/abs/2206.10482} {arXiv:2206.10482 [quant-ph]} \BibitemShut {NoStop}%
\bibitem [{\citenamefont {Hayden}\ \emph {et~al.}(2016)\citenamefont {Hayden}, \citenamefont {Nezami}, \citenamefont {Qi}, \citenamefont {Thomas}, \citenamefont {Walter},\ and\ \citenamefont {Yang}}]{Hayden:2016cfa}%
  \BibitemOpen
  \bibfield  {author} {\bibinfo {author} {\bibfnamefont {P.}~\bibnamefont {Hayden}}, \bibinfo {author} {\bibfnamefont {S.}~\bibnamefont {Nezami}}, \bibinfo {author} {\bibfnamefont {X.-L.}\ \bibnamefont {Qi}}, \bibinfo {author} {\bibfnamefont {N.}~\bibnamefont {Thomas}}, \bibinfo {author} {\bibfnamefont {M.}~\bibnamefont {Walter}},\ and\ \bibinfo {author} {\bibfnamefont {Z.}~\bibnamefont {Yang}},\ }\bibfield  {title} {\bibinfo {title} {{Holographic duality from random tensor networks}},\ }\href {https://doi.org/10.1007/JHEP11(2016)009} {\bibfield  {journal} {\bibinfo  {journal} {JHEP}\ }\textbf {\bibinfo {volume} {11}}\bibfield  {number} {\bibinfo  {number} { (009)}},\ }\Eprint {https://arxiv.org/abs/1601.01694} {arXiv:1601.01694 [hep-th]} \BibitemShut {NoStop}%
\bibitem [{\citenamefont {van Luijk}\ \emph {et~al.}(2024{\natexlab{b}})\citenamefont {van Luijk}, \citenamefont {Stottmeister}, \citenamefont {Werner},\ and\ \citenamefont {Wilming}}]{vanLuijk:2024nnx}%
  \BibitemOpen
  \bibfield  {author} {\bibinfo {author} {\bibfnamefont {L.}~\bibnamefont {van Luijk}}, \bibinfo {author} {\bibfnamefont {A.}~\bibnamefont {Stottmeister}}, \bibinfo {author} {\bibfnamefont {R.~F.}\ \bibnamefont {Werner}},\ and\ \bibinfo {author} {\bibfnamefont {H.}~\bibnamefont {Wilming}},\ }\href@noop {} {\bibinfo {title} {{Embezzlement of entanglement, quantum fields, and the classification of von Neumann algebras}}} (\bibinfo {year} {2024}{\natexlab{b}}),\ \Eprint {https://arxiv.org/abs/2401.07299} {arXiv:2401.07299 [math-ph]} \BibitemShut {NoStop}%
\bibitem [{\citenamefont {van Luijk}\ \emph {et~al.}(2025)\citenamefont {van Luijk}, \citenamefont {Stottmeister},\ and\ \citenamefont {Wilming}}]{vanLuijk:2025ufz}%
  \BibitemOpen
  \bibfield  {author} {\bibinfo {author} {\bibfnamefont {L.}~\bibnamefont {van Luijk}}, \bibinfo {author} {\bibfnamefont {A.}~\bibnamefont {Stottmeister}},\ and\ \bibinfo {author} {\bibfnamefont {H.}~\bibnamefont {Wilming}},\ }\href@noop {} {\bibinfo {title} {{The Large-Scale Structure of Entanglement in Quantum Many-body Systems}}} (\bibinfo {year} {2025}),\ \Eprint {https://arxiv.org/abs/2503.03833} {arXiv:2503.03833 [quant-ph]} \BibitemShut {NoStop}%
\bibitem [{\citenamefont {Dong}\ \emph {et~al.}(2019)\citenamefont {Dong}, \citenamefont {Harlow},\ and\ \citenamefont {Marolf}}]{Dong:2018seb}%
  \BibitemOpen
  \bibfield  {author} {\bibinfo {author} {\bibfnamefont {X.}~\bibnamefont {Dong}}, \bibinfo {author} {\bibfnamefont {D.}~\bibnamefont {Harlow}},\ and\ \bibinfo {author} {\bibfnamefont {D.}~\bibnamefont {Marolf}},\ }\bibfield  {title} {\bibinfo {title} {{Flat entanglement spectra in fixed-area states of quantum gravity}},\ }\href {https://doi.org/10.1007/JHEP10(2019)240} {\bibfield  {journal} {\bibinfo  {journal} {JHEP}\ }\textbf {\bibinfo {volume} {10}}\bibfield  {number} {\bibinfo  {number} { (240)}},\ }\Eprint {https://arxiv.org/abs/1811.05382} {arXiv:1811.05382 [hep-th]} \BibitemShut {NoStop}%
\bibitem [{\citenamefont {Soni}(2024)}]{Soni:2023fke}%
  \BibitemOpen
  \bibfield  {author} {\bibinfo {author} {\bibfnamefont {R.~M.}\ \bibnamefont {Soni}},\ }\bibfield  {title} {\bibinfo {title} {{A type I approximation of the crossed product}},\ }\href {https://doi.org/10.1007/JHEP01(2024)123} {\bibfield  {journal} {\bibinfo  {journal} {JHEP}\ }\textbf {\bibinfo {volume} {01}}\bibfield  {number} {\bibinfo  {number} { (123)}},\ }\Eprint {https://arxiv.org/abs/2307.12481} {arXiv:2307.12481 [hep-th]} \BibitemShut {NoStop}%
\bibitem [{\citenamefont {Donnelly}(2012)}]{Donnelly:2011hn}%
  \BibitemOpen
  \bibfield  {author} {\bibinfo {author} {\bibfnamefont {W.}~\bibnamefont {Donnelly}},\ }\bibfield  {title} {\bibinfo {title} {{Decomposition of entanglement entropy in lattice gauge theory}},\ }\href {https://doi.org/10.1103/PhysRevD.85.085004} {\bibfield  {journal} {\bibinfo  {journal} {Phys. Rev. D}\ }\textbf {\bibinfo {volume} {85}},\ \bibinfo {pages} {085004} (\bibinfo {year} {2012})},\ \Eprint {https://arxiv.org/abs/1109.0036} {arXiv:1109.0036 [hep-th]} \BibitemShut {NoStop}%
\bibitem [{\citenamefont {Donnelly}(2014)}]{Donnelly:2014gva}%
  \BibitemOpen
  \bibfield  {author} {\bibinfo {author} {\bibfnamefont {W.}~\bibnamefont {Donnelly}},\ }\bibfield  {title} {\bibinfo {title} {{Entanglement entropy and nonabelian gauge symmetry}},\ }\href {https://doi.org/10.1088/0264-9381/31/21/214003} {\bibfield  {journal} {\bibinfo  {journal} {Class. Quant. Grav.}\ }\textbf {\bibinfo {volume} {31}},\ \bibinfo {pages} {214003} (\bibinfo {year} {2014})},\ \Eprint {https://arxiv.org/abs/1406.7304} {arXiv:1406.7304 [hep-th]} \BibitemShut {NoStop}%
\bibitem [{\citenamefont {Aoki}\ \emph {et~al.}(2015)\citenamefont {Aoki}, \citenamefont {Iritani}, \citenamefont {Nozaki}, \citenamefont {Numasawa}, \citenamefont {Shiba},\ and\ \citenamefont {Tasaki}}]{Aoki:2015bsa}%
  \BibitemOpen
  \bibfield  {author} {\bibinfo {author} {\bibfnamefont {S.}~\bibnamefont {Aoki}}, \bibinfo {author} {\bibfnamefont {T.}~\bibnamefont {Iritani}}, \bibinfo {author} {\bibfnamefont {M.}~\bibnamefont {Nozaki}}, \bibinfo {author} {\bibfnamefont {T.}~\bibnamefont {Numasawa}}, \bibinfo {author} {\bibfnamefont {N.}~\bibnamefont {Shiba}},\ and\ \bibinfo {author} {\bibfnamefont {H.}~\bibnamefont {Tasaki}},\ }\bibfield  {title} {\bibinfo {title} {{On the definition of entanglement entropy in lattice gauge theories}},\ }\href {https://doi.org/10.1007/JHEP06(2015)187} {\bibfield  {journal} {\bibinfo  {journal} {JHEP}\ }\textbf {\bibinfo {volume} {06}}\bibfield  {number} {\bibinfo  {number} { (187)}},\ }\Eprint {https://arxiv.org/abs/1502.04267} {arXiv:1502.04267 [hep-th]} \BibitemShut {NoStop}%
\bibitem [{\citenamefont {Witten}(2022)}]{Witten:2021unn}%
  \BibitemOpen
  \bibfield  {author} {\bibinfo {author} {\bibfnamefont {E.}~\bibnamefont {Witten}},\ }\bibfield  {title} {\bibinfo {title} {{Gravity and the crossed product}},\ }\href {https://doi.org/10.1007/JHEP10(2022)008} {\bibfield  {journal} {\bibinfo  {journal} {JHEP}\ }\textbf {\bibinfo {volume} {10}}\bibfield  {number} {\bibinfo  {number} { (008)}},\ }\Eprint {https://arxiv.org/abs/2112.12828} {arXiv:2112.12828 [hep-th]} \BibitemShut {NoStop}%
\bibitem [{\citenamefont {Chandrasekaran}\ \emph {et~al.}(2023{\natexlab{a}})\citenamefont {Chandrasekaran}, \citenamefont {Longo}, \citenamefont {Penington},\ and\ \citenamefont {Witten}}]{Chandrasekaran:2022cip}%
  \BibitemOpen
  \bibfield  {author} {\bibinfo {author} {\bibfnamefont {V.}~\bibnamefont {Chandrasekaran}}, \bibinfo {author} {\bibfnamefont {R.}~\bibnamefont {Longo}}, \bibinfo {author} {\bibfnamefont {G.}~\bibnamefont {Penington}},\ and\ \bibinfo {author} {\bibfnamefont {E.}~\bibnamefont {Witten}},\ }\bibfield  {title} {\bibinfo {title} {{An algebra of observables for de Sitter space}},\ }\href {https://doi.org/10.1007/JHEP02(2023)082} {\bibfield  {journal} {\bibinfo  {journal} {JHEP}\ }\textbf {\bibinfo {volume} {02}}\bibfield  {number} {\bibinfo  {number} { (082)}},\ }\Eprint {https://arxiv.org/abs/2206.10780} {arXiv:2206.10780 [hep-th]} \BibitemShut {NoStop}%
\bibitem [{\citenamefont {Chandrasekaran}\ \emph {et~al.}(2023{\natexlab{b}})\citenamefont {Chandrasekaran}, \citenamefont {Penington},\ and\ \citenamefont {Witten}}]{Chandrasekaran:2022eqq}%
  \BibitemOpen
  \bibfield  {author} {\bibinfo {author} {\bibfnamefont {V.}~\bibnamefont {Chandrasekaran}}, \bibinfo {author} {\bibfnamefont {G.}~\bibnamefont {Penington}},\ and\ \bibinfo {author} {\bibfnamefont {E.}~\bibnamefont {Witten}},\ }\bibfield  {title} {\bibinfo {title} {{Large N algebras and generalized entropy}},\ }\href {https://doi.org/10.1007/JHEP04(2023)009} {\bibfield  {journal} {\bibinfo  {journal} {JHEP}\ }\textbf {\bibinfo {volume} {04}}\bibfield  {number} {\bibinfo  {number} { (009)}},\ }\Eprint {https://arxiv.org/abs/2209.10454} {arXiv:2209.10454 [hep-th]} \BibitemShut {NoStop}%
\bibitem [{\citenamefont {Jensen}\ \emph {et~al.}(2023)\citenamefont {Jensen}, \citenamefont {Sorce},\ and\ \citenamefont {Speranza}}]{Jensen:2023yxy}%
  \BibitemOpen
  \bibfield  {author} {\bibinfo {author} {\bibfnamefont {K.}~\bibnamefont {Jensen}}, \bibinfo {author} {\bibfnamefont {J.}~\bibnamefont {Sorce}},\ and\ \bibinfo {author} {\bibfnamefont {A.}~\bibnamefont {Speranza}},\ }\href@noop {} {\bibinfo {title} {{Generalized entropy for general subregions in quantum gravity}}} (\bibinfo {year} {2023}),\ \Eprint {https://arxiv.org/abs/2306.01837} {arXiv:2306.01837 [hep-th]} \BibitemShut {NoStop}%
\bibitem [{\citenamefont {Klinger}\ and\ \citenamefont {Leigh}(2024)}]{Klinger:2023tgi}%
  \BibitemOpen
  \bibfield  {author} {\bibinfo {author} {\bibfnamefont {M.~S.}\ \bibnamefont {Klinger}}\ and\ \bibinfo {author} {\bibfnamefont {R.~G.}\ \bibnamefont {Leigh}},\ }\bibfield  {title} {\bibinfo {title} {{Crossed products, extended phase spaces and the resolution of entanglement singularities}},\ }\href {https://doi.org/10.1016/j.nuclphysb.2024.116453} {\bibfield  {journal} {\bibinfo  {journal} {Nucl. Phys. B}\ }\textbf {\bibinfo {volume} {999}},\ \bibinfo {pages} {116453} (\bibinfo {year} {2024})},\ \Eprint {https://arxiv.org/abs/2306.09314} {arXiv:2306.09314 [hep-th]} \BibitemShut {NoStop}%
\bibitem [{\citenamefont {Kudler-Flam}\ \emph {et~al.}(2023{\natexlab{a}})\citenamefont {Kudler-Flam}, \citenamefont {Leutheusser},\ and\ \citenamefont {Satishchandran}}]{Kudler-Flam:2023qfl}%
  \BibitemOpen
  \bibfield  {author} {\bibinfo {author} {\bibfnamefont {J.}~\bibnamefont {Kudler-Flam}}, \bibinfo {author} {\bibfnamefont {S.}~\bibnamefont {Leutheusser}},\ and\ \bibinfo {author} {\bibfnamefont {G.}~\bibnamefont {Satishchandran}},\ }\href@noop {} {\bibinfo {title} {{Generalized Black Hole Entropy is von Neumann Entropy}}} (\bibinfo {year} {2023}{\natexlab{a}}),\ \Eprint {https://arxiv.org/abs/2309.15897} {arXiv:2309.15897 [hep-th]} \BibitemShut {NoStop}%
\bibitem [{\citenamefont {Chen}\ and\ \citenamefont {Penington}(2024)}]{Chen:2024rpx}%
  \BibitemOpen
  \bibfield  {author} {\bibinfo {author} {\bibfnamefont {C.-H.}\ \bibnamefont {Chen}}\ and\ \bibinfo {author} {\bibfnamefont {G.}~\bibnamefont {Penington}},\ }\href@noop {} {\bibinfo {title} {{A clock is just a way to tell the time: gravitational algebras in cosmological spacetimes}}} (\bibinfo {year} {2024}),\ \Eprint {https://arxiv.org/abs/2406.02116} {arXiv:2406.02116 [hep-th]} \BibitemShut {NoStop}%
\bibitem [{\citenamefont {Gomez}(2022)}]{Gomez:2022eui}%
  \BibitemOpen
  \bibfield  {author} {\bibinfo {author} {\bibfnamefont {C.}~\bibnamefont {Gomez}},\ }\href@noop {} {\bibinfo {title} {{Cosmology as a Crossed Product}}} (\bibinfo {year} {2022}),\ \Eprint {https://arxiv.org/abs/2207.06704} {arXiv:2207.06704 [hep-th]} \BibitemShut {NoStop}%
\bibitem [{\citenamefont {Kudler-Flam}\ \emph {et~al.}(2023{\natexlab{b}})\citenamefont {Kudler-Flam}, \citenamefont {Leutheusser}, \citenamefont {Rahman}, \citenamefont {Satishchandran},\ and\ \citenamefont {Speranza}}]{Kudler-Flam:2023hkl}%
  \BibitemOpen
  \bibfield  {author} {\bibinfo {author} {\bibfnamefont {J.}~\bibnamefont {Kudler-Flam}}, \bibinfo {author} {\bibfnamefont {S.}~\bibnamefont {Leutheusser}}, \bibinfo {author} {\bibfnamefont {A.~A.}\ \bibnamefont {Rahman}}, \bibinfo {author} {\bibfnamefont {G.}~\bibnamefont {Satishchandran}},\ and\ \bibinfo {author} {\bibfnamefont {A.~J.}\ \bibnamefont {Speranza}},\ }\href@noop {} {\bibinfo {title} {A covariant regulator for entanglement entropy: proofs of the bekenstein bound and qnec}} (\bibinfo {year} {2023}{\natexlab{b}}),\ \Eprint {https://arxiv.org/abs/2312.07646} {arXiv:2312.07646 [hep-th]} \BibitemShut {NoStop}%
\bibitem [{\citenamefont {De~Vuyst}\ \emph {et~al.}(2024)\citenamefont {De~Vuyst}, \citenamefont {Eccles}, \citenamefont {Hoehn},\ and\ \citenamefont {Kirklin}}]{DeVuyst:2024uvd}%
  \BibitemOpen
  \bibfield  {author} {\bibinfo {author} {\bibfnamefont {J.}~\bibnamefont {De~Vuyst}}, \bibinfo {author} {\bibfnamefont {S.}~\bibnamefont {Eccles}}, \bibinfo {author} {\bibfnamefont {P.~A.}\ \bibnamefont {Hoehn}},\ and\ \bibinfo {author} {\bibfnamefont {J.}~\bibnamefont {Kirklin}},\ }\href@noop {} {\bibinfo {title} {{Crossed products and quantum reference frames: on the observer-dependence of gravitational entropy}}} (\bibinfo {year} {2024}),\ \Eprint {https://arxiv.org/abs/2412.15502} {arXiv:2412.15502 [hep-th]} \BibitemShut {NoStop}%
\bibitem [{\citenamefont {Ali~Ahmad}\ and\ \citenamefont {Klinger}(2024)}]{AliAhmad:2024saq}%
  \BibitemOpen
  \bibfield  {author} {\bibinfo {author} {\bibfnamefont {S.}~\bibnamefont {Ali~Ahmad}}\ and\ \bibinfo {author} {\bibfnamefont {M.~S.}\ \bibnamefont {Klinger}},\ }\href@noop {} {\bibinfo {title} {{Emergent Geometry from Quantum Probability}}} (\bibinfo {year} {2024}),\ \Eprint {https://arxiv.org/abs/2411.07288} {arXiv:2411.07288 [hep-th]} \BibitemShut {NoStop}%
\bibitem [{\citenamefont {Leutheusser}(2023)}]{Leutheusser:2021frk}%
  \BibitemOpen
  \bibfield  {author} {\bibinfo {author} {\bibfnamefont {S.~A.~W.}\ \bibnamefont {Leutheusser}},\ }\bibfield  {title} {\bibinfo {title} {{Emergent Times in Holographic Duality}},\ }\href@noop {} {\bibfield  {journal} {\bibinfo  {journal} {Phys. Rev. D}\ }\textbf {\bibinfo {volume} {108}},\ \bibinfo {pages} {086020} (\bibinfo {year} {2023})},\ \Eprint {https://arxiv.org/abs/2112.12156} {arXiv:2112.12156 [hep-th]} \BibitemShut {NoStop}%
\bibitem [{\citenamefont {Haco}\ \emph {et~al.}(2018)\citenamefont {Haco}, \citenamefont {Hawking}, \citenamefont {Perry},\ and\ \citenamefont {Strominger}}]{Haco:2018ske}%
  \BibitemOpen
  \bibfield  {author} {\bibinfo {author} {\bibfnamefont {S.}~\bibnamefont {Haco}}, \bibinfo {author} {\bibfnamefont {S.~W.}\ \bibnamefont {Hawking}}, \bibinfo {author} {\bibfnamefont {M.~J.}\ \bibnamefont {Perry}},\ and\ \bibinfo {author} {\bibfnamefont {A.}~\bibnamefont {Strominger}},\ }\bibfield  {title} {\bibinfo {title} {{Black Hole Entropy and Soft Hair}},\ }\href {https://doi.org/10.1007/JHEP12(2018)098} {\bibfield  {journal} {\bibinfo  {journal} {JHEP}\ }\textbf {\bibinfo {volume} {12}}\bibfield  {number} {\bibinfo  {number} { (098)}},\ }\Eprint {https://arxiv.org/abs/1810.01847} {arXiv:1810.01847 [hep-th]} \BibitemShut {NoStop}%
\bibitem [{\citenamefont {von Neumann}(1939)}]{vonNeumann1939}%
  \BibitemOpen
  \bibfield  {author} {\bibinfo {author} {\bibfnamefont {J.}~\bibnamefont {von Neumann}},\ }\bibfield  {title} {\bibinfo {title} {On infinite direct products},\ }\href@noop {} {\bibfield  {journal} {\bibinfo  {journal} {Compositio Mathematica}\ }\textbf {\bibinfo {volume} {6}},\ \bibinfo {pages} {1} (\bibinfo {year} {1939})}\BibitemShut {NoStop}%
\bibitem [{\citenamefont {Jordan}\ and\ \citenamefont {Wigner}(1928)}]{jordan1928}%
  \BibitemOpen
  \bibfield  {author} {\bibinfo {author} {\bibfnamefont {P.}~\bibnamefont {Jordan}}\ and\ \bibinfo {author} {\bibfnamefont {E.}~\bibnamefont {Wigner}},\ }\bibfield  {title} {\bibinfo {title} {{{\"U}ber das Paulische {\"A}quivalenzverbot}},\ }\href {https://doi.org/10.1007/BF01331938} {\bibfield  {journal} {\bibinfo  {journal} {Zeitschrift f\"ur Physik}\ }\textbf {\bibinfo {volume} {47}},\ \bibinfo {pages} {631} (\bibinfo {year} {1928})}\BibitemShut {NoStop}%
\end{thebibliography}%
\end{document}